\documentclass[usenatbib,usegraphicx]{mn2e}
\usepackage{amssymb}
\usepackage{epsfig}
\usepackage{epsf}
\usepackage{amsmath}
\usepackage{natbib}
\usepackage{verbatim}

\def\Lya{Ly$\alpha\ $}
\def\Lyb{Ly$\beta\ $}

\def\HI{\hbox{H~$\scriptstyle\rm I\ $}}

\def\ltsima{$\; \buildrel < \over \sim \;$}
\def\lsim{\lower.5ex\hbox{\ltsima}}
\def\gtsima{$\; \buildrel > \over \sim \;$}
\def\gsim{\lower.5ex\hbox{\gtsima}}

\def\spose#1{\hbox to 0pt{#1\hss}}
\def\lta{\mathrel{\spose{\lower 3pt\hbox{$\mathchar"218$}}
     \raise 2.0pt\hbox{$\mathchar"13C$}}}
\def\gta{\mathrel{\spose{\lower 3pt\hbox{$\mathchar"218$}}
     \raise 2.0pt\hbox{$\mathchar"13E$}}}

\newcommand{\dd}{\mathrm{d}}   %% Does a differential `d' correctly (use as \dd T rater that dT)

\journal{Preprint-00}

\title{The scattering of \Lya radiation in the intergalactic medium:\
  numerical methods and solutions}

\author[J. Higgins, A. Meiksin]{Jonathan Higgins, Avery Meiksin \\
SUPA\thanks{Scottish Universities Physics Alliance},
Institute for Astronomy, University of Edinburgh,
Blackford Hill, Edinburgh\ EH9\ 3HJ, UK}

\begin{document}

\maketitle
\begin{abstract}
  Two methods are developed for solving the steady-state spherically
  symmetric radiative transfer equation for resonance line radiation
  emitted by a point source in the Intergalactic Medium, in the
  context of the Wouthuysen-Field mechanism for coupling the hyperfine
  structure spin temperature of hydrogen to the gas temperature. One
  method is based on solving the ray and moment equations using finite
  differences. The second uses a Monte Carlo approach incorporating
  methods that greatly improve the accuracy compared with previous
  approaches in this context. Several applications are presented
  serving as test problems for both a static medium and an expanding
  medium, including inhomogeneities in the density and velocity
  fields. Solutions are obtained in the coherent scattering limit and
  for Doppler RII redistribution with and without recoils. We find
  generally that the radiation intensity is linear in the cosine of
  the azimuthal angle with respect to radius to high accuracy over a
  broad frequency region across the line centre for both linear and
  perturbed velocity fields, yielding the Eddington factors
  $f_\nu\simeq1/3$ and $g_\nu\simeq3/5$. The radiation field produced
  by a point source divides into three spatial regimes for a uniformly
  expanding homogeneous medium. The regimes are governed by the
  fraction of the distance $r$ from the source in terms of the
  distance $r_*$ required for a photon to redshift from line centre to
  the frequency needed to escape from the expanding gas. For a
  standard cosmology, before the Universe was reionized $r_*$ takes on
  the universal value independent of redshift of 1.1~Mpc, depending
  only on the ratio of the baryon to dark matter density. At
  $r/r_*<1$, the radiation field is accurately described in the
  diffusion approximation, with the scattering rate declining with the
  distance from the source as $r^{-7/3}$, except at $r/r_*\ll1$ where
  frequency redistribution nearly doubles the mean intensity around
  line centre. At $r/r_*>1$, the diffusion approximation breaks down
  and the decline of the mean intensity near line centre and the
  scattering rate approach the geometric dilution scaling $1/r^2$. The
  mean intensity and scattering rate are found to be very sensitive to
  the gradient of the velocity field, growing exponentially with the
  amplitude of the perturbation as the limit of a vanishing velocity
  gradient is approached near the source. We expect the 21cm signal
  from the Epoch of Reionization to thus be a sensitive probe of both
  the density and the peculiar velocity fields.

  The solutions for the mean intensity are made available in
  machine-readable format.
\end{abstract}


\section{Introduction}
\label{sect:intro}

The nature of formation of the first radiating objects in the Universe
is one of the paramount unsolved problems in cosmological structure
formation. Searches for the earliest galaxies have broken the
spectroscopically confirmed redshift barrier of $z=7$
\citep{2011ApJ...730L..35V, 2012ApJ...744...83O, 2012ApJ...744..179S},
with plausible candidates identified photometrically up to $z\lta9$
\citep{2011MNRAS.418.2074M}, and possibly as high as $z\simeq10$
\citep{2011Natur.469..504B}. These systems may well reside within the
Epoch of Reionization (EoR) of hydrogen, as Cosmic Microwave
Background (CMB) measurements suggest the epoch, if a sudden event,
occurred at $z_r=10.4\pm1.2$ ($1\sigma$)
\citep{2011ApJS..192...18K}. If so, 21cm emission from the diffuse
Intergalactic Medium (IGM) would become visible near this same epoch,
as the same sources would provide sufficient UV continuum to excite
the line through the Wouythuysen-Field effect (WFE)
\citep{1952AJ.....57R..31W, 1958PROCIRE.46..240F, MMR97}.

The prospect of discovering the EoR through the associated 21cm
signature from the diffuse IGM has inspired the development of a new
generation of radio telescopes, such as the LOw Frequency Array
(LOFAR)\footnote{www.lofar.org}, upgrades to the Giant Metrewave Radio
Telescope (GMRT)\footnote{gmrt.ncra.tifr.res.in}, the Murchison
Widefield Array (MWA) \footnote{www.haystack.mit.edu/ast/arrays/mwa},
the Primeval Structure Telescope/21 Centimeter Array (PaST/21CMA)
\footnote{web.phys.cmu.edu/$\sim$past}, the Precision Array to Probe
EoR (PAPER)\footnote{astro.berkeley.edu/$\sim$dbacker/eor}, and a
possible Square Kilometre Array (SKA)\footnote{www.skatelescope.org}.
A recent review of this rapidly growing area is provided by
\citet{2011arXiv1109.6012P}.

The interpretation of the signal will require modelling the radiative
transfer of the Lyman resonance line photons, primarily Ly$\alpha$,
that drive the WFE. Most estimates have presumed a homogeneous
expanding medium. Early modelling neglected the effects of atomic
recoil and of spatial diffusion about the emitting sources, assuming
the sources were homogeneously and isotropically distributed
throughout the Universe \citep{1959ApJ...129..536F, MMR97}. Allowing
for recoil somewhat suppresses the \Lya photon scattering rate by an
amount depending on the local temperature and expansion rate of the
IGM \citep{2004ApJ...602....1C, 2006MNRAS.372.1093F}. Using Monte
Carlo solutions to the radiative transfer equation to include spatial
diffusion, it was found that near an emitting source the scattering
rate varies as $r^{-7/3}$ \citep{2007ApJ...670..912C,
  2007A&A...474..365S}, more rapidly than the geometric dilution
factor $r^{-2}$ predicted without spatial diffusion. It will be shown
below that the steeper dependence arises generally over all distances
for which the radiative transfer of the \Lya photons may be treated in
the diffusion approximation.

In reality the IGM is clumpy, with structures breaking away from the
cosmological expansion. Simple analytic estimates suggest that the
scattering rate will be substantially modified not only by density and
temperature fluctuations, but by gradients in the velocity field of
the gas around individual sources \citep{2009MNRAS.393..949H}. More
sophisticated methods are required for accurate solutions of the
spatial and frequency dependent radiative transfer equation. Monte
Carlo codes were developed for this purpose
\citep{2002ApJ...578...33Z, 2006ApJ...645..792T, 2007ApJ...670..912C,
  2007A&A...474..365S}. The method has recently been applied to
estimate the expected cosmological 21cm signal
\citep{2009A&A...495..389B, 2011A&A...532A..97V}. A difficulty with
the Monte Carlo technique is the limited resolution imposed by the
restricted number of photon packets that may be practically
followed. An alternative grid-based method for solving the
spherically-symmetric radiative transfer equation in the Eddington
approximation has been developed by \citet{2009ApJ...703.1992R} for
linear flow fields.

Scattering in the deuterium \Lya resonance and the addition of \Lya
photons produced in radiative cascades following the scattering of
higher order Lyman resonance line photons will modify the \Lya mean
intensity and scattering rate near a source. As these are smaller,
secondary effects \citep{2007ApJ...670..912C, 2007A&A...474..365S},
they are not included in this paper.

A further complication is the time required to establish a
steady-state radiation field. The \Lya photon scattering rate scales
like $t_s\simeq3.2 n_{\rm HI}^{-1}T^{1/2}\,{\rm s}$ for a gas with
neutral hydrogen density $n_{\rm HI}$ and temperature
$T$. Time-dependent radiative transfer computations suggest it may
take $10^4-10^{12}$ scatterings to establish a steady-state radiation
field, depending on the internal structure of the scattering system,
including internal velocity gradients \citep{2009MNRAS.393..949H,
  2009ApJ...694.1121R}. Timescales of $10^6-10^9\,{\rm yr}$,
comparable to or longer than the evolutionary timescale of starbursts
and quasars, may be required for structures that have broken away from
the cosmic expansion, particularly in the presence of substantial
x-ray heating.

In this paper, we present two algorithms for solving the radiative
transfer equation of an inhomogeneous medium, one based on
finite-differences on a grid and the second Monte Carlo based. The
grid-based method combines solutions to the ray equation and the
moment equation based on the method of \citet{1975ApJ...202..465M,
  1976ApJ...210..419M, 1977ApJ...214..337M}. The inclusion of
solutions to the ray equation allows the sequence of moment equations
to be closed without imposing the Eddington approximation. Since the
method was developed for stellar atmospheres, modifications to the
approach are described necessary to adapt the method to the problem of
a source in the IGM. The method has a few restrictions:\ 1.\ it is
implemented assuming spherical symmetry, 2.\ it assumes a steady
state, and 3.\ it requires the velocity field around the source to be
monotonically increasing or decreasing. In practice the latter is not
a severe restriction because the peculiar velocity field only
modulates the Hubble flow except within the turnaround radius very
near the source. The assumption of spherical symmetry may not be very
restrictive either, since the photons do not diffuse very far compared
with the coherence length of cosmological structures. A solution along
a ray in a 3D computation may therefore not differ much from that
assuming the medium is isotropic with the radial properties of the
ray. We are not able to test this, however, without a fully 3D
solution to the radiative transfer equation, which is beyond the scope
of this paper. As noted above, situations may arise in which the
steady-state approximation will break down.

The Monte Carlo method is similar to existing algorithms for resonance
line radiation, but with two improvements. It incorporates RII
frequency redistribution by interopolating on the RII redistribution
function rather than directly computing collisions with atoms. This
improves the speed of the computations by a factor of a few, but at
the cost of requiring a frequency grid tailored to the particular
problem. The second improvement is to compile the specific mean
intensity based on the path lengths traversed by the photon packets
rather than on frequency and position bin crossings. This improvement
is general and substantially reduces the noise in the specific mean
intensity for a fixed number of photon packets. The method also serves
as an independent check on the solutions obtained through the ray and
moments method.

The paper is organised as follows. The basic framework used to solve
the radiative transfer equation in spherical symmetry using the ray
and moments method on a grid is presented in the next section. In
Section~\ref{sect:MC_method}, we summarise the Monte Carlo method
developed. We present the results of validation tests of both methods
against analytic solutions for a homogeneous medium in
Section~\ref{sect:HomogTest}. Both methods are applied to scattering
problems in an inhomogeneous medium in
Section~\ref{sect:InhomogTest}. A summary and conclusions
follow. Details of the source term used for the ray and moment
solution are provided in Appendix~\ref{ap:RMsource}. The Monte Carlo
method is described in Appendix~\ref{ap:MC}. The solution to the
radiative transfer problem for a point continuum source in a uniformly
expanding homogeneous medium in the diffusion approximation is derived
in Appendix~\ref{ap:hem}. Tables of solutions to a test suite of
problems using the ray and moments method are provided online in
machine-readable format, as summarised in Appendix~\ref{ap:solutions}.

\section{Numerical Radiative Transfer in Spherical Symmetry I: Ray and Moment Equations}
\label{sect:rme}

\subsection{Methodology}

\label{subsect:grid_method}

We developed a grid-based method for obtaining precise numerical
solutions to the radiative transfer equation in spherical symmetry
based on the method of \citet{1975ApJ...202..465M,
  1976ApJ...210..419M, 1977ApJ...214..337M}. This method was developed
for stellar atmospheres and consequently we have had to formulate the
boundary conditions for a non-blackbody source. Here we provide a
summary of our implementation.

The radiative transfer equation for the specific intensity $I_\nu$ in
the comoving frame for a spherically symmetric scattering medium with
specific volume emissivity $\eta_\nu$, inverse attenuation length
$\chi_\nu$ and radial velocity $V(r)$ is given by
\citep{1978stat.book.....M}
\begin{equation}
\begin{split}
\mu \frac{\partial I_{\nu}}{\partial r} + \frac{1-\mu^2}{r} \frac{\partial I_{\nu}}{\partial \mu} - \alpha(r) \left[ 1-\mu^2 + \mu^2 \beta(r) \right] \frac{\partial I_{\nu}}{\partial \nu} \\ = -\chi_{\nu}I_{\nu} + \eta_{\nu},\end{split}
\label{eq:sph_comoving}
\end{equation} 
\noindent where $\alpha(r)$ and $\beta(r)$ are given by
\begin{eqnarray}
\alpha(r) &\equiv& (\nu_0/c)\frac{V(r)}{r}, \\
\label{eq:alpha}
\beta(r) &\equiv& \frac{\dd \ln{V(r)}}{\dd \ln{r}},
\label{eq:beta}
\end{eqnarray}
\noindent and $\mu$ is the cosine of the inclination relative to the
radial direction of the radiation field described by $I_\nu(r,\mu)$,
and $\nu_0$ is the resonance line frequency. We limit our study to
expanding velocity fields $V(r)$ satisfying $V(r) > 0$ and $V'(r) > 0$
for all $r$. (The method also may be applied to a monotonically
decreasing radial velocity field. Non-monotonic fields would require
combining solutions from piecewise monotonic fields.)  The method
requires solving a system of angular moment equations derived from
equation~(\ref{eq:sph_comoving}), together with a solution of
equation~(\ref{eq:sph_comoving}) itself along a specific set of
rays. The approach is to solve the ray and moment equations
simultaneously, iterating between the two until adequate convergence
is achieved. In solving the ray equations, we treat the angular
dependence of the radiation field exactly, however we assume that the
source function $S_{\nu}(r)$ is a known quantity. In solving the
moment equations we assume angular information in the form of variable
Eddington factors in order to close the system of equations, however
we are able to treat the dependence of the source function on the
radiation field exactly. The iteration procedure is implemented by
first seeking a solution to the ray equations utilising an estimate of
the source function, before evaluating the variable Eddington factors
from the ray solution and using them to obtain a solution to the
moment equations. It may then be necessary to improve on the estimate
of the Eddington factors by repeating the ray solution steps with an
improved source function estimate derived from the moment solution, a
process that may be iterated until convergence. At each step the
solution of the ray/moment equations is obtained from
finite-difference forms which are written as matrix equations. The
matrix equations are provided by \citet{2012HigginsPhD}. We next
outline the method in greater detail.

\subsection{Ray Equations} 
\label{subsect:reqs}

In spherical symmetry the radiation field may be described as a
function of radius $r$ and the angle $\theta$ to the local radius
vector. A pair of variables more appropriate for describing the
properties of the radiation field along a particular ray is given by
the impact parameter, $p \equiv r \sin{\theta} = r (1-\mu^2)^{1/2}$,
and the distance along the ray from the point of closest approach to
the origin, $z \equiv r \cos{\theta} = r \mu$ where $\mu \equiv
\cos{\theta}$. Radiative transfer along a ray of a given impact
parameter $p$ is described by the equations
\begin{equation}
\begin{split} \pm \frac{\partial I_{\nu}^{\pm}(p,z)}{\partial z} - \alpha(r)
\left[1-\mu^2+\beta(r) \mu^2 \right]\frac{\partial
  I_{\nu}^{\pm}(p,z)}{\partial \nu}\\
 = \eta_{\nu}(r) - \chi_{\nu}(r)
I_{\nu}^{\pm}(p,z)\end{split}
\label{eq:sph_ray_com}
\end{equation}
where $I_{\nu}^{\pm} (p,z)$ is the specific intensity of radiation a
distance $z$ along the ray, in the direction of increasing $z$ for
$I_{\nu}^{+}$ or decreasing $z$ for $I_{\nu}^{-}$. In spherical
symmetry the intensities are equivalently written as $I_{\nu}^{+}(p,z)
= I_{\nu}(r,\mu)$ and $I_{\nu}^{-}(p,z) = I_{\nu}(r,-\mu)$ where $\mu
> 0$. It is convenient to work with the following linear combinations
of the specific intensities:
\begin{eqnarray}
u_{\nu}(p,z) &\equiv& \frac{1}{2}[I_{\nu}^+(p,z) + I_{\nu}^-(p,z)], 
\label{eq:u_def}\\
v_{\nu}(p,z) &\equiv& \frac{1}{2}[I_{\nu}^+(p,z) - I_{\nu}^-(p,z)]. 
\label{eq:v_def}
\end{eqnarray}
We may obtain equations for the development of $u_{\nu}$ and $v_{\nu}$
along a ray by summing and differencing, respectively, the transfer
equations for $I_{\nu}^{+}$ and $I_{\nu}^{-}$. If we then change
variables to the optical depth along the ray, defined by $\dd
\tau_{\nu} = -\chi_{\nu} \, \dd z$, we arrive at
\begin{eqnarray}
\frac{\partial u_{\nu}(z)}{\partial \tau_{\nu}} &+& \gamma_{\nu}(z)
\frac{\partial v_{\nu}(z)}{\partial \nu} = v_{\nu}(z), 
\label{eq:ray_one}\\
\frac{\partial v_{\nu}(z)}{\partial \tau_{\nu}} &+& \gamma_{\nu}(z)
\frac{\partial u_{\nu}(z)}{\partial \nu} = u_{\nu}(z) - S_{\nu}(r),
\label{eq:ray_two}
\end{eqnarray}
where we have defined coefficients
\begin{equation}
\gamma_{\nu}(z) \equiv \frac{\alpha(r)}{\chi_{\nu}(r)} \left[1 - \mu^2 +
  \beta(r) \mu^2 \right],
\end{equation}
and $S_{\nu}(r) \equiv \eta_{\nu}(r)/\chi_{\nu}(r)$ is the source
function of the scattering medium, which is here assumed to be a known
function.

We specify a maximum and minimum radius of the spherical system under
consideration, given respectively by $R$ and $R_{\rm C}$. (The
subscript `C' here denotes the `core', the central object/region of
the spherical system.) The boundary conditions are:

\noindent (a)\ $r = R, z = z_{\rm max}$: We assume that $R$ has been
chosen to be sufficiently large that there is no radiation scattered
back into the system from this radius, i.e. $I_{\nu}(R, \mu) = 0$ for
$\mu < 0$, and thus along the ray $I_{\nu}^-(p, z_{\rm max}) = 0$
where $z_{\rm max} = (R^2 - p^2)^{1/2}$. From the definitions given by
equations~(\ref{eq:u_def}) and (\ref{eq:v_def}) we see that
$v_{\nu}(z_{\rm max}) = u_{\nu}(z_{\rm max})$. We arrive at a suitable
boundary condition by substituting this into
equation~(\ref{eq:ray_one}):
\begin{equation}
\frac{\partial u_{\nu}(z_{\rm max})}{\partial \tau_{\nu}} + \gamma_{\nu}(z_{\rm max})
\frac{\partial u_{\nu}(z_{\rm max})}{\partial \nu} = u_{\nu}(z_{\rm max}). 
\label{eq:ray_zmax}
\end{equation}
(b)\ $r = R_{\rm C}, z = z_{\rm min}$: For the inner boundary
condition we need to consider two distinct cases applicable depending
on the impact parameter of the ray chosen:

\noindent (${\rm i}$) $p \geq R_{\rm C}$: We compute the radiative
transfer along the ray which is considered to originate from $z_{\rm
  min} = 0$. Due to the symmetry of the problem we may impose
$I_{\nu}^+ = I_{\nu}^-$ at $z = 0$ and therefore $v_{\nu}(p, z_{\rm
  min}) = 0$. Thus equation~(\ref{eq:ray_one}) becomes
\begin{equation}
\frac{\partial u_{\nu}(z=0)}{\partial \tau_{\nu}} = 0.
\label{eq:ray_zzero}
\end{equation}
(${\rm ii}$) $p < R_{\rm C}$: The ray intersects the core, and the
minimum value of $z$ denotes the distance along the ray at which the
intersection occurs: $z_{\rm min} = (R_{\rm C}^2 - p^2)^{1/2}$. The
boundary condition is written as:
\begin{equation}
\frac{\partial u_{\nu}(z_{\rm min})}{\partial \tau_{\nu}} = v_{\nu}(z_{\rm min}) - \gamma_{\nu}(z_{\rm min})\frac{\partial v_{\nu}(z_{\rm min})}{\partial \nu}
\label{eq:ray_zmin}
\end{equation}
where we assume the function $v_{\nu}(R_{\rm C}, \mu)$ is known. 

If the optical depth at the core radius is sufficiently high then the
intensity will display the angular behaviour found in the diffusion
limit, $I_{\nu}(R_{\rm C}, \mu) = J_{\nu, {\rm C}} + 3H_{\nu, {\rm
    C}}\mu$, where $J_{\nu, {\rm C}}$ is the core mean intensity and
$H_{\nu, {\rm C}}$ is the core flux. From the definition of $v_{\nu}$
and the equation for $I_{\nu}(R_{\rm C}, \mu)$ we find
\begin{equation}
v_{\nu}(R_{\rm C}, \mu) = 3|\mu|H_{\nu, {\rm C}}. 
\label{eq:v_core_diff}
\end{equation}
In this limit we should take $H_{\nu, {\rm C}}$ to be a solution valid
in the diffusion limit, if such a solution is known. If the optical
depth at the core radius is significantly less than unity then we may
assume a free-streaming boundary condition of the form
\begin{equation}
v_{\nu}(R_{\rm C}, \mu) = H_{\nu, {\rm C}} \delta_{\rm D}(\mu - 1)
\label{eq:v_core_flux}
\end{equation}
where $\delta_{\rm D}$ is the Dirac delta function and thus the radiation field is directed radially outward from the
source ($\mu = 1$). The flux and the mean intensity are equal in this limit and
trivially related to the source spectrum $L_{\nu}$ by $H_{\nu, {\rm
    C}} = J_{\nu, {\rm C}} = L_{\nu}/(4\pi R_{\rm C})^2$.

The medium is subject to a radial velocity $V(r)$ that satisfies $V(r)
\geq 0$, $\dd V/\dd r \geq 0$ for all $r$; this ensures that
$\gamma_{\nu} \geq 0$ for all $(p,z)$. From subsequent analysis of the
characteristics of the system given by equations~(\ref{eq:ray_one})
and (\ref{eq:ray_two}), we need to enforce an initial condition at
high frequency; we denote this frequency by $\nu_{\rm max}$. The
properties of $V(r)$ tell us that every point moves away from every
other point and all radiation intercepted at a specific point from
elsewhere in the system is redshifted. From this argument we see that
if $\nu_{\rm max}$ lies sufficiently blueward of the local line
profile, such that no line photons ever reach $\nu_{\rm max}$, then
\begin{equation}
\left. I_{\nu}(r, \mu)\right |_{\nu_{\rm max}} = 0,
\label{eq:zero_photons} 
\end{equation}
\begin{equation}
\left. \frac{\partial I_{\nu}(r, \mu)}{\partial \nu} \right
|_{\nu_{\rm max}} = 0.
\label{eq:zero_line_photons}
\end{equation} 
Note that equation~(\ref{eq:zero_photons}) holds for a system
interacting with resonance-line photons only, while the condition of
zero frequency gradient given by equation~(\ref{eq:zero_line_photons})
also applies in the more general case where we allow for continuum
radiation as the continuum varies slowly with frequency. The equations
are solved ray-by-ray, frequency-by-frequency using a matrix equation
approach \citep{1975ApJ...202..465M}.

\subsection{Moment Equations}
\label{subsect:meqs}

From equation~(\ref{eq:sph_comoving}) we derive angular moment
equations with respect to $\mu$. We obtain the following for the
zeroth- and first-order moment equations:
\begin{equation}
\begin{split}
\frac{1}{r^2} \frac{\partial (r^2 H_{\nu})}{\partial r} -
\alpha \left[ \frac{\partial (J_{\nu}-K_{\nu})}{\partial \nu} +
  \beta \frac{\partial K_{\nu}}{\partial \nu} \right] \\ =
\eta_{\nu} - \chi_{\nu}J_{\nu}, \end{split}
\label{eq:moment_orig_zero}
\end{equation}
\begin{equation}
\begin{split}
\frac{\partial K_{\nu}}{\partial r} + \frac{3K_{\nu}-J_{\nu}}{r} - \alpha \left[ \frac{\partial (H_{\nu}-N_{\nu})}{\partial \nu} +
  \beta \frac{\partial N_{\nu}}{\partial \nu} \right] \\ =
- \chi_{\nu}H_{\nu} \end{split}
\label{eq:moment_orig_one}
\end{equation}
where we have made use of the definitions of the first four moments of
the specific intensity:
\begin{equation}
[J_{\nu}, H_{\nu}, K_{\nu}, N_{\nu}] = \frac{1}{2} \int_{-1}^1 \dd \mu
\, I_{\nu}(\mu) [1, \mu, \mu^2, \mu^3].
\label{eq:moment_def}
\end{equation}
It is necessary to utilise additional constraints relating the four
angular moments of the radiation intensity to obtain a closed system
of moment equations. We assume a relationship between the moments
dictated by the `variable Eddington factor' $f_{\nu}(r)$, defined as:
\begin{equation}
f_{\nu}(r) \equiv \frac{K_{\nu}(r)}{J_{\nu}(r)}.
\label{eq:f_edd}
\end{equation}
We define a further variable Eddington factor $g_{\nu}(r)$ linking the
first and third order moments $H_{\nu}(r)$ and $N_{\nu}(r)$:
\begin{equation}
g_{\nu}(r) \equiv \frac{N_{\nu}(r)}{H_{\nu}(r)}.
\label{eq:g_edd}
\end{equation}
The values of $f_{\nu}(r)$ and $g_{\nu}(r)$ are dependent on the
angular character of the radiation field at $(r,\nu)$, and are
determined by the solution of the ray equations.

We estimate the $n$th-order angular moments of the intensity
$I_{\nu}(r,\mu)$ for $n = 0,1,2,3$ which may be rewritten as
\begin{eqnarray}
  J_{\nu}(r) &\equiv& \frac{1}{2} \int_{-1}^1 I_{\nu}(r,\mu) \, \dd \mu = \int_0^1 u_{\nu}(r, \mu) \, \dd \mu,
\label{eq:J_def} \\
H_{\nu}(r) &\equiv& \frac{1}{2} \int_{-1}^1 \mu I_{\nu}(r,\mu) \, \dd \mu = \int_0^1 \mu v_{\nu}(r, \mu) \, \dd \mu,
\label{eq:H_def}\\
K_{\nu}(r) &\equiv& \frac{1}{2} \int_{-1}^1 \mu^2 I_{\nu}(r,\mu) \, \dd \mu = \int_0^1 \mu^2 u_{\nu}(r, \mu) \, \dd \mu,
\label{eq:K_def} \\
N_{\nu}(r) &\equiv& \frac{1}{2} \int_{-1}^1 \mu^3 I_{\nu}(r,\mu) \, \dd \mu = \int_0^1 \mu^3 v_{\nu}(r, \mu) \, \dd \mu
\label{eq:N_def}
\end{eqnarray}
where the second equality in each case is reached by utilising the
definitions of $u_{\nu}(r,\mu)$ and $v_{\nu}(r,\mu)$ in terms of
$I_{\nu}(r, \mu)$ and $I_{\nu}(r, -\mu)$. We evaluate the integrals by
a gaussian quadrature formula where the integral over $\mu$ is
replaced by a sum over different impact parameter values. For
numerical accuracy, we split the integral into the domains
$0<\mu<\mu_C$ and $\mu_C<\mu<1$, where $\mu_c=(r^2+R_C^2)^{1/2}/r$
corresponds to the core radius opening angle subtended at $r$. We then
obtain $f_{\nu}(r)$ and $g_{\nu}(r)$ by evaluating the relevant ratios
as described in equations~(\ref{eq:f_edd}) and (\ref{eq:g_edd}).

We define the `sphericality factor' $q_{\nu}(r)$ by the relation
\citep{Auer_1971}
\begin{equation}
\frac{\partial \ln{(r^2 q_{\nu})}}{\partial r} = \frac{3f_{\nu} -
    1}{f_{\nu}r}.
\label{eq:q_def}
\end{equation} 
We calculate $q_{\nu}(r)$ given the variable Eddington factor
$f_{\nu}(r)$ and the arbitrary normalisation $q_{\nu}(R_{\rm C}) = 1$
from
\begin{equation}
q_{\nu}(r) = \left(\frac{r}{R_{\rm C}} \right)^{-2} \exp{\left[
    \int_{R_{\rm C}}^r \frac{3 f_{\nu}(r')-1}{r' f_{\nu}(r')} \, \dd
    r' \right]}.
\label{eq:q_calc}
\end{equation}
This quantity allows us to make a change of variable in the radial
derivative to the dimensionless quantity
$X_{\nu}(r)$ defined through
\begin{equation}
 \dd X_{\nu}(r) = -\chi_{\nu}(r) q_{\nu}(r) \dd r.
\label{eq:X_def}
\end{equation}
The terms on the LHS of equations~(\ref{eq:moment_orig_zero}) and
(\ref{eq:moment_orig_one}) may then be expressed as
\begin{equation}
\frac{\partial (r^2 H_{\nu})}{\partial r} = -q_{\nu} \chi_{\nu}
\frac{\partial (r^2 H_{\nu})}{\partial X_{\nu}},
\label{eq:factor_1} 
\end{equation}
\begin{equation}
\frac{\partial(f_{\nu}J_{\nu})}{\partial r} + \frac{3f_{\nu}-1}{r}
J_{\nu} = - r^{-2} \chi_{\nu} \frac{\partial (r^2 q_{\nu} f_{\nu}
  J_{\nu})}{\partial X_{\nu}}.
\label{eq:factor_2} 
\end{equation}
These relations and the definitions of $f_{\nu}$ and $g_{\nu}$ may be
used to recast the zeroth and first order moment equations as
\begin{equation}
\begin{split}
q_{\nu} \frac{\partial (r^2 H_{\nu})}{\partial X_{\nu}} + \varGamma_{\nu}
\left[ \frac{\partial (1-f_{\nu})r^2J_{\nu}}{\partial \nu} + \beta
  \frac{\partial (f_{\nu}r^2 J_{\nu})}{\partial \nu} \right] \\ =
r^2(J_{\nu} - S_{\nu}), \end{split} 
\label{eq:moment_zero}
\end{equation}
\begin{equation}
\begin{split}
\frac{\partial (f_{\nu}q_{\nu}r^2 J_{\nu})}{\partial X_{\nu}} + \varGamma_{\nu}
\left[ \frac{\partial (1-g_{\nu})r^2H_{\nu}}{\partial \nu} + \beta
  \frac{\partial (g_{\nu}r^2 H_{\nu})}{\partial \nu} \right] \\ =
r^2H_{\nu} \end{split}
\label{eq:moment_one}
\end{equation}
where $\varGamma_{\nu}(r) = \alpha(r)/\chi_{\nu}(r)$ and once again
$S_{\nu}(r) = \eta_{\nu}(r)/\chi_{\nu}(r)$ is the source function,
although in solving the moment equations we account for the dependence
of the source function on the radiation field and will not assume a
given estimate as for the ray equations. Assuming the isotropic forms
of the opacity and emissivity for resonance-line scattering, a good
approximation in the comoving frame, the source function is given by
\begin{equation}
S_{\nu}(r) \equiv \frac{\eta_{\nu}(r)}{\chi_{\nu}(r)} = \frac{1}{\varphi(\nu, r)} \int R(\nu', \nu, r)
J_{\nu'}(r) \, \dd \nu'
\label{eq:S_res}
\end{equation}
where we have allowed for the redistribution function and absorption
profile to vary with radius due to any variation in the temperature
profile of the medium, $T(r)$. The integral is expanded as a numerical
quadrature sum as described in Appendix~\ref{ap:RMsource}.

We note that boundary conditions appropriate for the solution of the
system of moment equations are given by taking
equation~(\ref{eq:moment_one}) and (i) directly specifying
$H_{\nu}(R_{\rm C}) = H_{\nu,{\rm C}}$ for the inner boundary, and
(ii) at the outer boundary, specifying the ratio $h_{\nu} =
H_{\nu}(R)/J_{\nu}(R)$ as calculated from the formal ray solution
similarly to the variable Eddington factor computation. The equations
are solved using the Feautrier matrix equation approach
\citep{1976ApJ...210..419M}.

\section{Numerical Radiative Transfer in Spherical Symmetry II: Monte Carlo}

\label{sect:MC_method}

We implement a Monte Carlo method to obtain the mean intensity of \Lya
photons and the corresponding scattering rate in a spherically
symmetric medium with a general radial velocity profile. It differs
from existing schemes in a few aspects, leading to an order of
magnitude increase in speed for a given accuracy, so we describe it in
some detail. In addition to the assumption of spherical symmetry, we
sample the redistribution function directly and compute the energy
density based on the path lengths traversed by the photons.

Here we summarise the method developed and refer the reader to
Appendix~\ref{ap:MC} for details. We impose a grid of ${\rm ND}$
radius values $\{r_{\rm ND}, r_{\rm ND-1}, ..., r_1\}$ where $r_{\rm
  ND}$ is the innermost radius at which we compute the intensity and
$r_1$ is the radius of the outer boundary. We assume the neutral
hydrogen is confined to $r < r_1$ and thus there is no scattering for
$r > r_1$. The algorithm, in outline, is as follows: \\(i) A photon
packet is emitted at some radius $r_{\rm em}$ and comoving frequency
$x_{\rm em}$ determined from the properties of the source. The packet
is assigned a direction typically taken to be isotropic,
i.e. $\mu_{\rm em} = 2R-1$, and an optical depth $\tau = -\ln{R}$,
where $R$ is a random deviate uniformly distributed over $0\le R\le1$
(drawn separately for $\mu_{\rm em}$ and $\tau$), after which it will
be scattered by the hydrogen. \\ (ii) We follow the potential path of
the photon packet defined by $r_{\rm em}$ and $\mu_{\rm em}$,
originating at $\lambda = 0$, where $\lambda$ describes the path
length, to the first intersection with a shell boundary and compute
the distance $\lambda_{\rm s}$ and optical depth $\tau_{\rm s}$ from
equations~(\ref{eq:lambda_pm}) and (\ref{eq:tau_shell}),
respectively. \\ (iii) If $\tau > \tau_{\rm s}$ then the photon packet
will cross the boundary, if not we skip to (iv). We update the
position by setting $(\lambda + \lambda_{\rm s}) \rightarrow \lambda$
and compute the comoving frequency between the boundaries using
equations~(\ref{eq:mu_pm}) and (\ref{eq:x_shift}). We bin the
frequency and add the appropriate $\delta \lambda$(s) to the sum in
equation~(\ref{eq:J_path}) for the corresponding frequency bin(s) and
volume cell. We take $(\tau-\tau_{\rm s}) \rightarrow \tau$ and
determine values of $\lambda_{\rm s}$ and $\tau_{\rm s}$ from the
current position to the next shell boundary along the path; this step
is repeated until $\tau_{\rm s} > \tau$ indicating the packet scatters
before reaching the next boundary. If the packet crosses the boundary
at $r_1$ then it escapes the \HI medium and cannot be scattered back
to $r < r_1$. In this case we return to step (i) for a new photon
packet if required. \\ (iv) We assume the packet travels a further
distance $\Delta \lambda = (\tau/\tau_{\rm s})\lambda_{\rm s}$ along
the path and add $\delta \lambda = \Delta \lambda$ to the sum to
compute $J_{\nu}(r)$ for the appropriate volume cell, where again the
comoving frequency is binned along this section of path and $\Delta
\lambda$ must be broken into multiple values of $\delta \lambda$ if
the range in comoving frequency is not contained in a single frequency
bin. We obtain the total distance travelled by the photon packet from
$\mathbf r_{\rm em}$ before scattering from $\lambda + \Delta \lambda
\rightarrow \lambda$. The radius $r$ is updated according to
equation~(\ref{eq:radius_update}) with $\lambda$ in place of
$\lambda_{\max}$. The comoving frequency from which the packet is
scattered, $x'$, is determined from equation~(\ref{eq:x_shift}) with
$\mu(r) = (\mu_{\rm em} r_{\rm em} + \lambda)/r$.  \\ (v) The photon
packet is re-emitted at $r_{\rm em} = r$ with a comoving frequency
$x_{\rm em} = x_{\rm em}(x', T)$ determined from the appropriate
redistribution function, e.g. type ${\rm II}$ redistribution (RII)
with or without recoil, or even coherent scattering ($x_{\rm em} =
x'$). We assign an isotropically distributed direction $\mu_{\rm em} =
2R-1$ and optical depth to next scatter $\tau = -\ln{R}$. We return to
step (ii).

This sequence of steps is repeated for as many photon packets as are
necessary to reduce the statistical noise in $J_{\nu}(r)$ to
acceptable levels. The normalisation in equation~(\ref{eq:J_path}) is
enforced by relating the $\Delta t$ for the Monte Carlo simulation to
the number of photons followed, using the assumed physical properties
of the source as described later for the particular examples we study.

\section{Test Problems for a Homogeneous Medium}
\label{sect:HomogTest}

\subsection{Static Sphere}
\label{subsect:statsph}

\citet{DHS_06} obtained an analytic solution for the mean intensity of
\Lya radiation in a static uniform sphere allowing for RII frequency
redistribution analogous to that for a plane-parallel slab
\citep{1973MNRAS.162...43H}. They considered a system of radius $R$
and line-centre opacity $\kappa_0$, with a source term of unit
strength, frequency distribution $\phi(x)$ and arbitrary radial
distribution $j(r)$, and assumed an outer boundary condition
applicable in the Eddington limit:
\begin{equation}
\left. H_{\nu}(r) \right|_{R} = \frac{1}{2} \left. J_{\nu}(r)
\right|_{R}. 
\label{eq:BC_outer_sphere}
\end{equation}
They found the solution
\begin{equation}
J(x, r) = \frac{\sqrt{6 \pi} \kappa_0}{16\pi^2 R}
\displaystyle\sum_{n = 1}^{\infty} Q_n \frac{\sin{(\lambda_n
    r)}}{\lambda_n r}
\exp{\left(-\sqrt{\frac{2\pi}{27}}\frac{\lambda_n}{a \kappa_0} |x^3|
    \right)},
\label{eq:DHS_sphere}
\end{equation}
where the factors $\{Q_n\}$ are related to the radial distribution
function $j(r)$ by
\begin{equation}
Q_n \equiv \int_0^R 4\pi r^2 \frac{\sin{(\lambda_n r)}}{r} j(r) \, \dd r,
\label{eq:DHS_Q}
\end{equation}
and the radial `eigenvalues' $\{\lambda_n \}$ are
constrained by the outer boundary condition to the frequency-dependent values:
\begin{equation}
\lambda_n(x) \approx \frac{n \pi}{R} \left[ 1 - \frac{1}{1+
    1.5\pi^{1/2}\kappa_0 R \phi(x)} \right].
\label{eq:DHS_lambda}
\end{equation}
In the diffusion approximation assumed by \citet{DHS_06}, the
radiation flux is related to the average intensity by the first order
angular moment equation, $H(r,x) = -1/(3\chi_{\nu})\partial
J(r,x)/\partial r$ where the opacity is $\chi_{\nu} =
\sqrt{\pi}\kappa_0 \phi(x)$. The corresponding solution for the flux,
derived by taking the radial derivative, is
\begin{equation}
\begin{split}
H(x, r) = \frac{\sqrt{6}}{48\pi^2 R \phi(x)}
\displaystyle\sum_{n = 1}^{\infty} \frac{Q_n}{\lambda_n} \left[ \left( \frac{\sin{(\lambda_n r)}}{r^2} - \frac{\lambda_n \cos{(\lambda_n r)}}{r} \right) \right. \\
\left. \times \exp{\left(-\sqrt{\frac{2\pi}{27}}\frac{\lambda_n}{a \kappa_0} |x^3| \right)} \right].
\end{split}
\label{eq:DHS_flux}
\end{equation}

\citet{DHS_06} suggest a source term describing a thin shell at some
source radius $r_{\rm S}$ with the radial distribution and
corresponding $Q_{n}$ values
\begin{equation}
j(r) = \frac{1}{\sqrt{\pi} \kappa_0} \frac{\delta_{\rm D}(r-r_{\rm S})}{4\pi r^2} \implies Q_n = \frac{1}{\sqrt{\pi} \kappa_0} \frac{\sin{(\lambda_n r_{\rm S})}}{r_{\rm S}},
\label{eq:j_shell}
\end{equation}
which corresponds to a point source in the limit $r_{\rm S}
\rightarrow 0$ or $r \gg r_{\rm S}$. The source term is $S(x, r) =
\chi_{\nu} j(r)/(4\pi)$. The normalisation ensures the condition $\int
\, \dd x \int \, \dd V \oint \, \dd \Omega \, S(x, r) = 1$ is met.

We now consider the application of the moment and ray equation
solution methods to this particular problem. The inner boundary
condition is applied at a `core radius' $R_{\rm C}$. As the moment
equations that we solve do not account for a source term, it is
important that the source is contained within the core, $R_{\rm C} >
r_{\rm S}$, such that the properties of the source are described by
the core boundary condition. This boundary condition is given by
assuming the diffusion limit $v_{\nu}(R_{\rm C}, \mu) = 3 |\mu|
H_{\nu}(R_{\rm C})$ in solving the ray equations. In both cases we
take the analytic solution equation~(\ref{eq:DHS_flux}) for
$H_{\nu}(R_{\rm C})$ with the $Q_n$ coefficients given by
equation~(\ref{eq:j_shell}). In estimating the source function assumed when
solving the ray equations, it is sufficient to evaluate the integral
equation~(\ref{eq:S_res}) using the analytic solution
equation~(\ref{eq:DHS_sphere}) as an estimate of the mean intensity, while
the source function in the moment equations follows from
equations~(\ref{eq:S_quad}) and (\ref{eq:Rfunk_diff_easy}) with $\epsilon =
0$ for the coefficients. We take a linear grid in radius spanning
$[R_{\rm C}, R]$ and a linear grid of frequency values $\{x_k\}$
spanning the range over which the analytic solution is
non-negligible. As a first step we attempt to solve the problem in the
Eddington approximation by solving the moment equations with
$f_{\nu}(r) \equiv 1/3$.

\begin{figure}
\begin{center}
\leavevmode
\epsfig{file = 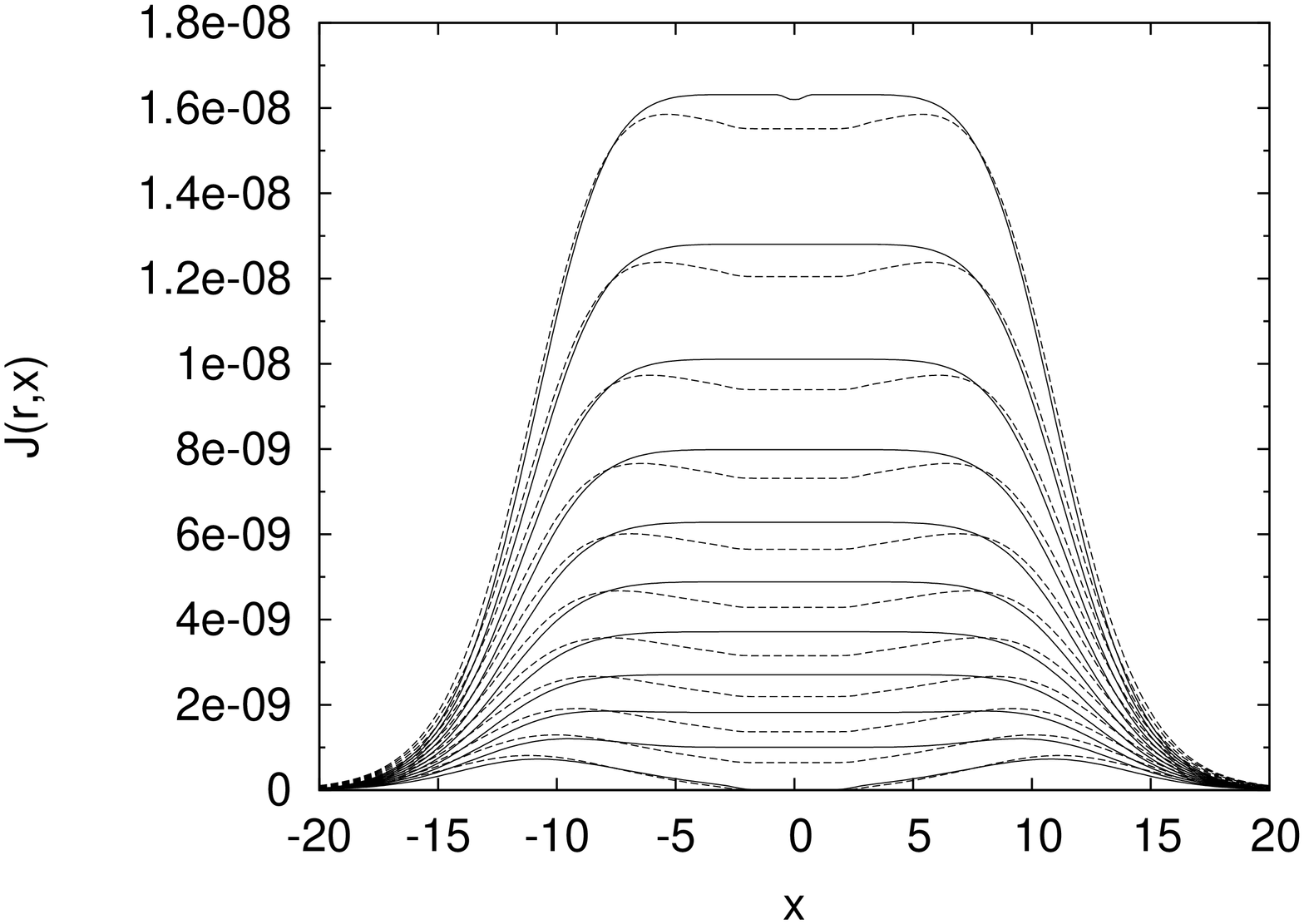, height = 7cm, angle = -90}
\epsfig{file = 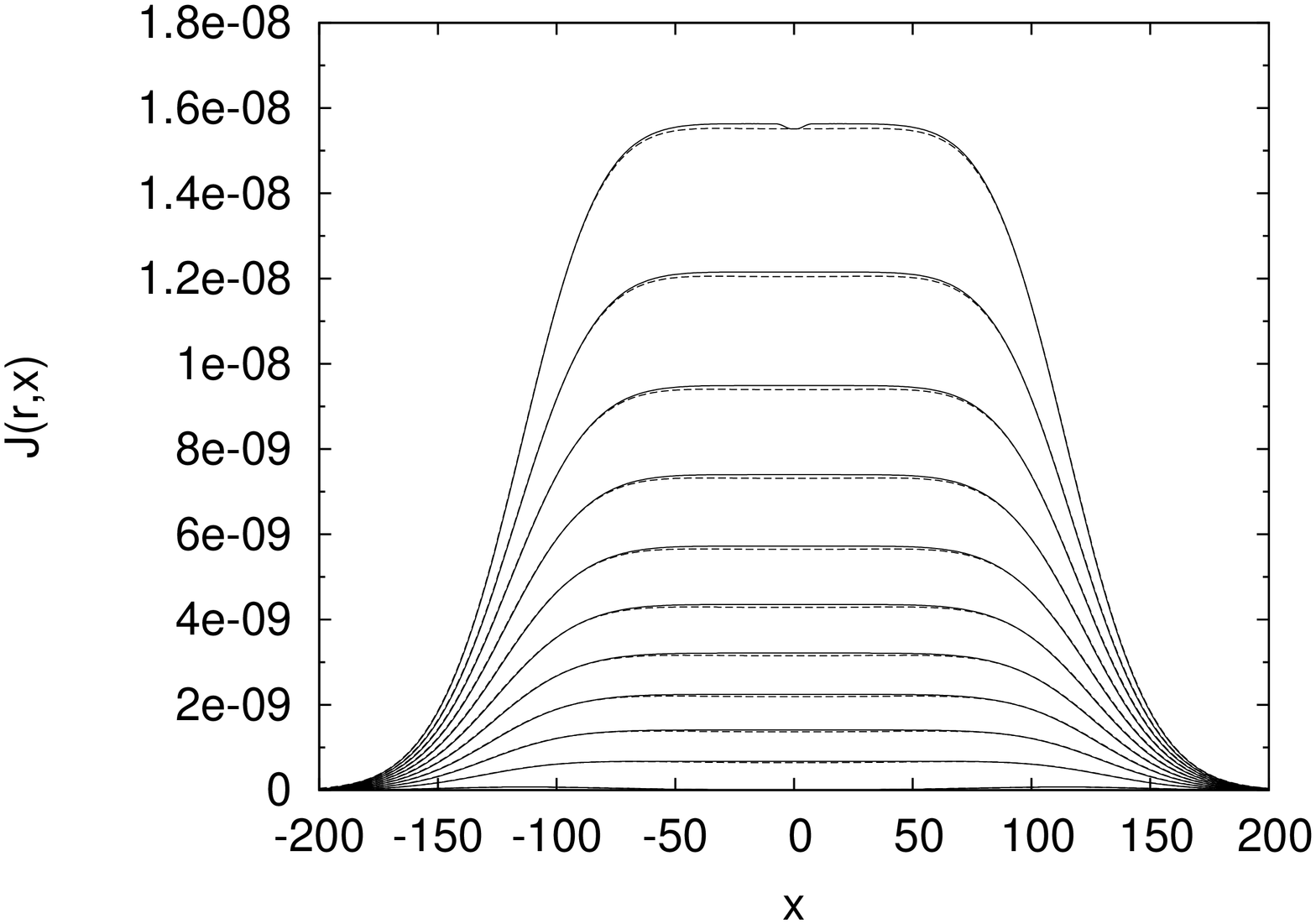, height = 7cm, angle = -90}
\end{center}
\caption{The mean intensity $J(r,x)$ for \Lya RII scattering in a
  static, uniform sphere in the Eddington approximation. Solutions
  assume a source radial dependence of the form
  equation~(\ref{eq:j_shell}) for $r_{\rm S} = 5$, a temperature $T =
  10 \, {\rm K}$, and a line-centre opacity $\kappa_0 = 100$ (upper
  panel) and $\kappa_0 = 1.2 \times 10^5$ (lower panel). Frequency
  profiles are given from $r = 500$ (highest values of $J$) to $r =
  1000$ (lowest values of $J$) in steps of $50$. In each panel we
  display the solutions of the moment equations in the Eddington
  approximation (solid line) together with the corresponding analytic
  solution in the diffusion approximation of \citet{DHS_06} (dashed
  line).}
\label{figure:DHS_diff}
\end{figure}

Examples of the resulting moment equation solution are shown in
Fig.~\ref{figure:DHS_diff}. We find our results agree with the
analytic solution of Dijkstra et al. apart from differences across the
line centre in the lower optical depth case; the moment equation
solution is the more accurate of the two as the analytic solution is
an exact solution in the Eddington approximation only in the limit
$(a\tau_0)^{1/3} \gg 1$.
  
\begin{figure}
\begin{center}
\leavevmode
\epsfig{file = 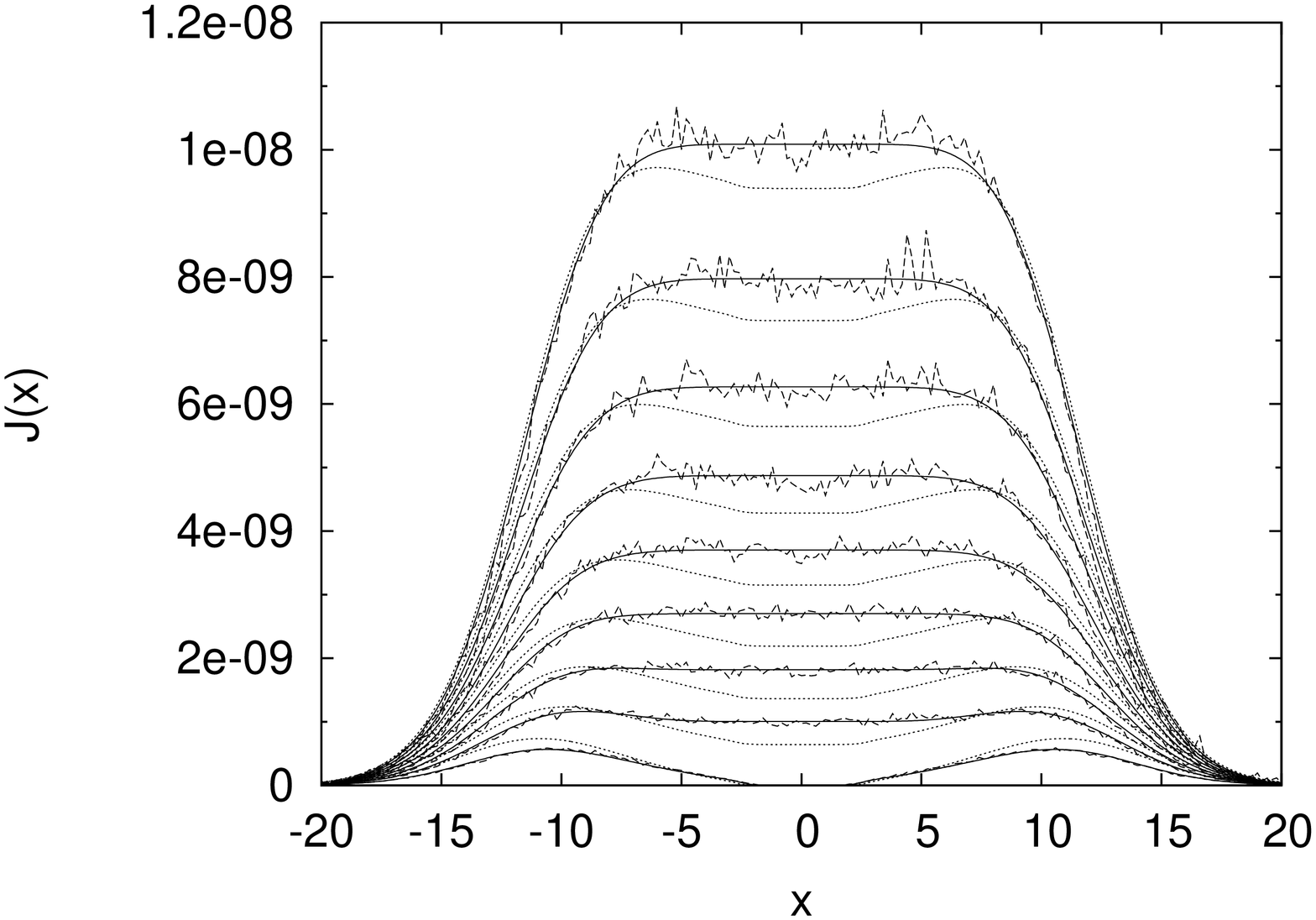, height = 7cm, angle = -90}
\epsfig{file = 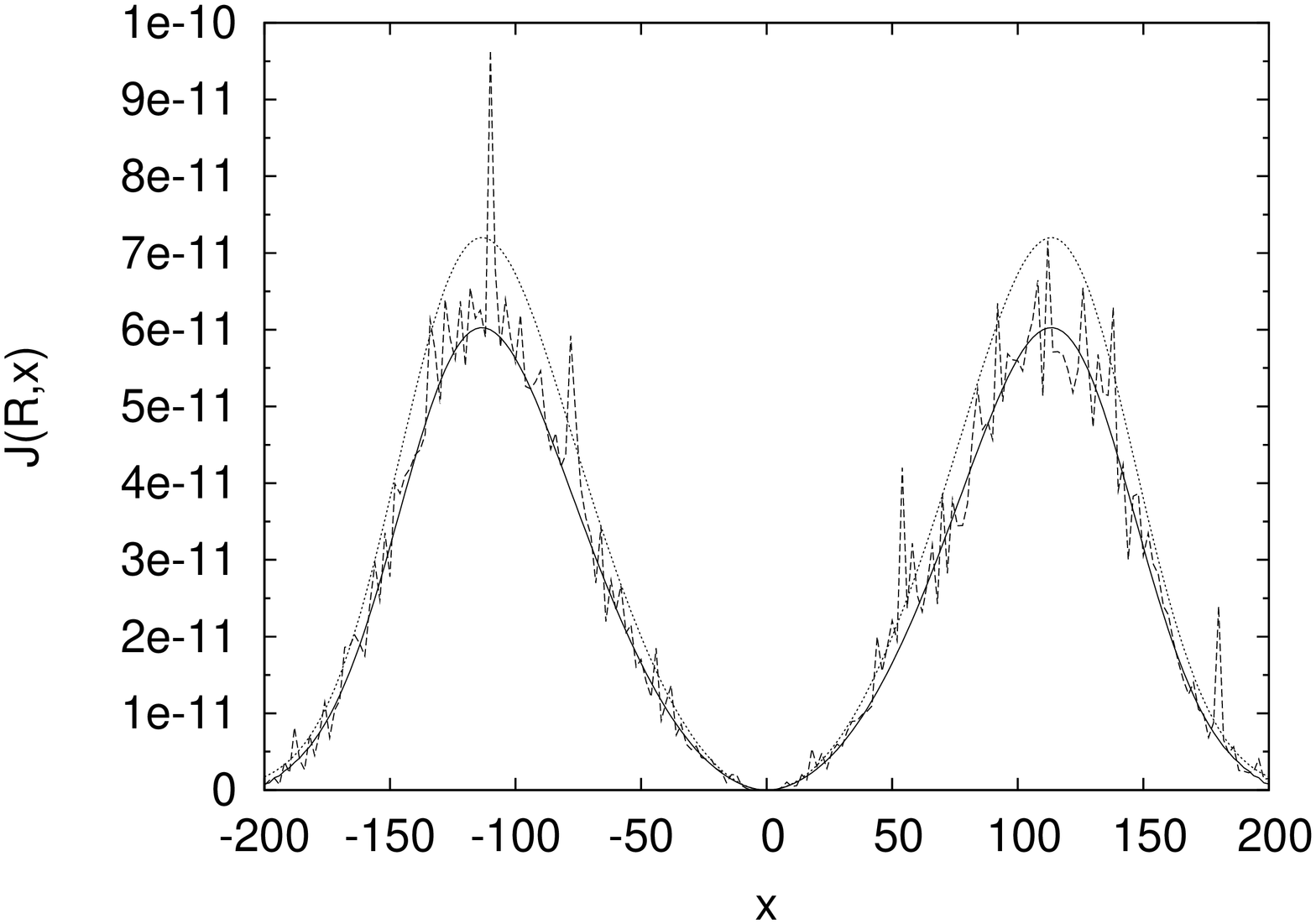, height = 7cm, angle = -90}
\end{center}
\caption{The mean intensity $J(r,x)$ for \Lya RII scattering in a
  static, uniform sphere. Solutions assume a source radial dependence
  of the form equation~(\ref{eq:j_shell}) for $r_{\rm S} = 5$, a
  temperature $T = 10 \, {\rm K}$, and a line-centre opacity $\kappa_0
  = 100$ (upper panel) and $\kappa_0 = 1.2 \times 10^5$ (lower
  panel). In the upper panel, frequency profiles are given from $r =
  600$ (highest values of $J$) to $r = 1000$ (lowest values of $J$) in
  steps of $50$ while we display only $J(R, x)$ in the lower panel. We
  compare our ray/moment equation solutions (solid line) with the
  corresponding Monte Carlo solution (dashed line) and the analytic
  solution in the diffusion approximation (dotted line). While noisy
  near line centre, the Monte Carlo solutions agree well with the
  ray/moment equations solutions in the wings.}
\label{figure:DHS_exact}
\end{figure}

A numerical solution without assuming the Eddington approximation may
be constructed by solving the moment equations with Eddington factors
determined from solving the ray equations as described in
Section~\ref{subsect:grid_method}. In Fig.~\ref{figure:DHS_exact}, we
compare the ray/moment equations solutions with the corresponding
Monte Carlo solutions. The Monte Carlo solution in the upper panel of
Fig.~\ref{figure:DHS_exact} was obtained according to the summed
path-length method of Section~\ref{sect:MC_method}, while in the lower
panel we obtained the mean intensity at the outer boundary surface
using the definition of specific intensity. The Monte Carlo solutions
in each case were normalised by computing the simulation timescale
$\Delta t$ as required by equation~(\ref{eq:J_path}) or
equation~(\ref{eq:I_shell}) as $\Delta t = N/(1 \ {\rm photon} \ {\rm
  s}^{-1})$, which follows from the unit source strength imposition.
The solutions were obtained by following $N = 2 \times 10^5$ photon
packets for the case shown in the upper panel, and $N = 2 \times 10^4$
photon packets for the lower. The frequency redistribution was treated
using the lookup table method in the low optical depth case, while in
the optically thick case we used the scattering atom velocity method
with a bias to skip over core scatterings. From
Fig.~\ref{figure:DHS_exact}, we find that our ray/moment equation
solutions agree closely with the Monte Carlo solutions for both low
and high optical depths, with only small deviations from the analytic
solution in both cases. We found no significant alterations to the
ray/moment solutions upon re-solving the ray and moment equations with
an estimated source function obtained from the moment solution,
suggesting that for this problem the combined method converges in a
single iteration.

\subsection{Homogeneous Expanding Medium: \Lya Source}
\label{subsect:LR_sols}

As a simple test of the ray/moment equation and Monte Carlo methods
for a non-static, spherically symmetric medium, we use the problem of
a \Lya point source in a homogeneous expanding \HI medium
\citep{1999ApJ...524..527L}. Loeb \& Rybicki added an extra source
term $S(\nu, r)$ to the RHS of equation~(\ref{eq:sph_comoving}) to
model a point source of \Lya photons at the origin:
\begin{equation}
  S(\nu, r) = \dot{N}_{\alpha}\delta(\nu - \nu_{\alpha}) \frac{\delta(r)}{(4\pi r)^2},
\label{eq:LR_source}
\end{equation} 
where $\dot{N}_{\alpha}$ is the rate of emission of \Lya photons with
frequency $\nu_\alpha$. They introduced a characteristic frequency
$\nu_*$, defined such that photons in the red Lorentz wing with
$\chi_{\nu} = \eta (\nu-\nu_{\alpha})^{-2}$, where $\eta \equiv n_{\rm
  H}\sigma \Gamma_{\alpha}/(4\pi^2)$, accumulate an optical depth of
unity in redshifting from the source with frequency $\nu = \nu_*$ to
an observer at infinity:
\begin{eqnarray}
  \nu_* &\equiv& \frac{\eta}{\alpha} = \frac{\sigma \Gamma_{\alpha} \lambda_{\alpha} n_{\rm H}(z)}{4\pi^2 H(z)} \simeq \frac{\sigma \Gamma_{\alpha} \lambda_{\alpha} n_{\rm H}(0)}{4\pi^2 \Omega_{\rm m}^{1/2} H_0} (1+z)^{3/2} \nonumber \\
  &\simeq& 5.52 \times 10^{12} \left( \frac{\Omega_{\rm b} h}{\Omega_{\rm m}^{1/2}} \right) (1+z)^{3/2} \, {\rm Hz}
\label{eq:nustar_def}
\end{eqnarray}
where we have assumed a neutral medium with $n_{\rm H}(z) = n_{\rm
  H}(0) (1+z)^{3}$, $H(z) \simeq H_0 \Omega_{\rm m}^{1/2}
(1+z)^{3/2}$, as the universe is matter-dominated at the redshifts of
interest, and $\alpha=H(z)\nu_\alpha/c$. Loeb \& Rybicki defined the
characteristic radius $r_*$ as the radius at which a photon free
streaming from the source redshifts from $\nu_\alpha$ to
$(\nu_\alpha-\nu_*)$, i.e. $\alpha r_* = \nu_*$:
\begin{eqnarray}
  r_* &\equiv& \frac{\nu_*}{\alpha} = \frac{\sigma \Gamma_{\alpha} \lambda_{\alpha}^2 n_{\rm H}(z)}{4\pi^2 H^2(z)} \simeq \frac{\sigma \Gamma_{\alpha} \lambda_{\alpha}^2 n_{\rm H}(0)}{4 \pi^2 \Omega_{\rm m} H_0^2} \nonumber \\
  &\simeq& 6.72 \left(\frac{\Omega_{\rm b}}{\Omega_{\rm m}} \right) \, {\rm Mpc}.
\label{eq:rstar_def}
\end{eqnarray}
Dimensionless frequency and radius variables are then given by
$\tilde{\nu} = (\nu_\alpha-\nu)/\nu_*$ and $\tilde{r} = r/r_*$. In
these units, a photon emitted at frequency $\tilde\nu_{\rm em}$ will
redshift over a distance $\tilde r$ to $\tilde\nu(\tilde r) =
{\tilde\nu_{\rm em}} + {\tilde r}$. It is also useful to define a
dimensionless radiation intensity $\tilde{I} = I_{\nu}/I_*^l$ or
$\tilde{J} = J_{\nu}/I_*^l$, where $I_*^l = \dot{N_{\alpha}}/(r_*^2
\nu_*)$. Loeb \& Rybicki derived an analytic solution for the mean
intensity arising from a \Lya point source in the homogeneous
expanding medium, applicable in the diffusion limit:
\begin{equation}
\tilde{J}(\tilde{r}, \tilde{\nu}) = \frac{1}{4\pi} \left(\frac{9}{4\pi
  \tilde{\nu}^3} \right)^{3/2} \exp{\left(-\frac{9 \tilde{r}^2}{4
      \tilde{\nu}^3} \right)}.
\label{eq:LR_diff}
\end{equation}
The corresponding scattering rate per atom is
\begin{equation}
P_{\alpha} \equiv 4\pi \sigma \int J_{\nu}(r) \varphi(\nu) \, \dd
\nu = \sigma I_*^l {\tilde P}_\alpha,
\label{eq:P_LR99}
\end{equation}
where the dimensionless scattering rate ${\tilde P}_\alpha$ is given by
\begin{equation} {\tilde
    P}_\alpha=\frac{1}{3\pi^{3/2}}\left(\frac{4}{9}\right)^{1/3}\Gamma\left(\frac{11}{6}\right)\gamma{\tilde
    r}^{-11/3}.
\label{eq:Pt_LR99}
\end{equation}
Here $\gamma=H(z)/[\sigma\lambda_\alpha n_{\rm H}(z)]$ is the Sobolev
parameter for a uniform velocity gradient $H(z)$. The flux in the
diffusion limit is straightforwardly derived from $\tilde J$:
\begin{equation}
  \tilde{H}(\tilde{r}, \tilde{\nu}) = -\frac{1}{3 \tilde{\chi}}
  \frac{\partial \tilde{J}}{\partial \tilde{r}} = \frac{3\tilde{r}}{8\pi\tilde{\nu}} \left(\frac{9}{4\pi \tilde{\nu}^3} \right)^{3/2} \exp{\left(-\frac{9 \tilde{r}^2}{4 \tilde{\nu}^3} \right)}
\label{eq:LR_diff_flux}
\end{equation}
where $\tilde{\chi} \equiv r_* \chi_{\nu} = \tilde{\nu}^{-2}$. Details
of the derivation are provided in Appendix~\ref{ap:hem}.

\begin{figure}
\begin{center}
\leavevmode
\epsfig{file = 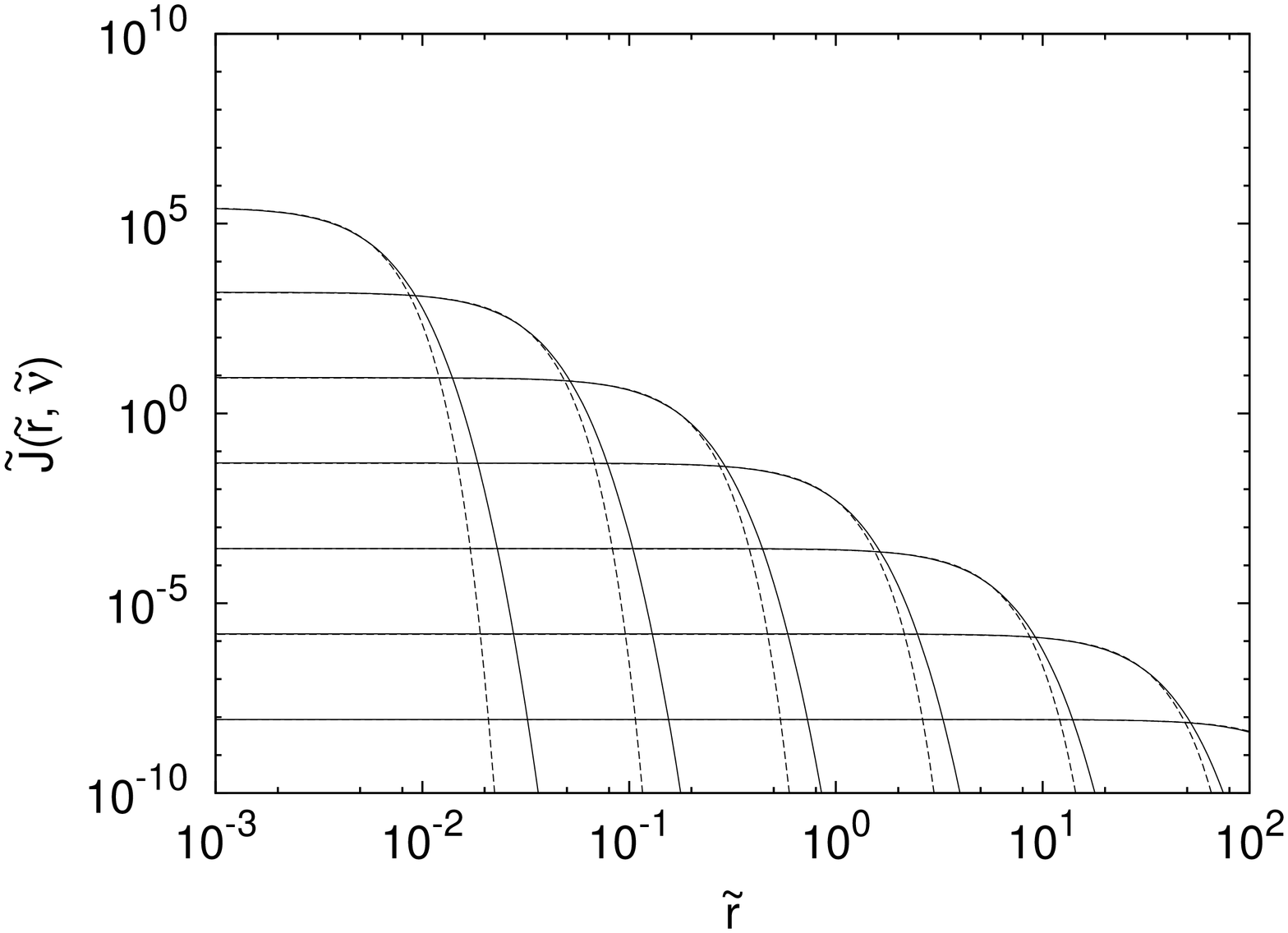, height = 7cm, angle = -90}
\epsfig{file = 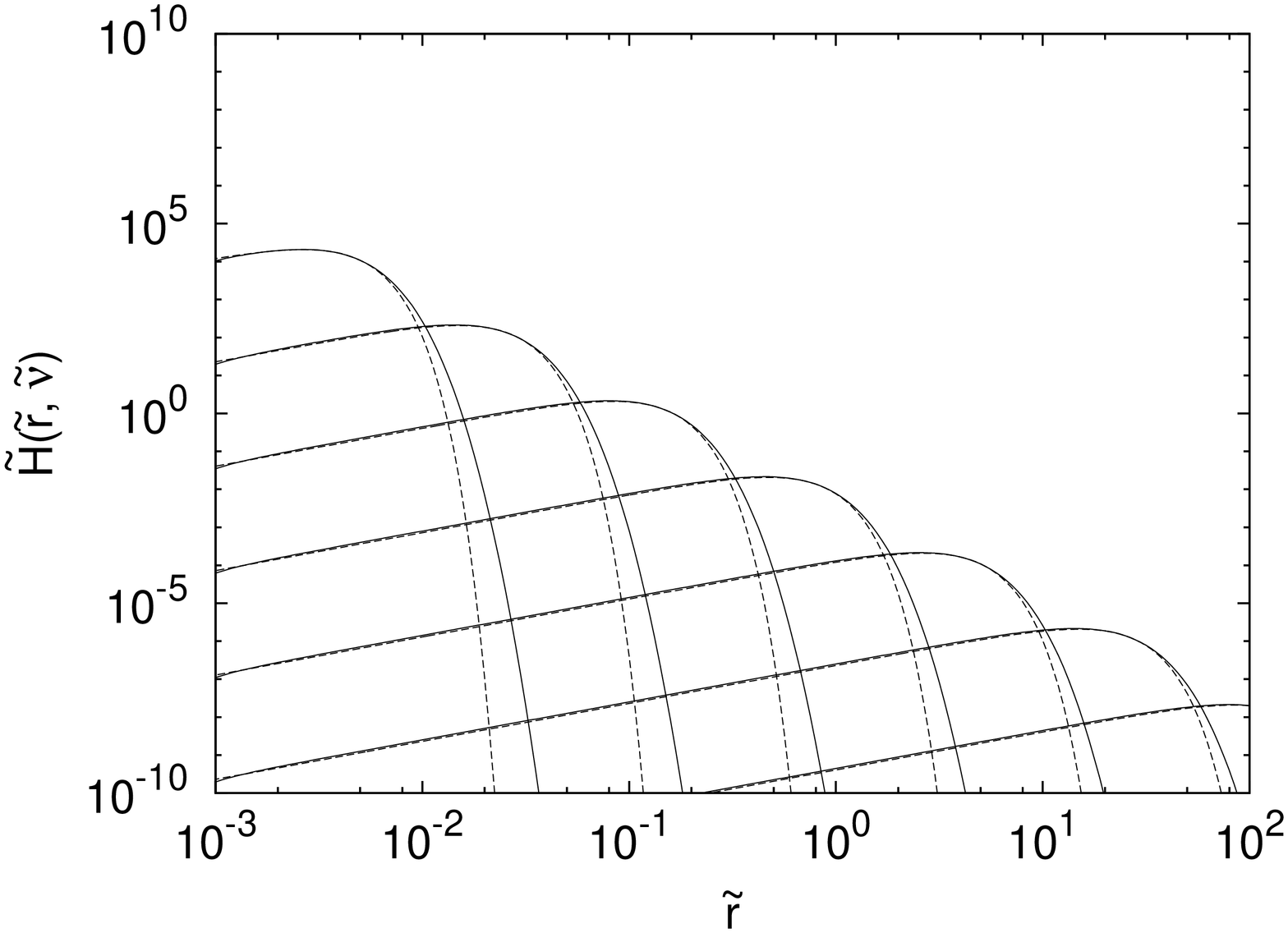, height = 7cm, angle = -90}
\end{center}
\caption{The mean intensity $\tilde{J}(\tilde{r}, \tilde{\nu})$ (upper
  panel) and flux $\tilde{H}(\tilde{r}, \tilde{\nu})$ (lower panel)
  for a point emission line source in a uniformly expanding
  homogeneous medium, from the moment equations in the diffusion
  approximation (solid lines) together with the analytic diffusion
  solutions (dashed lines). Coherent scattering is assumed. Profiles
  are given for $\log_{10}{\tilde{\nu}} = -1.5$, $-1.0$, $-0.5$,
  $0.0$, $0.5$, $1.0$, $1.5$ in order of decreasing values of
  $\tilde{J}$ or $\tilde{H}$.}
\label{figure:LR_diff}
\end{figure}

In this section, we attempt to reproduce the analytic solution using
the moment equation solution method, before solving the problem more
generally (outwith the diffusion limit) using the ray and moment
equation methods and comparing with the Monte Carlo solution. We
continue to assume coherent scattering. We take our grid in radius to
be spaced logarithmically from $\tilde{R}_{\rm C}$ to
$\tilde{R}$. Anticipating results comparable to those shown by Loeb \&
Rybicki, we adopt $\tilde{R}_{\rm C} = 10^{-3}$ and $\tilde{R} =
10^2$. The inner boundary condition for the moment equations is given
by directly specifying the flux at the boundary according to
equation~(\ref{eq:LR_diff_flux}), while in the ray equations we assume
the same form for the flux with an angular dependence given by the
diffusion limit as in equation~(\ref{eq:v_core_diff}). The outer
boundary condition results from assuming no photons are scattered back
into the system from $\tilde{r} \geq \tilde{R}$. This is technically
not guaranteed to apply in an infinite scattering medium, but we
expect no significant effect on the solution at distances well within
the outer surface. Our frequency grid is subject to the restriction
that the highest frequency (and therefore smallest value of
$\tilde{\nu}$) lies bluewards of any frequency at which there are
photons, in order to ensure compatibility with the frequency initial
condition equation~(\ref{eq:zero_photons}). This may be ensured by
taking $\tilde{\nu}_1$ to lie blueward of the frequency of the
least-redshifted photons emitted by the source as arises from free
streaming, which requires that $\tilde{\nu}_1 < \tilde{R}_{\rm C}$. We
use a logarithmically spaced grid spanning the range from
$\tilde{\nu}_1 = 10^{-3.5}$ to $\tilde{\nu}_{\rm NF} = 10^{1.5}$.

A zero-temperature medium is assumed that provides no Doppler
broadening of the \Lya line, giving a Lorentz profile for the opacity
which in the wings takes the form $\tilde{\chi} =
1/\tilde{\nu}^2$. The frequency redistribution function results in
coherent scattering, so that the source function is simply $S_{\nu}(r)
= J_{\nu}(r)$. From equation~(\ref{eq:S_quad}), the redistribution
coefficients for coherent scattering simplify to $\mathcal{R}_{k',k,
  d} \rightarrow \delta_{k',k}$. When estimating the source function
in order to solve the ray equations it is sufficient to take $S_{k,d}
= J_{k,d}$ and assume the analytic solution
equation~(\ref{eq:LR_diff}). The analytic expressions
equations~(\ref{eq:LR_diff}) and (\ref{eq:LR_diff_flux}) apply in the
diffusion approximation where the Eddington approximation is used, and
the frequency derivative of the flux is assumed to be negligible. The
former point requires $f_{\nu}(r) \equiv 1/3$ (note that as the
velocity law is linear, $\beta \equiv 1$ and the equations are
independent of $g_{\nu}(r)$), while the latter is enforced explicitly
by setting to zero corresponding elements in the matrices used to
solve the problem (details are given in \citet{2012HigginsPhD}). The
resulting solutions of the moment equations for $\tilde{J}$ and
$\tilde{H}$ are given in Fig.~\ref{figure:LR_diff}. They agree very
well with the analytic solutions, deviating only beyond the cutoff
radius where the finiteness of the frequency grid restricts the
solution from following the steep decline with radius.

\begin{figure}
\begin{center}
\leavevmode
\epsfig{file = 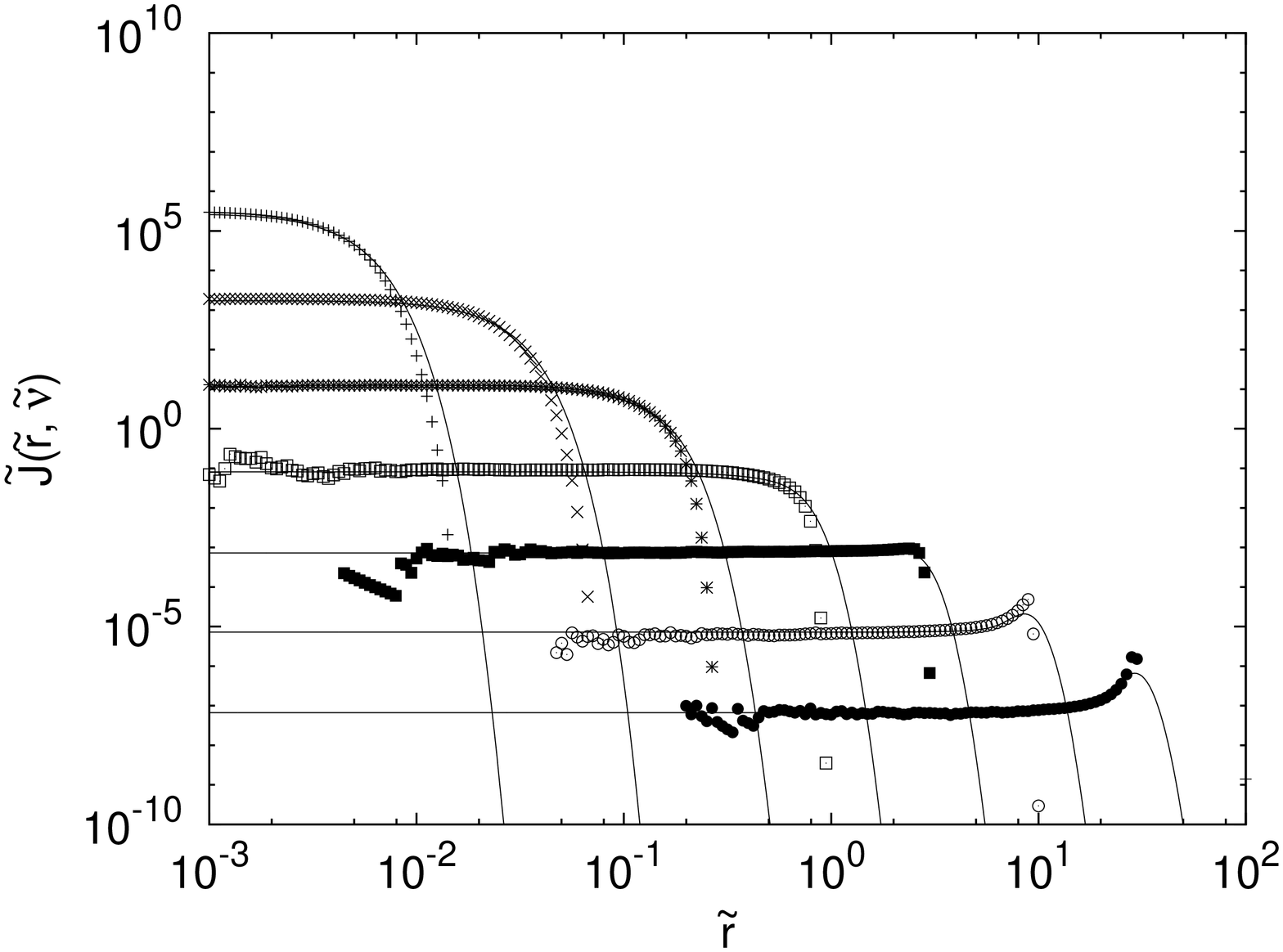, height = 7cm, angle = -90}
\epsfig{file = 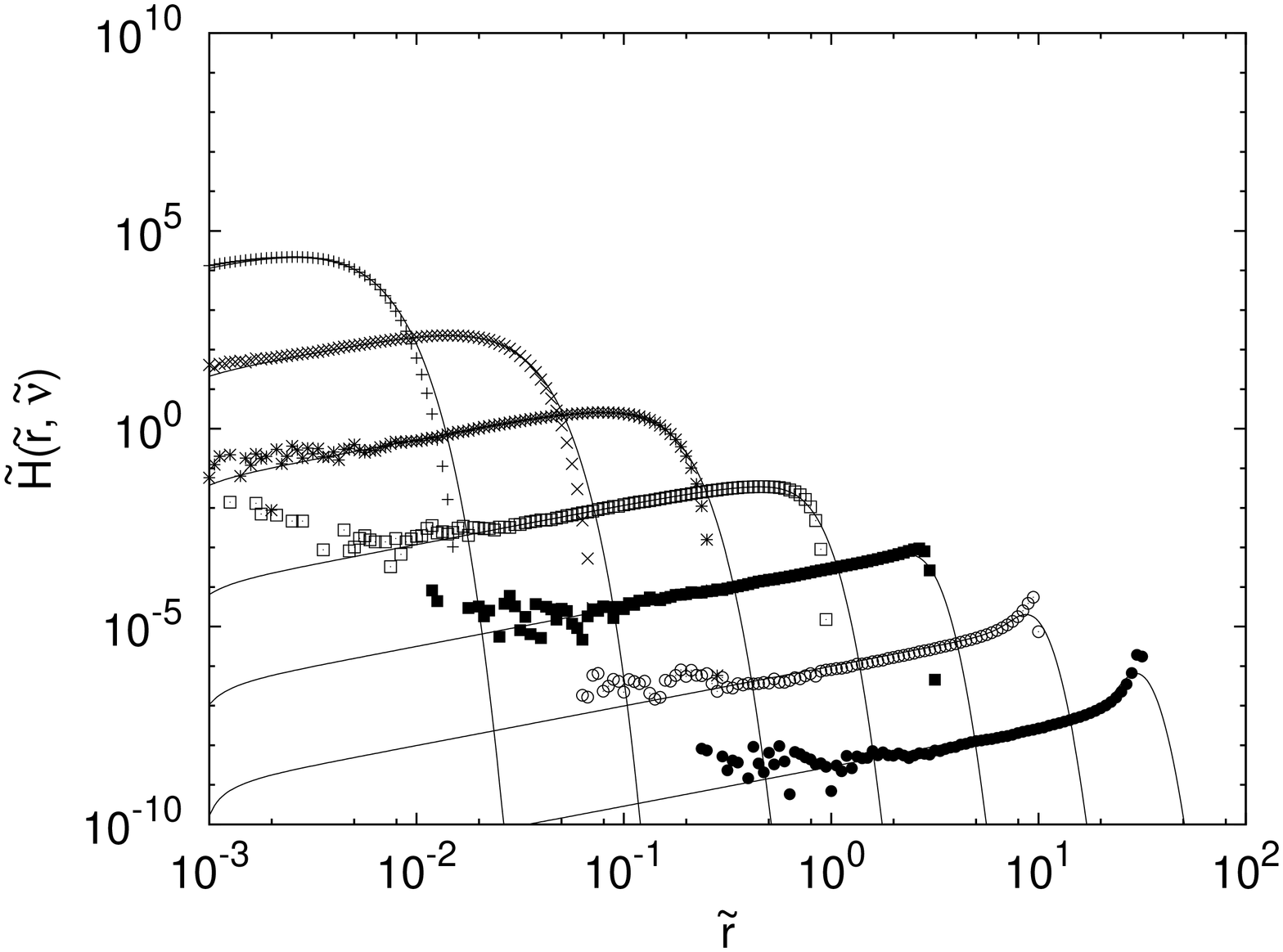, height = 7cm, angle = -90}
\end{center}
\caption{The mean intensity $\tilde{J}(\tilde{r}, \tilde{\nu})$ (upper
  panel) and flux $\tilde{H}(\tilde{r}, \tilde{\nu})$ (lower panel)
  for a point emission line source in a uniformly expanding
  homogeneous medium, obtained from two iterations of the combined
  ray/moment method with a diffusion solution inner boundary condition
  (solid lines), compared with the respective Monte Carlo solutions
  (various points). Coherent scattering is assumed. Profiles are given
  for $\log_{10}{\tilde{\nu}} = -1.5$, $-1.0$, $-0.5$, $0.0$, $0.5$,
  $1.0$, $1.5$ in order of decreasing values of $\tilde{J}$ or
  $\tilde{H}$.}
\label{figure:LR_MCtest}
\end{figure}

A more comprehensive illustration of the utility of the combined ray
and moment equation solution method is provided by seeking an exact
solution to the homogeneous expanding medium problem and comparing our
results with the Monte Carlo solutions. We take as an inner boundary
condition the analytic solution equation~(\ref{eq:LR_diff_flux}); this
is physically reasonable as the moment equations are expected to
converge to the diffusion limit at great optical depths. The
corresponding Monte Carlo solution for the mean intensity $\tilde{J}$
is obtained from the summed path-length method of
Section~\ref{sect:MC_method}, while the solution for the flux
$\tilde{H}$ is obtained from the definition of specific intensity and
the appropriate angular integration. In Fig.~\ref{figure:LR_MCtest} we
compare the results of the combined ray/moment equation solution
method and the Monte Carlo solution method. The Monte Carlo solutions
are normalised by expressing the timescale $\Delta t$ in
equation~(\ref{eq:J_path}) and equation~(\ref{eq:I_shell}) as $\Delta
t = N/\dot{N}_{\alpha}$. The solutions are based on $N = 10^7$ photon
packets. Both the mean intensity and the flux obtained from the
ray/moment method closely match the corresponding solutions determined
from the Monte Carlo method. The cusps at large $\tilde r$ for the
higher values of $\log_{10}\tilde\nu$ agree with those found by Loeb
\& Rybicki, who attributed them to causal limitation at the surface
$\tilde{r}=\tilde{\nu}$. The solutions were obtained from two
iterations:\ the moment solution, determined using Eddington factors
from the ray solution, determines the source function estimate for a
second ray solution, which determines the Eddington factors for a
second moment solution. Upon further iterations we found that both
methods had converged upon the same solution, independent of the
initial estimate of the source function used in the first ray
solution, as we tested by using a free streaming source function
estimate of the form $\tilde{J} = \delta_{\rm D}(\tilde{\nu} -
\tilde{r})/(4\pi \tilde{r})^2$. We also solved the problem using an
inner free streaming boundary condition of the form $\tilde{H} =
\delta_{\rm D}(\tilde{\nu} - \tilde{R}_{\rm C})/(4\pi \tilde{R}_{\rm
  C})^2$. Except on the inner boundary, the solutions are identical to
those produced by the diffusion boundary condition.

\subsection{Homogeneous Expanding Medium:\ Analytic Results for a
  Continuum Source}
\label{subsect:hmecont_an}

We consider the problem of \Lya scattering about a continuum point
source of UV photons in the neutral hydrogen intergalactic medium at
high redshift, prior to the large-scale onset of reionisation. It is
assumed that the scattering medium may be treated as spherically
symmetric, at least in the vicinity of the source. The problem is then
identical to that treated by \citet{1999ApJ...524..527L}, with the
exception of the frequency dependence of the source, which is here
taken to be spectrally flat. The source term analogous to
equation~(\ref{eq:LR_source}) is
\begin{equation}
S(\nu, r) = \dot{N}_{\nu} \frac{\delta(r)}{(4\pi r)^2}
\label{eq:flat_source}
\end{equation}
where the constant $\dot{N}_{\nu}$ is the rate of emission of photons
per unit frequency. The parametrization of the radiative transfer
equation proceeds identically to the case of a \Lya source, with the
exception of the characteristic intensity which is defined as $I_*^c =
\dot{N}_{\nu}/r_*^2$. It is useful to generalise the frequency
dependence of the source by assuming a limiting frequency $\nu_{\rm
  m}$ bluewards of the resonance frequency beyond which no photons are
emitted:
\begin{equation}
S(\nu, r) = \dot{N}_{\nu} \frac{\delta(r)}{(4\pi r)^2} \times \Theta(\nu_{\rm m} - \nu)
\label{eq:step_source}
\end{equation}
where the step function $\Theta(\nu_{\rm m}-\nu)$ is equal to unity
for $\nu < \nu_{\rm m}$ (or equivalently $\tilde{\nu} >
\tilde{\nu}_{\rm m}$) and is zero otherwise. We show in
Appendix~\ref{ap:hem} that the resulting radiation field in the
diffusion limit is described by the solutions
\begin{eqnarray}
  \tilde{J}(\tilde{r}, \tilde{\nu}) &=& \left(\frac{16}{3}\right)^{1/3}
\frac{2}{(4\pi)^{5/2}}\frac{1}{\tilde{r}^{7/3}}\nonumber\\
&&\times\int_{\frac{9\tilde{r}^2}{4(\tilde{\nu}^3-\tilde{\nu}_{\rm
      m}^3)}}^{\infty}\,\dd u\, \left(t + \frac{1}{u} \right)^{-2/3} u^{-1/2} e^{-u}, 
\label{eq:J_analytic_calc}
\end{eqnarray}
\begin{eqnarray}
  \tilde{H}(\tilde{r}, \tilde{\nu}) &=&
  \left(\frac{4^5}{3^4}\right)^{1/3}\frac{1}{(4\pi)^{5/2}}
  \frac{\tilde{\nu}^2}{\tilde{r}^{10/3}}\nonumber\\
  &&\times \int_{\frac{9\tilde{r}^2}{4(\tilde{\nu}^3-\tilde{\nu}_{\rm
        m}^3)}}^{\infty}\, \dd u\, \left(t+\frac{1}{u}\right)^{-2/3}
  u^{1/2} e^{-u},
\label{eq:H_analytic_calc}
\end{eqnarray}
where $t=-4\tilde{\nu}^3/9{\tilde r}^2$. Note that in the case of a
spectrally flat source with no blue cutoff frequency,
$\tilde{\nu}_{\rm m} \rightarrow -\infty$ and the lower limit in each
integral reduces to zero.

\begin{figure}
\begin{center}
\leavevmode
\epsfig{file = 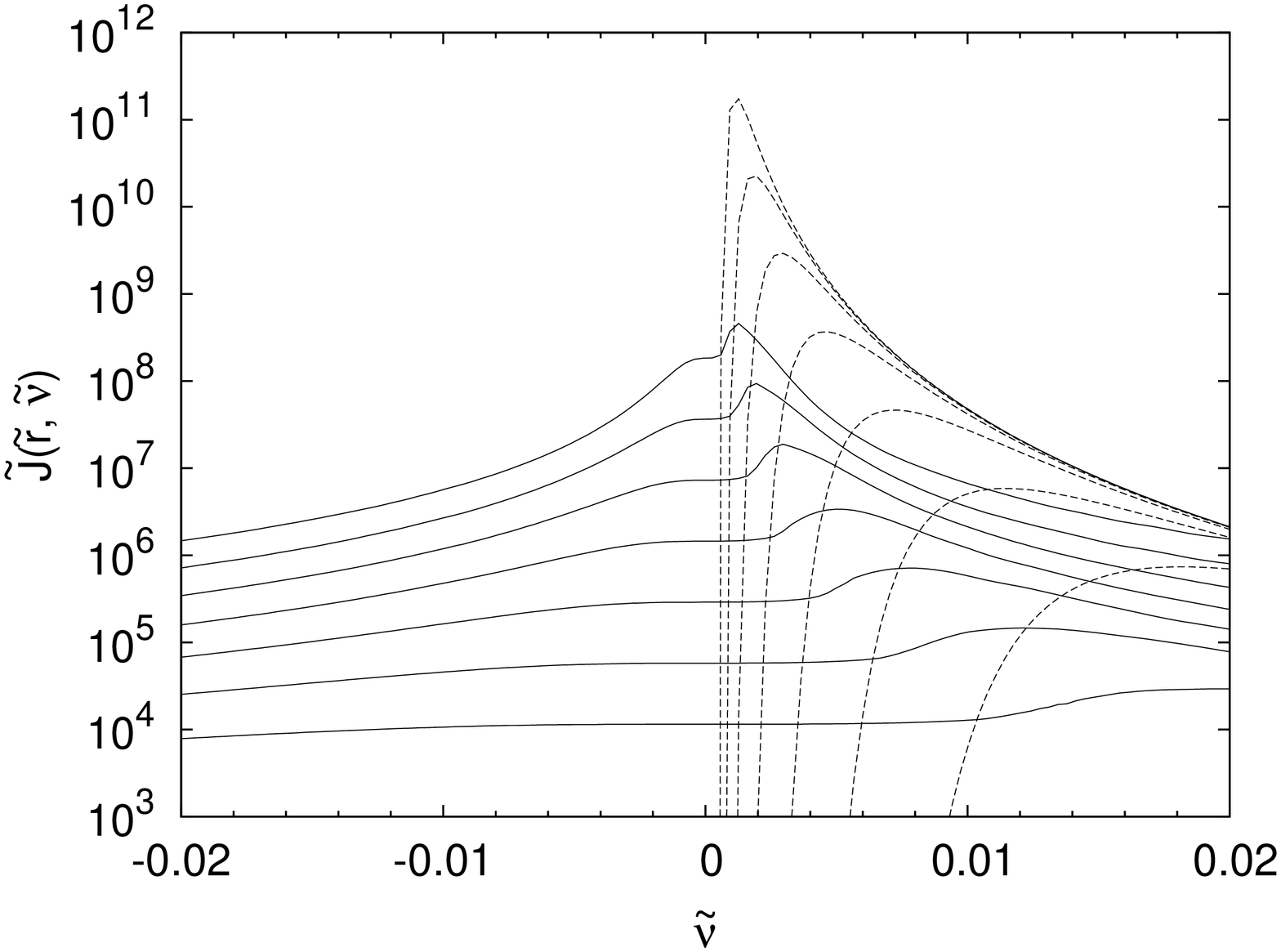, height = 7cm, angle = -90}\\
\epsfig{file = 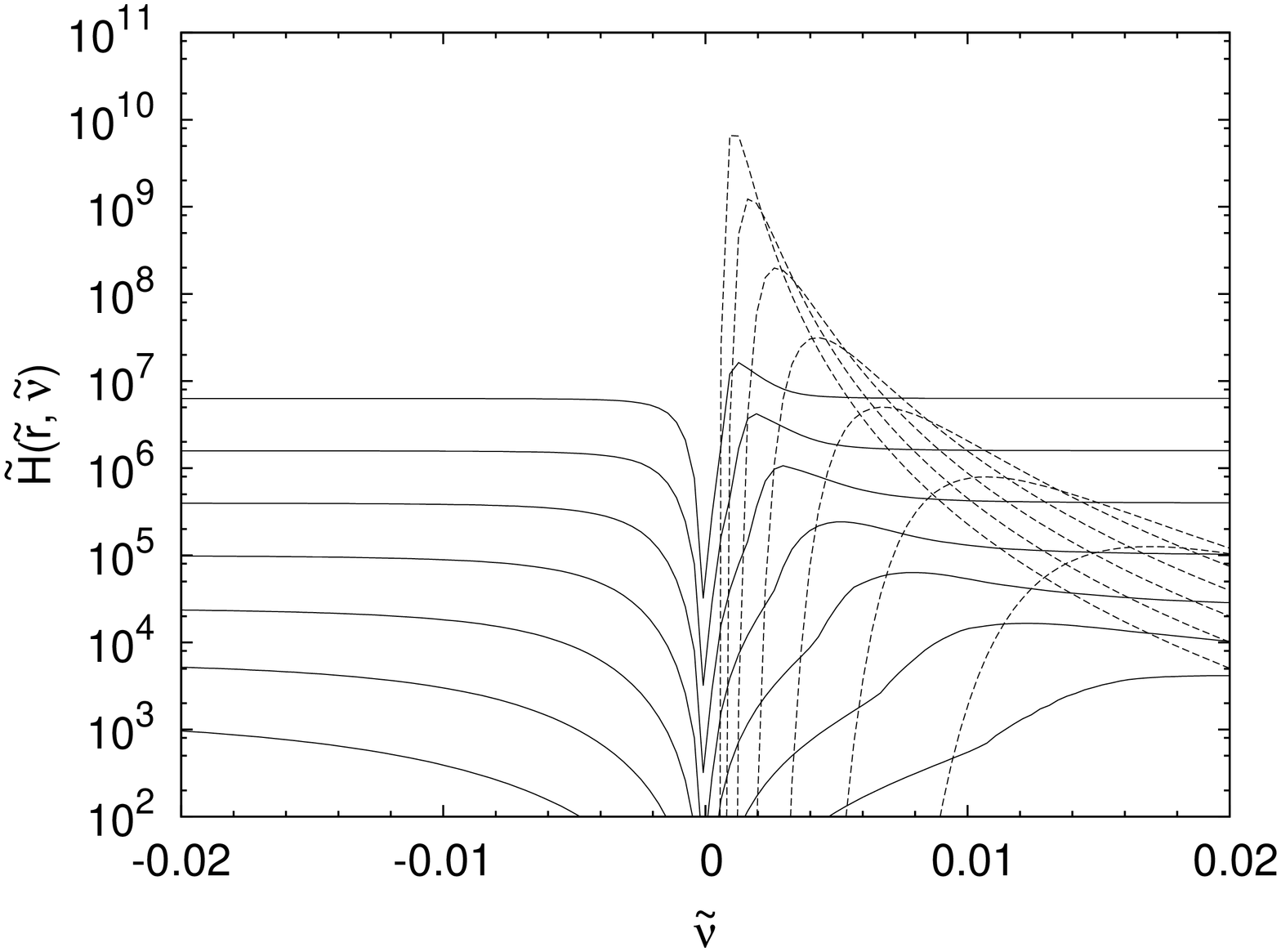, height = 7cm, angle = -90}
\end{center}
\caption{Analytic solution for $\tilde{J}$ (upper panel) and
  $\tilde{H}$ (lower panel) for a spectrally flat source in a
  uniformly expanding homogeneous medium in the zero-temperature
  diffusion approximation (solid lines), together with the
  corresponding analytic solution of Loeb \& Rybicki for a
  monochromatic \Lya source (dashed lines). Values of $\tilde{r}$ are
  given by $10^{-4.5}$, $10^{-4.2}$, $10^{-3.9}$, $10^{-3.6}$,
  $10^{-3.3}$, $10^{-3.0}$, $10^{-2.7}$ in order of decreasing
  $\tilde{J}$ or $\tilde{H}$.}
\label{figure:diff_analytic}
\end{figure}
 
The solutions are shown as a series of frequency profiles in
Fig.~\ref{figure:diff_analytic}. (Some small numerical artefacts
resulting from the numerical integration have been interpolated over
for the purpose of presentation.) We find a peak in both the mean
intensity and the flux that drifts redwards further from the source,
corresponding to the contribution from photons emitted by the source
at the line centre. This is seen from a comparison of the peak
frequency with that found from the analytic solutions for $\tilde{J}$
and $\tilde{H}$ for a monochromatic \Lya source, given by
equations~(\ref{eq:LR_diff}) and (\ref{eq:LR_diff_flux}). By solving
$\partial \tilde{J}/\partial \tilde{\nu} = 0$ for the analytic
solution of Loeb \& Rybicki, we find the frequency of the peak
satisfies $\tilde{\nu} = (3/2)^{1/3}\tilde{r}^{2/3}$.

The most physically relevant aspect of the \Lya solution to the 21cm
signature is the \Lya scattering rate. As shown in the upper panel of
Fig.~\ref{figure:diff_analytic}, $\tilde{J}(x)$ is approximately flat
across the line centre at radii exceeding some minimum distance from
the source. The scattering rate then varies as the average intensity
at line centre,
\begin{equation}
P_{\alpha} \equiv 4\pi \sigma \int J_{\nu}(r) \varphi(\nu) \, \dd
\nu \simeq 4\pi \sigma I_*^c \tilde{J}(\tilde{r},0)
\label{eq:P_flat}
\end{equation}
where
\begin{eqnarray}
 \tilde{J}(\tilde{r}, 0) &=& \frac{2^{7/3}}{3^{1/3}(4\pi)^{5/2}\tilde{r}^{7/3}} \int_{\frac{9\tilde{r}^2}{4|\tilde{\nu}_{\rm m}|^3}}^{\infty} t^{1/6} e^{-t} \, \dd t 
\nonumber\\
&=& \frac{2^{7/3}}{3^{1/3}(4\pi)^{5/2}\tilde{r}^{7/3}} 
\times \Gamma \left(\frac{7}{6}, \frac{9 \tilde{r}^2}{4|\tilde{\nu}_{\rm m}|^3} \right),
\label{eq:J_flat_zero_Gamma}
\end{eqnarray}
and $\Gamma(q, y) \equiv \int_y^{\infty} u^{q-1} e^{-u} \, \dd u$ is
the upper incomplete Gamma function. The solution saturates with
$\Gamma(7/6,y)\rightarrow\Gamma(7/6) \simeq 0.93$ for values of the
argument $y \lesssim 10^{-2}$ or, for our problem, $\tilde{r} \lesssim
0.1 \times (2/3)|\tilde{\nu}_{\rm m}|^{3/2}$. In this limit, the
dimensionless scattering rate $\tilde{P}_{\alpha} = P_{\alpha}/(\sigma
I_*^c)$ becomes
\begin{equation}
  \tilde{P}_{\alpha}^{\rm diff} \equiv \frac{2^{7/3} \Gamma(7/6)}{3^{1/3}(4\pi)^{3/2}\tilde{r}^{7/3}}. 
\label{eq:P_flat_diff}
\end{equation}
This form for the scattering rate applies for all $\tilde{r}$ if
$|\nu_{\rm m}| \rightarrow \infty$ and thus applies in the absence of
any cutoff in the source spectrum. For a finite $|\nu_{\rm m}|$ the
effect of the cutoff is to suppress $\tilde{J}(\tilde{r},0)$ for
sufficiently large $\tilde{r}$:\ physically this corresponds to the
redshifting past line centre of all photons at these radii.

\begin{figure}
\begin{center}
\leavevmode
\epsfig{file = 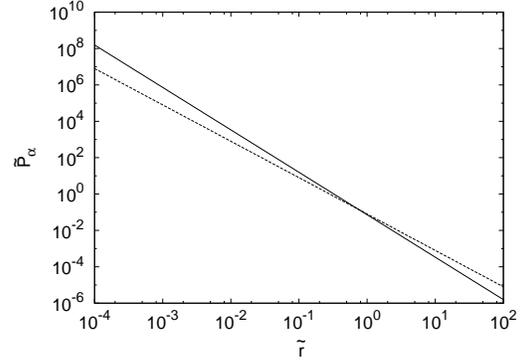, height = 7cm, angle = -90}
\end{center}
\caption{Scaled scattering rate for diffusion solution with flat
  spectrum extending infinitely bluewards, $\tilde{P}_{\alpha}^{\rm
    diff}$ (solid line) and expected result in the corresponding
  free-streaming limit, $\tilde{P}_{\alpha}^{\rm free}$ (dashed
  line).}
\label{figure:P_diff}
\end{figure}

A useful comparison may be made with the expected scattering rate in
the free streaming limit. The solution is $\tilde{J} =
\Theta(\tilde{\nu} - \tilde{\nu}_{\rm m}-\tilde{r})/(4\pi
\tilde{r})^2$ and the corresponding scattering rate for sufficiently
negative $\tilde{\nu}_{\rm m}$ is simply
\begin{equation}
\tilde{P}_{\alpha}^{\rm free} \equiv \frac{1}{4\pi \tilde{r}^2}.
\label{eq:P_flat_free}
\end{equation}
We show the radial dependence of the \Lya scattering rate induced
around the source in Fig.~\ref{figure:P_diff}. The diffusion limit
scattering rate exceeds the corresponding result in the free-streaming
limit close to the source and decreases to below the free-streaming
rate at $\tilde{r} \geq 2^7(\Gamma(7/6))^3/(3(4\pi)^{3/2}) \simeq
0.75$.

The size of the \Lya scattering region around the source is given by
the maximum radius at which the scattering rate is non-zero. This
radius will depend upon $\tilde{\nu}_{\rm m}$, which we take to be the
bluemost frequency at which the source emits photons that are capable
of redshifting into the \Lya line. Typically this cutoff will be the
\Lyb frequency, $\nu_{\beta} = (32/27)\nu_{\alpha}$, as photons
emitted bluewards of \Lyb will redshift into the \Lyb or higher-energy
resonance lines and do not produce \Lya photons (except as products of
radiative cascades which we do not consider here). Assuming a \Lya
source term of the form of equation~(\ref{eq:step_source}), the
scattering rate found in the diffusion limit given by
equation~(\ref{eq:J_flat_zero_Gamma}) is found to decrease below
one-tenth of its value in the absence of a cutoff when the second
argument $y$ of the upper incomplete Gamma function satisfies $y \geq
f \simeq 2.5$, or
\begin{equation}
  \tilde{r} \geq (2/3) f^{1/2} |\tilde{\nu}_{\beta}|^{3/2} \simeq 1.05  |\tilde{\nu}_{\beta}|^{3/2}.
\label{eq:r_max_diff}
\end{equation}
Outside of the diffusion regime, an upper limit for the size of the
\Lya scattering region is given by free streaming:\ all photons will
have redshifted past the line centre at radii $\tilde{r} >
|\tilde{\nu}_{\beta}|$.

The number of scatters a \Lya photon undergoes before escaping the IGM
may be computed as the ratio of the total rate of scatters through a
region of radius $r_{\rm max}$ and the emission rate:
\begin{eqnarray}
  N_{\rm scatt}&=&{{\dot N}_\alpha}^{-1}4\pi\int_0^{r_{\rm
      max}}\,dr\,r^2 n_{\rm H}(z)P_\alpha(r)\nonumber\\
  &\simeq&\left[\frac{{12}^{2/3}}{(4\pi)^{1/2}}\Gamma\left(\frac{7}{6}\right)
    {\tilde{r}_{\rm hor}}^{-1} - {\tilde{r}_{\rm
        hor}}^{-1}+\frac{\tilde{r}_{\rm max}}{\tilde{r}_{\rm
        hor}}\right]\gamma^{-1}\nonumber\\
  &\simeq&\gamma^{-1}\frac{\tilde{r}_{\rm max}}{\tilde{r}_{\rm hor}},
\label{eq:Nscatt}
\end{eqnarray}
where equation~(\ref{eq:P_flat_diff}) is used for $0<\tilde{r}<1$ and
equation~(\ref{eq:P_flat_free}) for $\tilde{r}>1$. Here $r_{\rm
  hor}=(5/27)c/H(z)$ is the `horizon' distance a photon emitted just
longward of the \Lyb resonance frequency may travel before redshifting
into the local \Lya resonance frequency, and the total \Lya photon
production rate is taken as ${\dot N}_\alpha={\dot
  N}_\nu(\nu_\beta-\nu_\alpha)=(5/27){\dot N}_\nu \nu_\alpha$.

For the Wouthuysen-Field effect to compete with the CMB and couple the
spin temperature to the gas kinetic temperature, a critical
thermalization \Lya scattering rate of $P_{\rm th}=27A_{10}T_{\rm
  CMB}/ 4T_*$ is required, where $A_{10}\simeq2.85\times10^{-15}\,
{\rm s}^{-1}$ is the 21-cm transition rate, $T_*\equiv
h\nu_{10}/k_{\rm B}\simeq0.068$~K and $T_{\rm CMB}$ is the temperature
of the Cosmic Microwave Background \citep{MMR97}. For a continuum
source, from equations~(\ref{eq:P_flat}) and (\ref{eq:P_flat_diff})
the critical thermalization photon production rate between the \Lya
and \Lyb resonance line frequencies is
\begin{equation}
{\dot N}_{\alpha, {\rm th}}^{\rm
  cont}\simeq4.39\times10^{55}\left(\frac{1+z}{11}\right) r_{\rm
  Mpc}^{7/3}\,{\rm s}^{-1}
\label{eq:Ndot_cont_diff}
\end{equation}
where $r_{\rm Mpc}$ is the distance from the source in
megaparsecs. Here we adopted properties corresponding to the mean
intergalactic medium at $z = 10$, taking the baryon and dark matter
density parameters $\Omega_b h^2 = 0.022$ and $\Omega_m=0.27$,
respectively, and Hubble constant $H_0=70\,{\rm
  km\,s^{-1}\,Mpc^{-1}}$, consistent with the results of
\citet{2011ApJS..192...18K}. These values give $\nu_* = 1.25 \times
10^{13}[(1+z)/11]^{3/2} \, {\rm Hz}$ and $r_* = 1.12 \, {\rm Mpc}$
(independent of redshift). The Sobolev parameter is $\gamma =
\nu_{\alpha} H/(\sigma c n_{\rm H}) \simeq 1.27 \times
10^{-6}[(1+z)/11]^{-3/2}$. Thus a source with a continuum luminosity
of $\sim2\times10^{11}\,L_\odot$ would be able to couple the spin
temperature to the gas kinetic temperature out to a distance of
1~Mpc. This is far less demanding than for an emission-line
source. From equations~(\ref{eq:P_LR99}) and (\ref{eq:Pt_LR99}), the
required \Lya photon production rate is
\begin{equation}
{\dot N}_{\alpha, {\rm th}}^{\rm
  line}\simeq1.38\times10^{60}\left(\frac{1+z}{11}\right)^4 r_{\rm
  Mpc}^{11/3}\,{\rm s}^{-1}.
\label{eq:Ndot_line_diff}
\end{equation}
An emission-line source with a \Lya luminosity as great as
$10^{12}\,L_\odot$ would be able to couple the spin temperature to the
gas temperature only out to a distance of $\sim90$~kpc. Indeed, at this
distance, if the source were a radio-loud AGN, its radio emission
would likely dominate the coupling of the spin temperature
\citep{MMR97}.

\subsection{Homogeneous Expanding Medium:\ Numerical Results for a
  Continuum Source}
\label{subsect:hmecont_num}

Outside of the zero-temperature regime treated by Loeb \& Rybicki for
a monochromatic source, it is necessary to specify the physical state
of the scattering medium. We adopt properties corresponding to the
mean intergalactic medium at $z = 10$ as above, with an assumed gas
temperature $T = 10 \, {\rm K}$.

\begin{figure}
\begin{center}
\leavevmode
\epsfig{file = 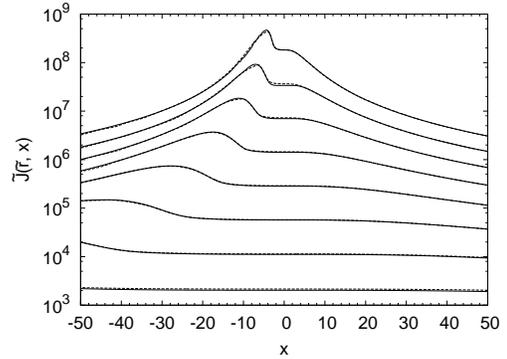, height = 7cm, angle = -90}
\end{center}
\caption{Diffusion approximation moment solution for $\tilde{J}$ in a
  uniformly expanding homogeneous medium assuming coherent scattering
  and a Voigt line profile (solid line), together with the
  corresponding analytic solution in the wing approximation (dashed
  line). The frequency dependence is shown as a function of $x$
  assuming a temperature $T = 10 \, {\rm K}$ for values of $\tilde{r}$
  given by $10^{-4.5}$, $10^{-4.2}$, $10^{-3.9}$, $10^{-3.6}$,
  $10^{-3.3}$, $10^{-3.0}$, $10^{-2.7}$, $10^{-2.4}$.}
\label{figure:flat_moment_diff}
\end{figure}

In solving the moment equations, the initial condition in frequency is
enforced by taking our frequency grid to start from $\tilde{\nu}_1 <
\tilde{\nu}_{\rm m}$, i.e. the frequency grid must extend bluewards of
the source emission cutoff frequency. Unless otherwise stated, we have
taken $x_{\rm m} = -(\nu_*/\Delta \nu_{\rm D}) \tilde{\nu}_{\rm m} =
1000$ for all solutions. We use two frequency grids, one with a small
frequency spacing across the line centre for $x$ lying in the range
$[-x_{\rm max}, x_{\rm max}]$ where $x_{\rm max} = 60$ (unless stated
otherwise), and a second grid with a larger frequency spacing across
$[x_{\rm max}, x_{1}]$. We adopt the analytic solutions to satisfy the
inner and outer boundary conditions for the moment equations in the
diffusion approximation. We solve the moment equations with the
addition of the full Voigt profile form for $\chi_{\nu}$ with an
opacity given by $\tilde{\chi} = \sqrt{\pi} r_* \kappa_0 \phi_{\rm
  V}(x)$. The diffusion limit is enforced by setting $f_{\nu}(r) =
1/3$, and the appropriate matrix elements representing the frequency
derivative of the flux in the solution method are set to zero to match
the diffusion equation. Our moment solution for $\tilde{J}$ is shown
in Fig.~\ref{figure:flat_moment_diff}. It nearly coincides with the
analytic solution, demonstrating that the assumption of the wing form
of the opacity does not affect the solution in this approximation. A
similarly identical solution was found for the flux. As a consequence,
the solutions as a function of frequency are very insensitive to the
temperature of the medium in the coherent scattering approximation.

\begin{figure}
\begin{center}
\leavevmode
\epsfig{file = 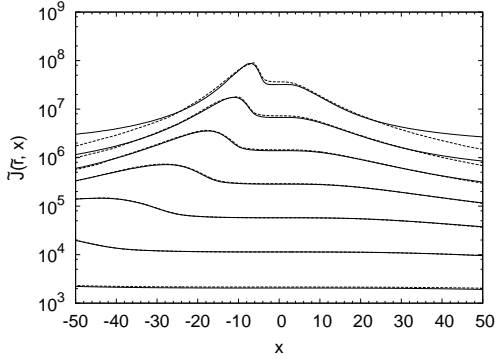, height = 7cm, angle = -90}
\end{center}
\caption{Coherent scattering solution for $\tilde{J}$ obtained from
  ray and moment equations (solid lines), compared with the analytic
  solutions in the zero-temperature diffusion approximation (dashed
  lines). Frequency profiles correspond to radii $\tilde{r} =
  10^{-4.2}, 10^{-3.9}, 10^{-3.6}, 10^{-3.3}, 10^{-3.0}, 10^{-2.7},
  10^{-2.4}$.}
\label{figure:ray_moment_diff}
\end{figure}

We solved the problem using the ray/moment method applied to the exact
forms of the ray and moment equations for the uniformly expanding
homogeneous medium problem. Again, we used the Voigt profile form for
$\tilde{\chi}$ and assumed coherent scattering. We assumed the
diffusion limit solutions given by
equations~(\ref{eq:J_analytic_calc}) and (\ref{eq:H_analytic_calc}) in
specifying the inner boundary condition and the initial estimate of
the source function used in solving the ray equations. Note that for
this and subsequent problems where the conditions do not match those
that determine the solution used for the boundary condition, the
ray/moment solution at the boundary is subject to an unphysical
constraint and we should not assume it is accurate on the boundary.

Our ray/moment solution is displayed in
Fig.~\ref{figure:ray_moment_diff}. For distances sufficiently far from
the source we find the ray/moment solution is a close match to the
diffusion approximation solution. Closer to the source we find both
solutions match reasonably well across the line centre, while away
from line centre there is some deviation. The ray/moment solution
approaches a flat profile closer to the free-streaming solution. This
is an expected consequence of the negligible optical depth out to
these radii away from line centre. We have checked the Eddington
factor and found $f_\nu\simeq1/3$ except in the wings, with the
deviation from $1/3$ matching the deviation in the wings between the
two solutions shown in Fig.~\ref{figure:ray_moment_diff}. We found
that the flux is altered from the analytic solution to a similarly
limited extent.

The accuracy of the solutions may be checked by use of an integral
constraint on the number of photons obtained from the radiative
transfer equation. It may be derived from the zeroth-order moment
equation obtained from equation~(\ref{eq:RT_gen_scaled}) with $V = Hr$
and $\delta(\tilde{\nu}) \rightarrow 1$ for a continuum source in a
homogeneous expanding medium by multiplying it by $\tilde{r}^2$,
integrating in radius from $\tilde{r} = 0$ to $\tilde{R}$ and in
frequency across the line centre from $-\tilde{\nu}_{\rm max}$ to
$+\tilde{\nu}_{\rm max}$, causing the terms corresponding to the
opacity and emissivity to cancel out due to radiative equilibrium. The
result is
\begin{equation}
\begin{split}
\int_0^{\tilde{R}} \tilde{r}^2 \left[ \tilde{J}(\tilde{r},\tilde{\nu}_{\rm
  max}) - \tilde{J}(\tilde{r}, -\tilde{\nu}_{\rm max}) \right] \, \dd \tilde{r} \hspace{2cm} \\ 
 \hspace{1.6cm} + \tilde{R}^2 \int_{-\tilde{\nu}_{\rm max}}^{\tilde{\nu}_{\rm max}}
\tilde{H}(\tilde{R},\tilde{\nu}) \, \dd \tilde{\nu} = \frac{2\tilde{\nu}_{\rm max}}{(4\pi)^2}.
\end{split}
\label{eq:HEM_check}
\end{equation} 
We have verified that all the solutions examined in this section
satisfy this constraint.

\begin{figure}
\begin{center}
\leavevmode
\epsfig{file = 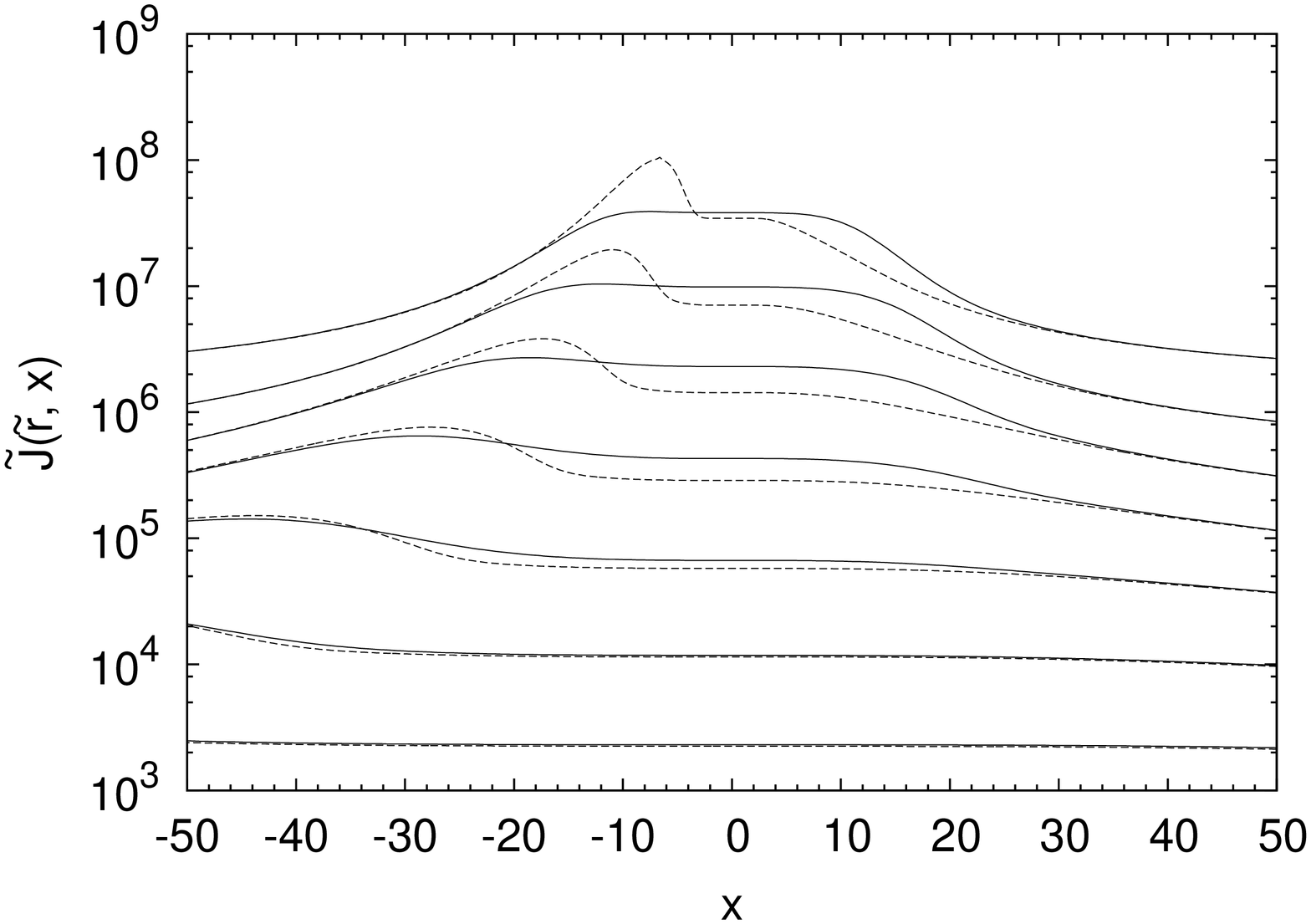, height = 7cm, angle = -90}
\end{center}
\caption{R${\rm II}$ redistribution solution for $\tilde{J}$ obtained
  from ray and moment equations (solid lines), compared with the
  corresponding solution for coherent scattering (dashed
  lines). Frequency profiles correspond to radii $\tilde{r} =
  10^{-4.2}, 10^{-3.9}, 10^{-3.6}, 10^{-3.3}, 10^{-3.0}, 10^{-2.7},
  10^{-2.4}$.}
\label{figure:ray_moment_RII}
\end{figure}

\begin{figure}
\begin{center}
\leavevmode
\epsfig{file = 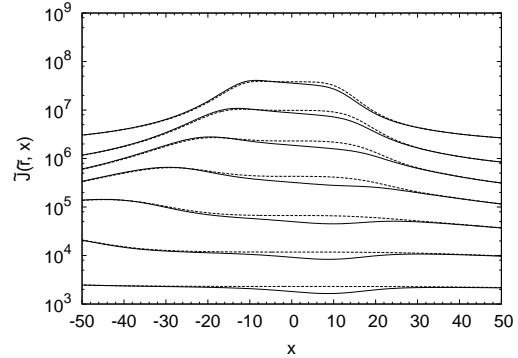, height = 7cm, angle = -90}
\end{center}
\caption{R${\rm II}$ redistribution solution with recoil obtained from
  ray and moment equations (solid lines), compared with the
  corresponding solution for RII redistribution without recoil (dashed
  lines). Frequency profiles correspond to radii $\tilde{r} =
  10^{-4.2}, 10^{-3.9}, 10^{-3.6}, 10^{-3.3}, 10^{-3.0}, 10^{-2.7},
  10^{-2.4}$.}
\label{figure:ray_moment_rec}
\end{figure}

The assumption of coherent scattering is valid for photons scattering
at frequencies far from line centre, but as we examine photons
redshifting across the line centre such an assumption is not
physically motivated. We now treat the added effect of Doppler shifts
due to the thermal- and recoil-induced atomic velocities. The thermal
velocities of the atoms are responsible for the R${\rm II}$ frequency
redistribution function for \Lya scattering. For convergence with
recoils, we found it necessary to increase $x_{\rm max}$ to 200. We
first consider the addition of RII redistribution in isolation by
ignoring the effect of recoils. Ray/moment equation solutions are
obtained similarly to the coherent solution with the R${\rm II}$ form
for the coefficients $\mathcal{R}_{k',k,d}$ with $\epsilon = 0$ as
given in equation~(\ref{eq:Rfunk_diff_easy}) in
Appendix~\ref{ap:RMsource}. We display our solution for $\tilde{J}$ in
Fig.~\ref{figure:ray_moment_RII}. There is some redistribution of
photons across the line centre relative to the red peak found in the
coherent scattering case, resulting in a small boost in the mean
intensity across the line centre over a limited range in radii. The
difference in the spectrum from the coherent scattering case is not
particularly significant at large radii.

We add the effect of atomic recoil by seeking ray/moment solutions
without setting the recoil parameter $\epsilon$ to zero. The resulting
solution is displayed in Fig.~\ref{figure:ray_moment_rec}. The
expected Boltzmann distribution gradient is recovered across the line
centre.

\begin{figure}
\begin{center}
\leavevmode
\epsfig{file = 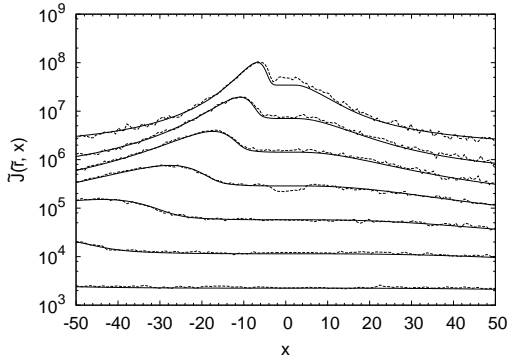, height = 7cm, angle = -90}
\end{center}
\caption{Monte Carlo solution (dashed lines) for coherent scattering,
  together with the corresponding ray/moment equation solution (solid
  lines). Frequency profiles correspond to radii $\tilde{r} =
  10^{-4.2}, 10^{-3.9}, 10^{-3.6}, 10^{-3.3}, 10^{-3.0}, 10^{-2.7},
  10^{-2.4}$.}
\label{figure:flat_MC}
\end{figure}

\begin{figure}
\begin{center}
\leavevmode
\epsfig{file = 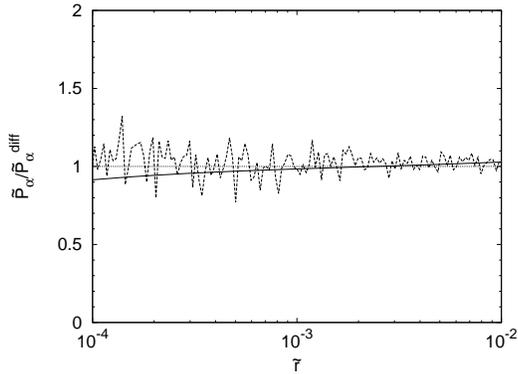, height = 7cm, angle = -90}
\end{center}
\caption{Scattering rates derived from numerical integration of
  ray/moment equation method (solid line) and Monte Carlo solution
  (long-dashed line), both for coherent scattering; normalised by the
  analytic result in the diffusion limit.}
\label{figure:P_MC_comp}
\end{figure}

We also apply the Monte Carlo method described in
Section~\ref{sect:MC_method} and Appendix~\ref{ap:MC} to the
problem. We assume a continuum source that emits $\dot{N}_{\nu}$
photons per second per Hz across the frequency range $[x_{\rm min},
x_{\rm max}]$ where $\dot{N}_{\nu}$ is independent of $\nu$,
\begin{equation}
\dot{N}_{\nu} \Delta \nu = \frac{N}{\Delta t} \times \left( \frac{\Delta x}{x_{\rm max} - x_{\rm min}} \right)
\label{eq:MC_cont_norm}
\end{equation}
where $\Delta \nu$ is the frequency bin width and the factor in
brackets is the fraction of the $N$ photons emitted within a single
frequency bin. In the case of a source cutoff at $x_{\rm m}$, we take
$x_{\rm max} = x_{\rm m}$. The solution is normalized by using $\Delta
t$ from equation~(\ref{eq:MC_cont_norm}) in
equation~(\ref{eq:J_path}). The resulting dependence on the source
strength $\dot{N}_{\nu}$ is scaled out when expressing the solution in
terms of $\tilde{J}$. We obtained a solution for coherent scattering
using $N = 2 \times 10^5$ photon packets. The solution is displayed in
Fig.~\ref{figure:flat_MC}. It closely matches the ray/moment
solution. The corresponding scattering rate is shown in
Fig.~\ref{figure:P_MC_comp}. It closely resembles the ray/moment
solution scattering rate, which itself is nearly coincident over this
range in radius with the diffusion approximation power law of
equation~(\ref{eq:P_flat_diff}). The noise level of the Monte Carlo
solution is a few to several percent per radial bin, decreasing with
increasing radius. Two hundred logarithmically spaced radial bins were
used to cover the range $-4.5<\log_{10}{\tilde r}<-1.5$.

\begin{figure}
\begin{center}
\leavevmode
\epsfig{file = 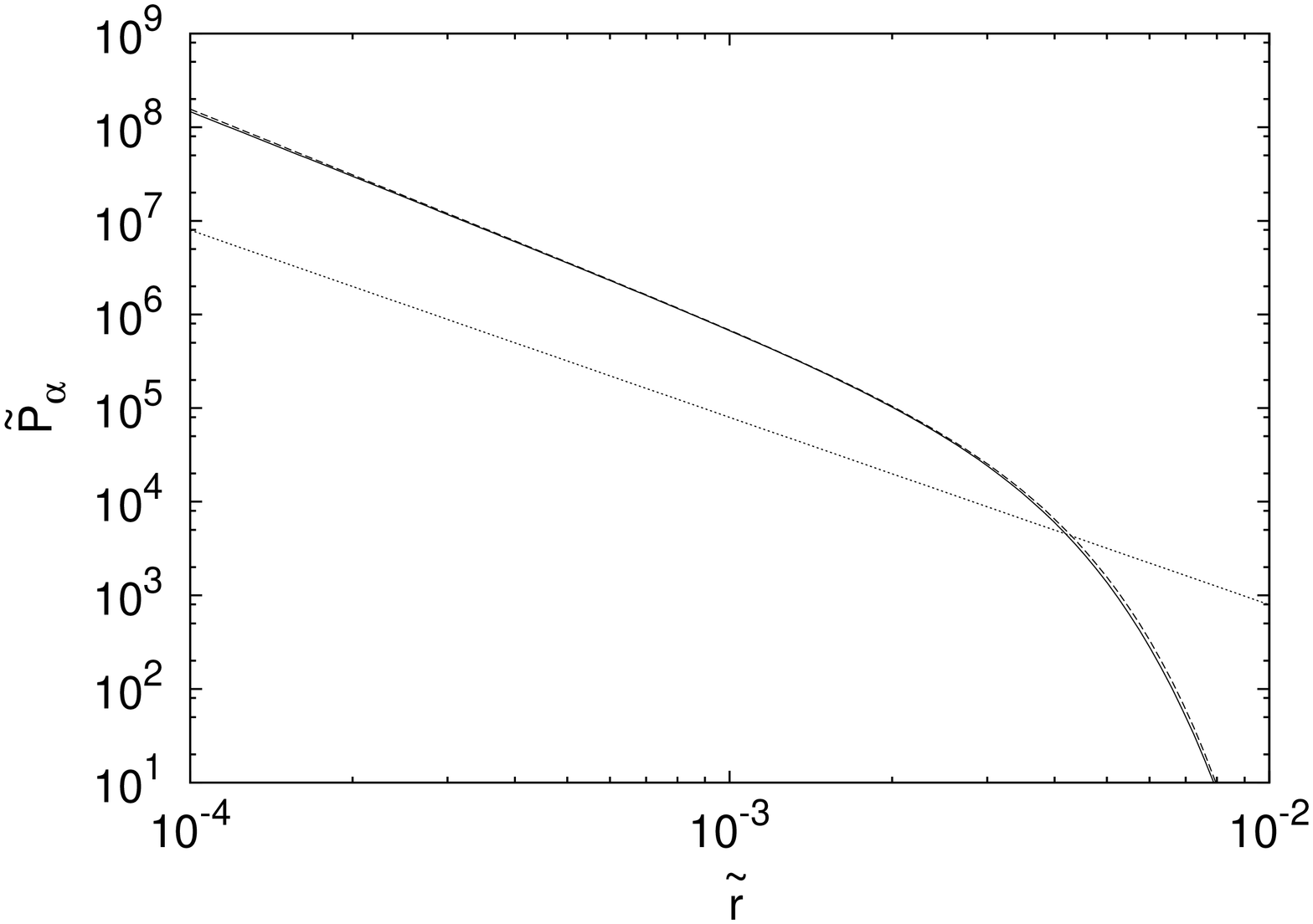, height = 7cm, angle = -90}\\
\epsfig{file = 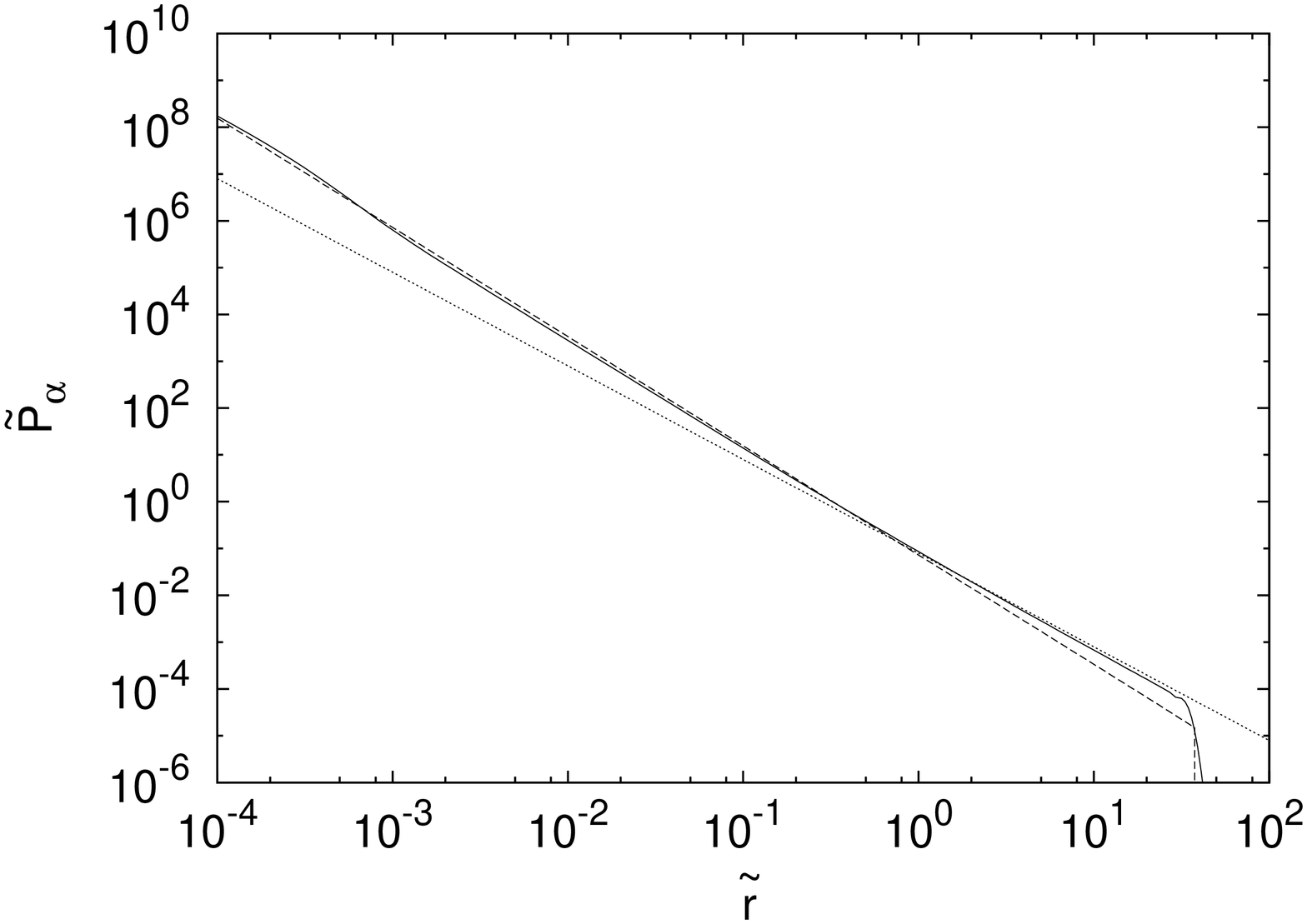, height = 7cm, angle = -90}
\end{center}
\caption{Scaled scattering rate for the IGM at $z=10$ with an assumed
  temperature $T=10$~K, derived from numerical integration of the
  scattering solution obtained from the ray/moment equations (solid
  line) and the analytic diffusion solution (dashed line, nearly
  coincident with the solid line in the upper panel), for source
  cutoff frequency $x_{\rm m} = 100$ and coherent scattering (upper
  panel) and $x_{\rm m} = 1.4 \times 10^5$ including RII
  redistribution and recoil (lower panel). For comparison we also show
  $\tilde{P}_{\alpha}^{\rm free}$ (dotted line).}
\label{figure:P_xinj}
\end{figure}

\begin{figure}
\begin{center}
\leavevmode
\epsfig{file = 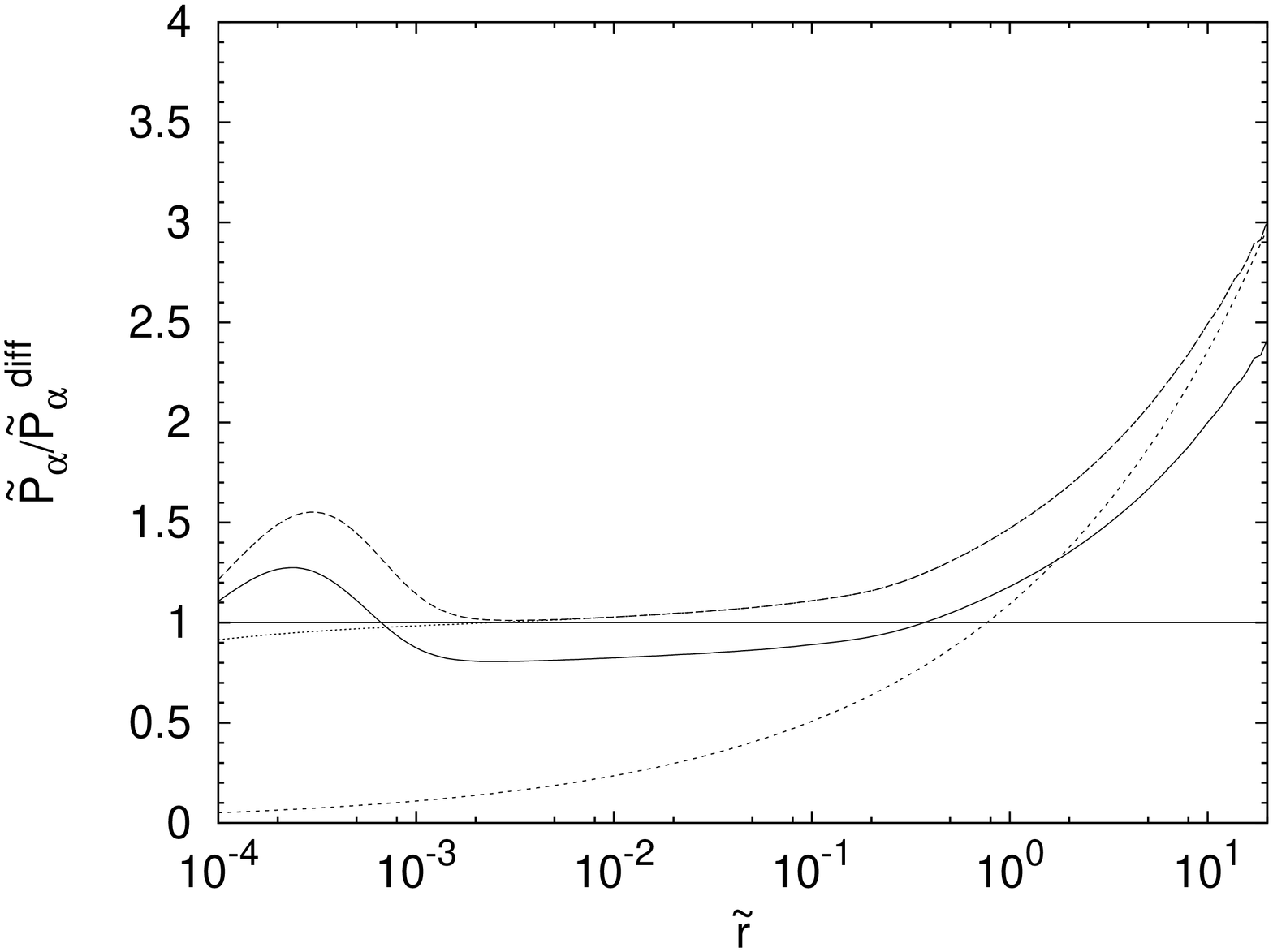, height = 7cm, angle = -90}
\epsfig{file = 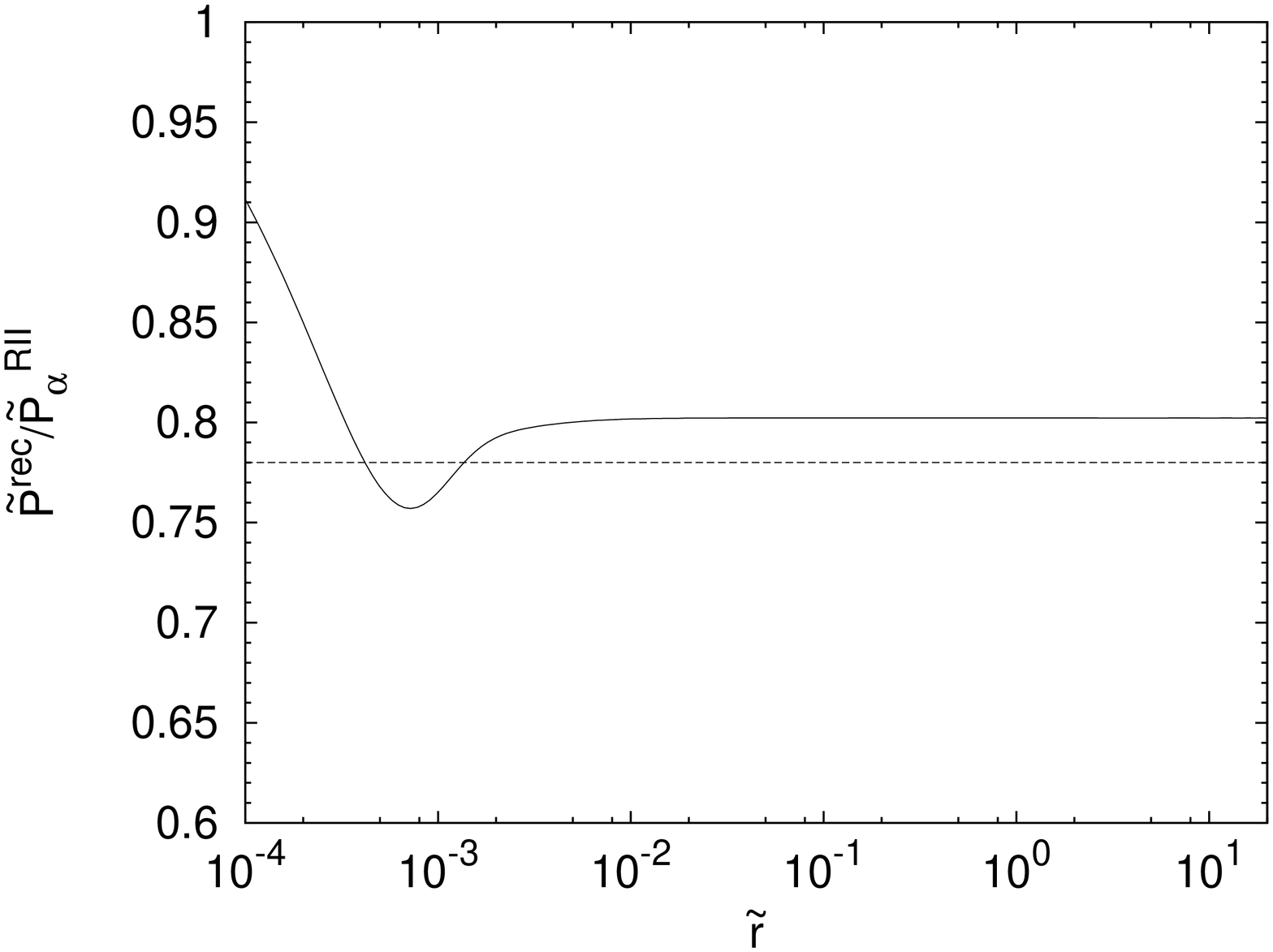, height = 7cm, angle = -90}
\end{center}
\caption{Upper panel: Ratio of scattering rate derived from numerical
  integration of the appropriate ray/moment solution, to the analytic
  diffusion limit result $\tilde{P}_{\alpha}^{\rm diff}$, for RII
  redistribution with recoil (solid line), RII redistribution without
  recoil (dashed line) and coherent scattering without recoil, which
  coincides with the RII redistribution without recoil solution for
  $\tilde{r}>10^{-2.5}$ (dotted line). Also shown is the
  free-streaming result (short-dashed line). All solutions assume $T =
  10 \, {\rm K}$ and $x_{\rm m} = 1.4\times10^5$. Lower panel:\ Ratio
  of scattering rate assuming RII redistribution with recoil to that
  assuming RII redistribution alone (solid line), compared with the
  estimate $S_{\alpha} = 0.78$ (dashed line) of the suppression factor
  in a uniformly expanding homogeneous medium for a uniform radiation
  field \citep{2006MNRAS.372.1093F}.}
\label{figure:P_hem_comp}
\end{figure}

We have determined the scattering rate profile from the ray/moment
solution. Fig.~\ref{figure:P_xinj} shows two examples with different
values of the source upper frequency cutoff $\nu_{\rm m}$. The upper
panel has $x_{\rm m} = 100$ and thus $|\tilde{\nu}_{\rm m}| \simeq
10^{-1.6}$. The `diffusion-limited' effective radius of the \Lya
scattering region, as given by equation~(\ref{eq:r_max_diff}) for
general $\tilde{\nu}_{\rm m}$, is $\tilde{r} = 1.05 |\tilde{\nu}_{\rm
  m}|^{3/2} \simeq 10^{-2.3}$, in approximate agreement with the
figure.

The lower panel has $x_{\rm m} = x_{\beta}$, where $x_{\beta} =
5\nu_{\alpha}/(27 \Delta \nu_{\rm D}) \simeq 1.4 \times 10^5$ denotes
the source cutoff corresponding to the \Lyb frequency. This
corresponds to $-\tilde\nu_{\rm m}\simeq10^{1.6}$. Photons with larger
values of $-\tilde\nu$ would not be able to redshift into the \Lya
resonance at any greater distance, as they would encounter the \Lyb or
a higher order Lyman resonance en route. The limiting distance the
\Lya photons may travel is ${\tilde r}_{\rm
  hor}\simeq38[(1+z)/11]^{-3/2}$ (the Ly$\alpha$ horizon). For
$\tilde{r} \lta 1$, the diffusion limit applies and the scattering
rate takes on the approximate $\tilde{r}^{-7/3}$ scaling expected. At
very small values, $\tilde{r}\lta10^{-3}$, frequency redistribution
substantially modifies the scattering rate, as shown in
Fig.~\ref{figure:P_hem_comp}. As discussed in Appendix~\ref{ap:hem},
the diffusion approximation breaks down at $\tilde{r}\gta1$ and the
scattering rate takes on a profile closer to the free-streaming value
$1/4\pi\tilde{r}^2$, as found in the Monte Carlo computations of
\citet{2007A&A...474..365S}. The scattering rate becomes vanishingly
small at $\tilde{r}>29$, as shown in Fig.~\ref{figure:P_xinj}. This is
slightly shorter than the \Lya horizon distance by the factor $0.77$
and appears to be a consequence of the causal limitation of the
radiation field (see Loeb \& Rybicki 1999).

The $1/r^2$ behaviour of the mean intensity at large distances has an
interesting implication for the global radiation field produced by a
uniform distribution of sources of spatial number density $n_0$. The
total scattering rate is limited by the maximum distance ${\tilde
  r}_{\rm max}$ the \Lya photons travel:
\begin{equation}
{\tilde P}_\alpha^{\rm tot} = 4\pi(n_0 r_*^3)\int_0^{{\tilde r}_{\rm
  max}}\,d{\tilde r}\,{\tilde r}^2\frac{1}{4\pi{\tilde
    r}^2}\simeq(n_0 r_*^3){\tilde r}_{\rm max}.
\label{eq:Pa_tot}
\end{equation}
The maximum distance is given by the causal limitation radius ${\tilde
  r}_{\rm max}={\tilde r}_{\rm causal}\simeq 29 \lta {\tilde r}_{\rm
  hor}$. Given that the scattering rate at large distances somewhat
exceeds $1/4\pi{\tilde r}^2$ within the horizon, however, taking the
maximum radius to be the horizon radius gives a very good
approximation to the total scattering rate, and corresponds to $N_{\rm
  scatt}\simeq\gamma^{-1}$.

It is instructive to compare this result with the estimate of
\citet{1959ApJ...129..536F}, who expressed the scattering rate as the
product of the production rate of \Lya photons per neutral hydrogen
atom and the number of scatters $N_{\rm scatt}$ a photon undergoes
before it redshifts sufficiently far from line centre to escape:\
$P_\alpha=(n_0 \dot N_\alpha/n_{\rm H})N_{\rm scatt}$. He argued that
for uniformly distributed sources in a homogeneous and isotropic
medium, $N_{\rm scatt}=\gamma^{-1}$ (cf. Higgins \& Meiksin
2009). Allowing for all photons emitted between the \Lya and \Lyb
frequency resonances, this corresponds to the dimensionless scattering
rate ${\tilde P}_\alpha^{\rm Field}=(n_0 r_*^3){\tilde r}_{\rm
  hor}$. For a uniform radiation field, the total scattering rate is
thus again given by equation~(\ref{eq:Pa_tot}), with the maximum
radius taking on the natural value ${\tilde r}_{\rm max}={\tilde
  r}_{\rm hor}$.

Allowing for recoils suppresses the scattering rate relative to the
R${\rm II}$ case without recoil, as shown in the lower panel of
Fig.~\ref{figure:P_hem_comp}. The suppression factor $S_{\alpha}$
computed in the diffusion approximation by \cite{2006MNRAS.372.1093F}
for an isotropic and homogeneous distribution of uniform sources in a
uniformly expanding homogeneous medium is given by $S_{\alpha} \simeq
0.78$. This is close to the average value of the radius-dependent
suppression factor determined from our solutions outwith the diffusion
approximation.

\section{Applications to Inhomogeneous Media}
\label{sect:InhomogTest}

\subsection{Overdense Shell}

\begin{figure}
\begin{center}
\leavevmode
\epsfig{file = 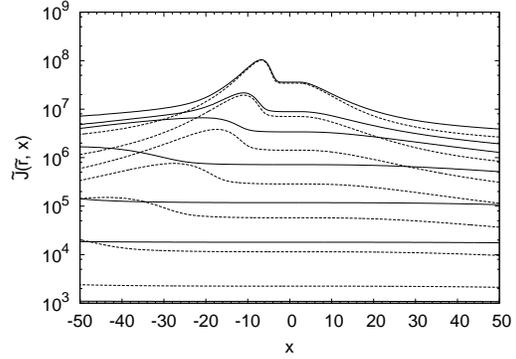, height = 7cm, angle = -90}
\end{center}
\caption{Coherent scattering solution obtained from ray and moment
  equations for the overdense shell given by
  equation~(\ref{eq:dens_clump}) (solid lines), compared with the
  corresponding solution for a uniform density $n_{\rm H}(z)$ (dashed
  lines). Frequency profiles correspond to radii $\tilde{r} =
  10^{-4.2}, 10^{-3.9}, 10^{-3.6}, 10^{-3.3}, 10^{-3.0}, 10^{-2.7},
  10^{-2.4}$.}
\label{figure:ray_moment_dens}
\end{figure}

\begin{figure}
\begin{center}
\leavevmode
\epsfig{file = 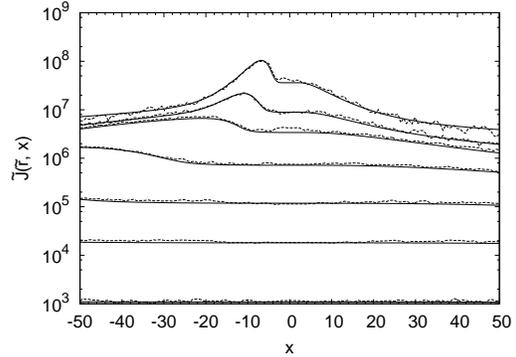, height = 7cm, angle = -90}
\end{center}
\caption{Monte Carlo solution (dashed lines) for coherent scattering
  with the density profile specified in
  equation~(\ref{eq:dens_clump}), compared with the corresponding
  ray/moment equation solution (solid lines). Frequency profiles
  correspond to radii $\tilde{r} = 10^{-4.2}, 10^{-3.9}, 10^{-3.6},
  10^{-3.3}$, $10^{-3.0}, 10^{-2.7}, 10^{-2.4}$.}
\label{figure:dens_MC}
\end{figure}

\begin{figure}
\begin{center}
\leavevmode
\epsfig{file = 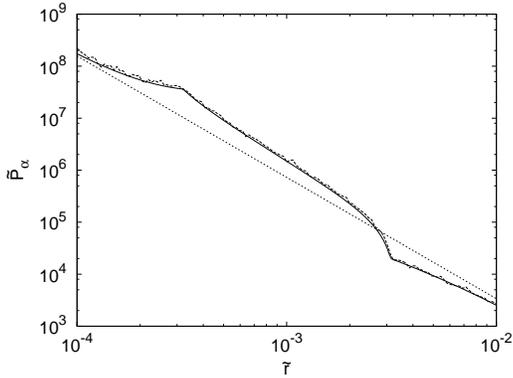, height = 7cm, angle = -90}
\end{center}
\caption{Scaled scattering rate derived from numerical integration of
  ray/moment equation solution (solid line) and Monte Carlo solution
  (long-dashed line) for coherent scattering in a medium with an
  overdense shell given by equation~(\ref{eq:dens_clump}). For
  comparison we also show $\tilde{P}_{\alpha}^{\rm diff}$
  (short-dashed line).}
\label{figure:P_MC_dens}
\end{figure}

As an example application for an inhomogeneous density profile, we
examine the continuum source \Lya scattering problem for a scattering
medium undergoing Hubble expansion but including an
overdense shell between $10^{-3.5} < \tilde{r} < 10^{-2.5}$:
\begin{equation}
  n_{\rm H}(r) = \begin{cases} 10 \times n_{\rm H}(z) & ; \quad 10^{-3.5} \leq \tilde{r} \leq 10^{-2.5}\\
    n_{\rm H}(z) & ; \quad \tilde{r} < 10^{-3.5}, \ \tilde{r} > 10^{-2.5}. \end{cases}
\label{eq:dens_clump}
\end{equation}
The IGM temperature is $T=10$~K. This affects the spatial dependence
of the opacity coefficients $\chi_{k,d}$ for the ray/moment equations
method. We show the ray/moment solution to the problem for coherent
scattering in Fig.~\ref{figure:ray_moment_dens}, while the
corresponding scattering rate is denoted by the solid line in
Fig.~\ref{figure:P_MC_dens}. Compared with the spectrum for a uniform
density scattering medium, the frequency peaks at the overdense radii
are displaced further redward, while the scattering rate across this
range is boosted. At the outer edge of the shell the scattering rate
dips below the uniform density value.

The corresponding Monte Carlo solution to the problem is shown in
Fig.~\ref{figure:dens_MC}. ($N=2\times10^5$ photon packets were used.)
Significantly more time was required to follow the photon packets
through the overdensity compared with the homogeneous density
case. The corresponding scattering rate is displayed in
Fig.~\ref{figure:P_MC_dens}.

\subsection{Quadratic Velocity Profile}

\begin{figure}
\begin{center}
\leavevmode
\epsfig{file = 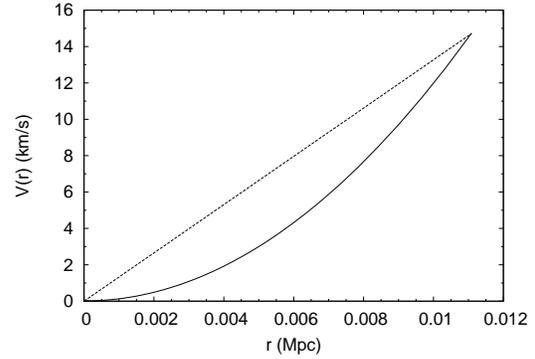, height = 7cm, angle = -90}
\end{center}
\caption{Velocity profile given by the quadratic relation of
  equation~(\ref{eq:V_quad}) with $\tilde{R}_{\rm min} = 10^{-5}$,
  $\tilde{R}_{\rm max} = 10^{-2}$ and $V_{\rm max/min} = H R_{\rm
    max/min}$ (solid line), compared with the Hubble law $V(r) = H r$
  (dashed line).}
\label{figure:V_quad_comp}
\end{figure}

\begin{figure}
\begin{center}
\leavevmode
\epsfig{file = 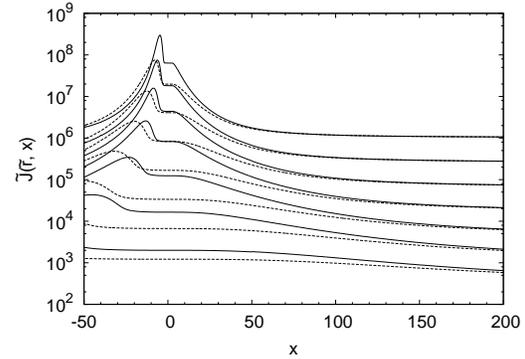, height = 7cm, angle = -90}
\end{center}
\caption{Coherent scattering solution obtained from ray and moment
  equations with the quadratic velocity profile of
  equation~(\ref{eq:V_quad}) (solid lines), compared with the
  corresponding solution for a Hubble velocity profile (dashed
  lines). Frequency profiles correspond to radii $\tilde{r} =
  10^{-4.1}, 10^{-3.8}, 10^{-3.5}, 10^{-3.2}, 10^{-2.9}, 10^{-2.6},
  10^{-2.3}$.}
\label{figure:ray_moment_quad}
\end{figure}

\begin{figure}
\begin{center}
\leavevmode
\epsfig{file = 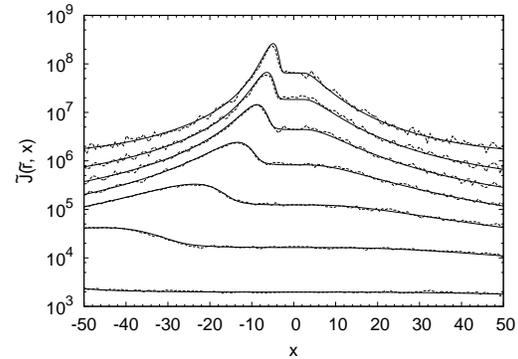, height = 7cm, angle = -90}
\end{center}
\caption{Monte Carlo solution (dashed lines) for coherent scattering
  with the quadratic velocity profile of equation~(\ref{eq:V_quad}),
  compared with the corresponding ray/moment equation solution (solid
  line). Frequency profiles correspond to radii $\tilde{r} =
  10^{-4.1}, 10^{-3.8}, 10^{-3.5}, 10^{-3.2}, 10^{-2.9}, 10^{-2.6},
  10^{-2.3}$.}
\label{figure:quad_MC}
\end{figure}

\begin{figure}
\begin{center}
\leavevmode
\epsfig{file = 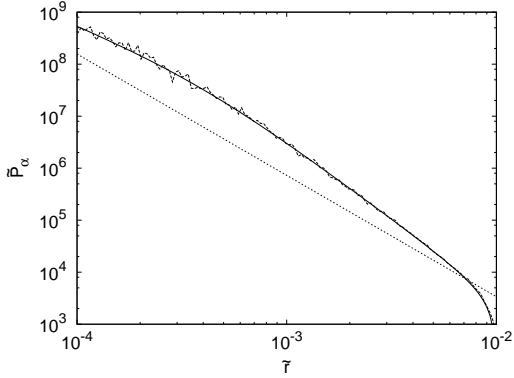, height = 7cm, angle = -90}
\end{center}
\caption{Scaled scattering rate for a uniform medium with a velocity
  profile given by equation~(\ref{eq:V_quad}), derived from numerical
  integration of the ray/moment equation method (solid line) for
  coherent scattering, together with the corresponding scattering rate
  obtained from the Monte Carlo method (long-dashed line). For
  comparison we also show $\tilde{P}_{\alpha}^{\rm diff}$
  (short-dashed line) given by equations~(\ref{eq:P_flat}) and
  (\ref{eq:J_flat_zero_Gamma}). The scattering rate decreases at large
  $\tilde r$ near the outer boundary.}
\label{figure:P_MC_quad_comp}
\end{figure}

In this section we consider a continuum source in a medium of uniform
density, however we move beyond the assumption of a simple Hubble
velocity profile $V = Hr$. To examine the specific effects of a
non-linear velocity profile alone, we consider the case of a quadratic
velocity law. We parametrize the radial velocity profile $V(r)$ for
$R_{\rm min} < r < R_{\rm max}$ by
\begin{equation}
  V(r) = (V_{\rm max} - V_{\rm min}) \left( \frac{r-R_{\rm min}}{R_{\rm max}-R_{\rm min}} \right)^2 + V_{\rm min}.
\label{eq:V_quad}
\end{equation} 
We show both the quadratic velocity profile given by
equation~(\ref{eq:V_quad}) and the corresponding Hubble profile in
Fig.~\ref{figure:V_quad_comp}, where we have taken $V_{\rm max} \equiv
V(R_{\rm max}) = H R_{\rm max}$ and $V_{\rm min} \equiv V(R_{\rm min})
= H R_{\rm min}$. Using this velocity profile we obtain \Lya radiative
transfer solutions from the ray/moment method. The IGM temperature is
$T=10$~K. The solution assuming coherent scattering is displayed in
Fig.~\ref{figure:ray_moment_quad}. In order to isolate the effect of
the velocity profile, the solution is compared with the equivalent
ray/moment solution for Hubble flow. The solution presents the same
qualitative features as the coherent scattering solution for the
Hubble velocity profile, in particular the peak in intensity that
drifts redwards from line centre further from the source, although
significant quantitative differences are apparent. The frequency
displacement of the peak increases more slowly with radius than in the
Hubble velocity case, a result of a reduced velocity across the range
in radius which is less efficient in redshifting photons.

The computations expended by the Monte Carlo code are significantly
reduced if we may assume the photon occupies only a single frequency
bin between scattering or boundary crossing events, which we take to
be that corresponding to the `final' frequency prior to scattering. In
a Hubble velocity field the comoving frequency changes with path
length $\lambda$ as $x = x_{\rm em} - (H/b)\lambda$, and the
approximate treatment is accurate in the limit that the path length
between scattering or boundary crossing events satisfies $\delta
\lambda \ll b \Delta x /H$, where $\Delta x$ is the frequency bin
size, a condition which is certainly satisfied in our examples for
distances between scattering events at line centre important in
determining the scattering rate. For a general velocity profile this
condition is replaced by $\delta \lambda \ll b \Delta x/|\dd V/\dd
r|$, and so this approximate treatment is valid as long as $V(r)$ does
not change too rapidly. We obtained a Monte Carlo solution for the
coherent scattering problem which is compared with our ray/moment
solution in Fig.~\ref{figure:quad_MC}. The Monte Carlo scattering rate
is compared with that determined from the equivalent ray/moment
solution in Fig.~\ref{figure:P_MC_quad_comp}. Well interior to the
outer boundary, the scattering rate is boosted by up to an order of
magnitude compared with the expected rate for a Hubble velocity
profile, given approximately by $\tilde{P}_{\alpha}^{\rm diff}$. Near
the outer boundary, where the velocity gradient well exceeds the
Hubble constant, the scattering rate dips below
$\tilde{P}_{\alpha}^{\rm diff}$.

\subsection{Spherical Perturbation in Density and Velocity}

In this section we examine the scattering rate resulting from a
perturbation to the uniform density and Hubble velocity profiles of a
homogeneous expanding medium. We assume a spherically symmetric
overdensity of the form
\begin{eqnarray}
  && \delta(r) = \Delta_0 j_0(kr) \nonumber \\ && \implies n_{\rm H}(r) = n_{\rm H}(z) [1+\Delta_0 j_0(kr)]
\label{eq:delta_sph}
\end{eqnarray}
where $j_0(x) = (\sin{x})/x$ is the zeroth-order spherical Bessel
function, $\Delta_0$ is the amplitude of the perturbation and the
value of $k$ is taken to be $2\pi/R$, where $R$ is the radius of the
outer boundary. A restriction on the amplitude follows from the
physical requirement that $\delta \geq -1$, which then requires that
$\Delta_0 \lesssim 5$. The corresponding self-consistent perturbation
in the peculiar velocity follows from the linear regime equation
$\delta = -(1/H)\mathbf{\nabla}\cdot\mathbf{v}_{\rm p}$ which in spherical
symmetry is solved for $\delta = \Delta_0 j_0(kr)$ and the boundary
condition that the velocity vanishes at $r = 0$ by
\begin{eqnarray}
  && v_{\rm p} = -\frac{\Delta_0 H(z)}{k} j_1(kr) \nonumber \\
  && \implies V(r) = Hr + v_{\rm p} = H(z) \left[r - \frac{\Delta_0}{k} j_1(kr) \right]
\label{eq:V_sph}
\end{eqnarray}
where $j_1(x) = (\sin{x})/x^2 - (\cos{x})/x$ is the first-order
spherical Bessel function.

\begin{figure}
\begin{center}
\leavevmode
\epsfig{file = 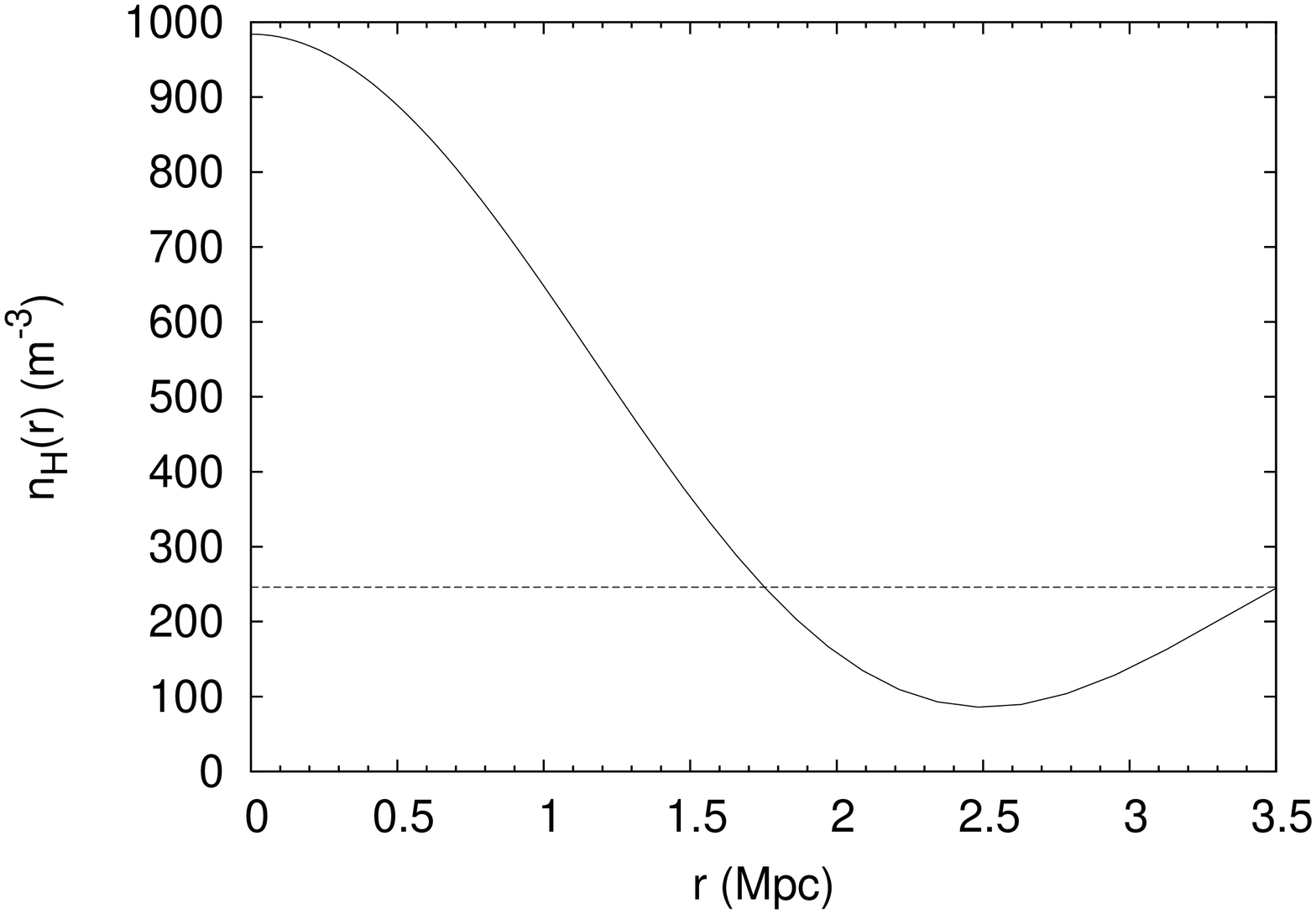, height = 7cm, angle = -90}
\epsfig{file = 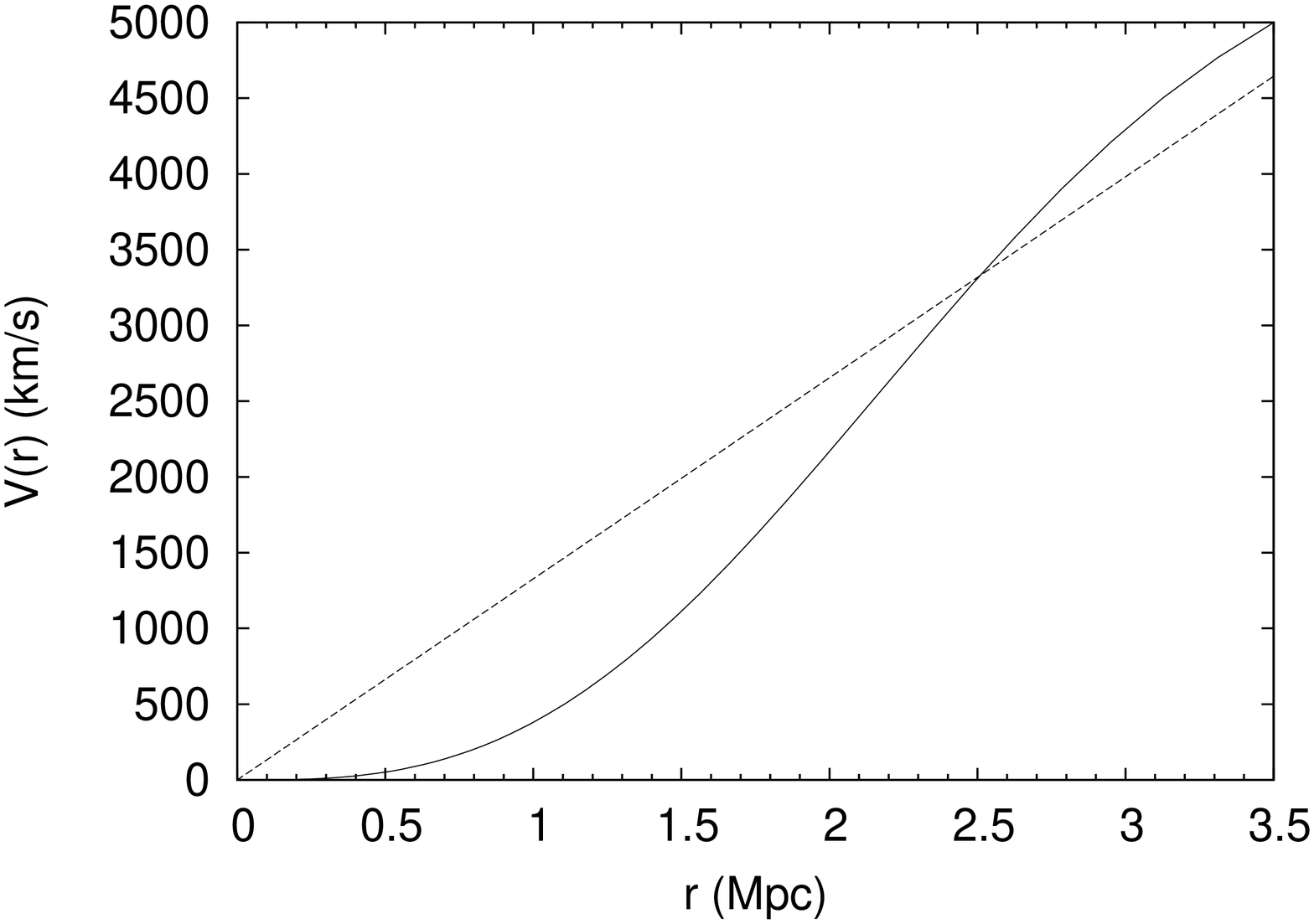, height = 7cm, angle = -90}
\end{center}
\caption{Density profile (solid line, upper panel) and velocity
  profile (solid line, lower panel) given by
  equations~(\ref{eq:delta_sph}) and (\ref{eq:V_sph}) respectively
  with amplitude $\Delta_0 = 3$ and compared with the corresponding
  uniform density or Hubble velocity profile at $z=10$ (dashed
  lines).}
\label{figure:bessel_comp}
\end{figure}

We solve the radiative transfer problem for a continuum source subject
to a scattering medium with the perturbed density and velocity
profiles of equations~(\ref{eq:delta_sph}) and (\ref{eq:V_sph}),
respectively, using the ray/moment equation solution method assuming
coherent scattering in a medium with temperature $T=10$~K. We note
that a further restriction on the perturbation amplitude $\Delta_0$
follows from requiring a monotonic velocity field, with $V(r) \geq 0$
and $V'(r) \geq 0$ for all $r$, necessary to apply the moment equation
solution method. The resulting constraint is $\Delta_0 \leq 3$,
corresponding to a vanishing velocity gradient at small $r$. While
such large values of $\Delta_0$ are no longer in the linear regime,
used to derive equation~(\ref{eq:V_sph}), the form permits an
exploration of the effect of the shape of the velocity field on the
transport of the \Lya photons.

\begin{figure}
\begin{center}
\leavevmode
\epsfig{file = 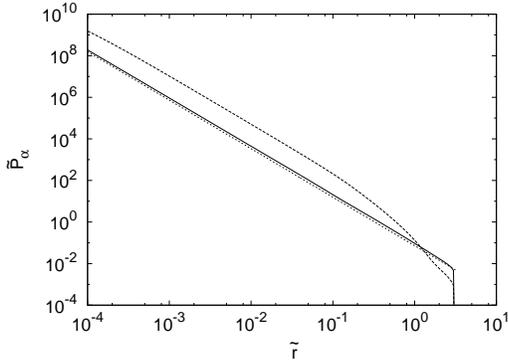, height = 7cm, angle = -90}
\end{center}
\caption{Scaled scattering rate for a spherical perturbation with the
  density profile equation~(\ref{eq:delta_sph}) and velocity profile
  equation~(\ref{eq:V_sph}), with $k = 2\pi/R$, $\tilde{R} = 10^{0.5}$
  and $x_{\rm m} = 6 \times 10^4$, for $\Delta_0=0.5$ (solid line) and
  $\Delta_0 = 2.9$ (dashed line). The scattering rates are derived
  from numerical integration of the ray/moment equation method for
  coherent scattering. For comparison the rate in the diffusion
  approximation for a uniformly expanding homogeneous medium is also
  shown (short-dashed line).}
\label{figure:P_moment_per}
\end{figure}

\begin{figure}
\begin{center}
\leavevmode
\epsfig{file = 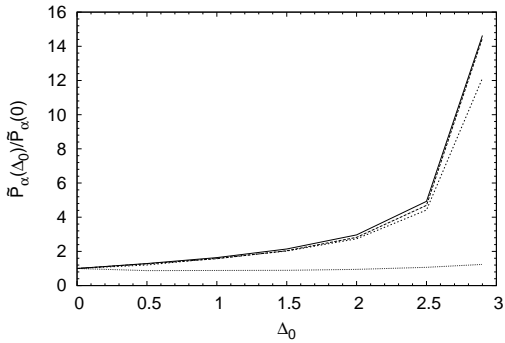, height = 7cm, angle = -90}
\end{center}
\caption{Scaled scattering rate for a spherical perturbation with the
  density profile equation~(\ref{eq:delta_sph}) and velocity profile
  equation~(\ref{eq:V_sph}), with $k = 2\pi/R$, $\tilde{R} = 10^{0.5}$
  and $x_{\rm m} = 6 \times 10^4$, as a function of the perturbation
  amplitude $\Delta_0$ and normalised by the value for $\Delta_0 = 0$,
  at different radii given by $\tilde{r} = 10^{-3}$ (solid line,
  $\tilde{P}_{\alpha}(0) = 7.2 \times 10^5$), $10^{-2}$ (long-dashed
  line, $\tilde{P}_{\alpha}(0) = 3.5 \times 10^3$), $10^{-1}$
  (short-dashed line, $\tilde{P}_{\alpha}(0) = 17$) and $10^0$ (dotted
  line, $\tilde{P}_{\alpha}(0) = 0.11$). The scattering rates are
  derived from numerical integration of the ray/moment equation method
  for coherent scattering.}
\label{figure:P_amp_ratio}
\end{figure}

The hydrogen density and velocity profiles corresponding to our
assumed cosmological parameters at $z = 10$, an outer boundary $R =
10^{0.5} r_* \simeq 3.5 \, {\rm Mpc}$ and a perturbation with
$\Delta_0 = 3$, are shown in Fig.~\ref{figure:bessel_comp}. The
scattering rate as a function of distance from the source is shown in
Fig.~\ref{figure:P_moment_per} for $\Delta_0=0.5$ and 2.9. While a
small amplitude perturbation has a correspondingly small effect on the
scattering rate, the rate is substantially boosted for
$\Delta_0\lta3$. The dependence of the scattering rate on the
perturbation amplitude for $\Delta_0 < 3$ is shown at various radii in
Fig.~\ref{figure:P_amp_ratio}. The scattering rate scales linearly
with the amplitude for small values, $\Delta_0 \lesssim 1$, but
increases more rapidly with larger values. It grows exponentially as
$\Delta_0\rightarrow3$, a result of a vanishing velocity gradient for
$r\rightarrow0$ (cf. Figs~\ref{figure:ray_moment_quad} and
\ref{figure:P_MC_quad_comp}). Clearly a small velocity gradient near
the source is able to substantially boost the scattering rate over a
wide range of radii compared with the case of uniform Hubble flow.

\begin{figure}
\begin{center}
\leavevmode
\epsfig{file = 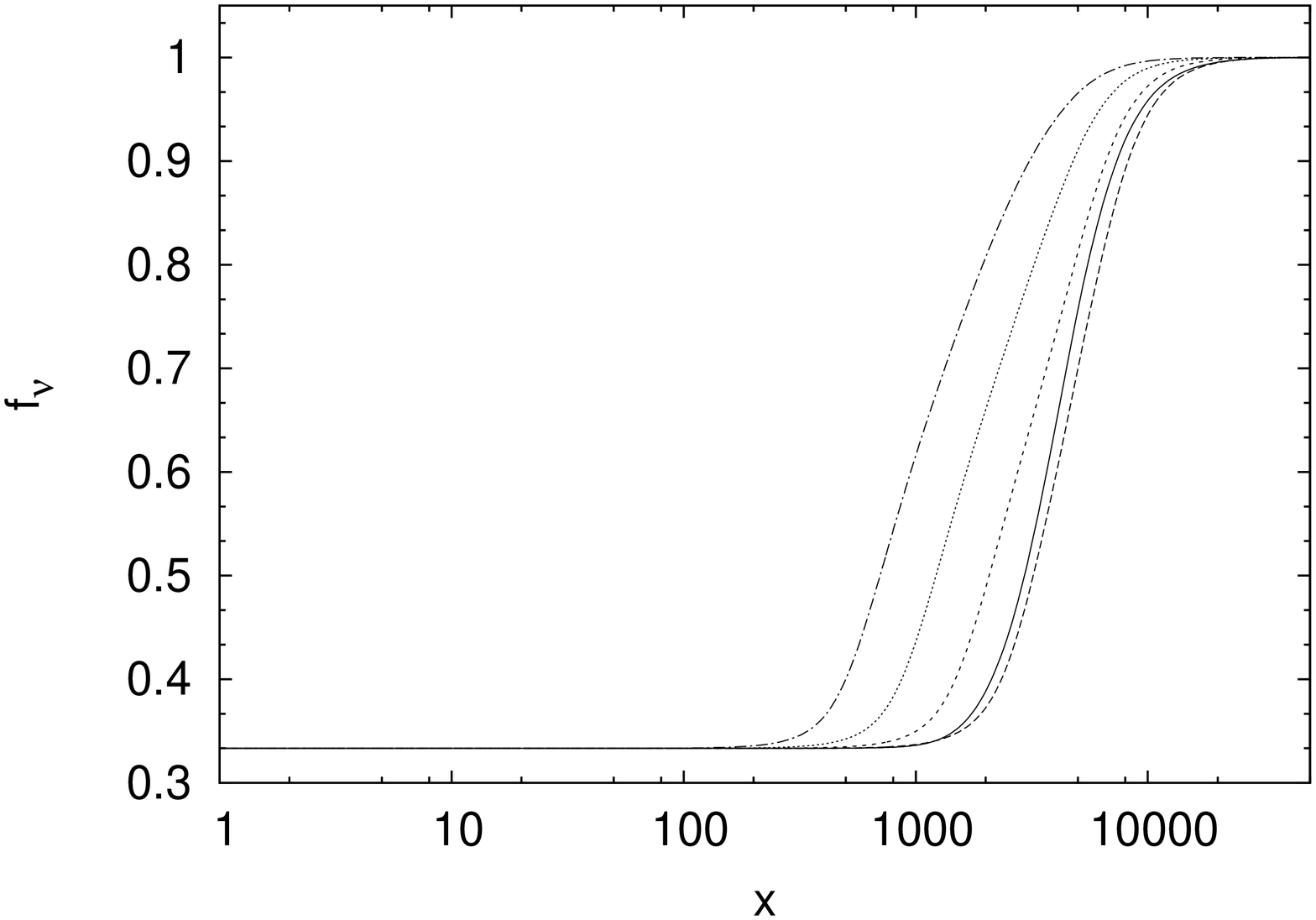, height = 7cm, angle=-90}
\epsfig{file = 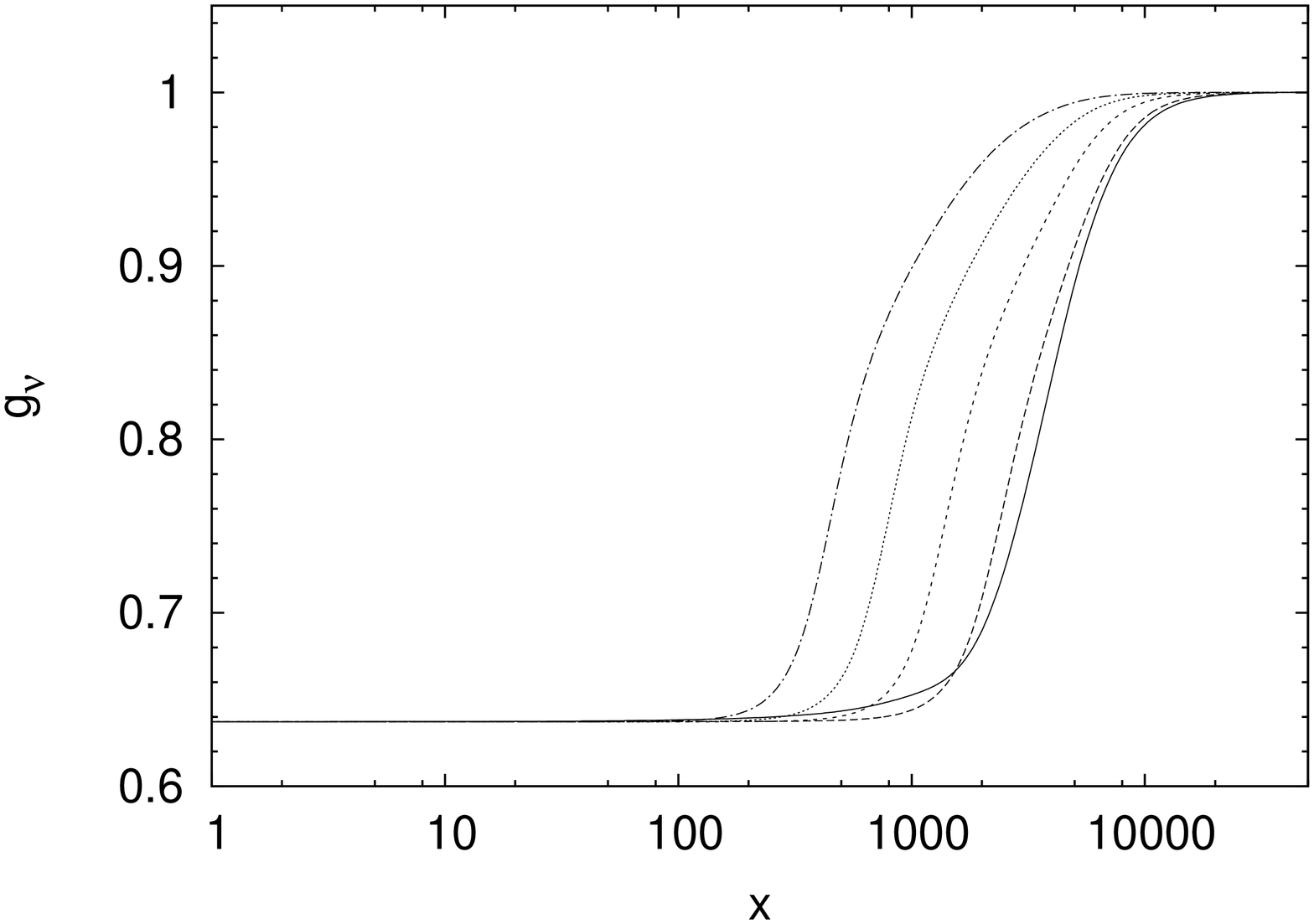, height = 7cm, angle=-90}
\end{center}
\caption{Eddington factors $f_\nu$ (upper panel) and $g_\nu$ (lower
  panel) for a spherical perturbation with the density profile
  equation~(\ref{eq:delta_sph}) and velocity profile
  equation~(\ref{eq:V_sph}) for $\Delta_0=2.9$, derived from numerical
  integration of the ray/moment equations assuming coherent
  scattering. The factors are shown at $\log_{10}\tilde{r}=0$ (solid
  line), $\log_{10}\tilde{r}=-0.5$ (long-dashed line),
  $\log_{10}\tilde{r}=-1.0$ (short-dashed line),
  $\log_{10}\tilde{r}=-1.5$ (dotted line) and
  $\log_{10}\tilde{r}=-2.0$ (dot-dashed line). The factors take on the
  values $f_\nu\simeq1/3$ and $g_\nu\gta3/5$, close to the values
  expected for the linear approximation $I_\nu\simeq J_\nu +
  3H_\nu\mu$, across the line centre, and $f_\nu=g_\nu=1$ far in the
  wings expected for free-streaming radiation.}
\label{figure:P_Edd_facts}
\end{figure}

The solutions obtained from the solving the ray/moment equations have
generally justified the linear approximation $I_\nu\simeq
J_\nu+3H_\nu\mu$ in $\mu$, for which the first two Eddington factors
are $f_\nu=1/3$ (the Eddington approximation) and $g_\nu=3/5$. For a
linear velocity field, only $f_\nu$ is required, as the terms
involving $N_\nu$ in equation~(\ref{eq:moment_orig_one}) cancel. The
perturbed flow here allows a check on $g_\nu$. The Eddington factors
are shown in Fig.~\ref{figure:P_Edd_facts}. They take on the values
$f_\nu=1/3$ and $g_\nu\gta3/5$ over a broad frequency region across
the line centre, and the values $f_\nu=g_\nu=1$ far in the wings, as
expected for free-streaming radiation. On the outer boundary,
$h_\nu\gta0.7$ over a similarly broad region across the line centre,
allowing closure of the moment equations. The value corresponds to a
single dominant escaping stream of radiation with
$\mu\simeq2^{-1/2}$. This would correspond to $f_\nu=g_\nu=1/2$, close
to the computed values on the outer boundary of $f_\nu\simeq0.5$ and
$g_\nu\gta0.6$.

\section{Conclusions}
\label{sect:concl}

The $21 \, {\rm cm}$ signatures of the first luminous objects are
dependent upon the \Lya scattering rate, which is necessary to
decouple the \HI spin temperature from the CMB temperature through the
Wouthuysen-Field effect in the diffuse IGM during the early stages of
reionization and so render the hydrogen detectable against the CMB via
$21 \, {\rm cm}$ observations. Realistic studies of \Lya scattering
caused by the first sources during the EoR will need to make use of
three-dimensional density and velocity fields derived from
cosmological simulations, and treat the time-dependent radiative
transfer of photons contributed from multiple sources of finite
lifetime, the detailed properties and spatial distributions of which
are largely uncertain and will have to be extrapolated from
lower-redshift observations or estimated from galaxy formation
simulations. Together with the extensive effort currently underway in
removing radio foregrounds and isolating the signal of the EoR, these
detailed analyses will be critical in predicting and interpreting
future $21 \, {\rm cm}$ tomographic observations made by existing and
future radio interferometric arrays, such as LOFAR and SKA.

We develop two methods for solving the radiative transfer equation for
resonance line photons in spherically symmetric systems, and apply
them to idealized problems relevant to the 21cm signature of the IGM.
We consider five classes of problems corresponding to a point source
emitting either emission line or continuum radiation in an expanding
medium, allowing for a uniformly expanding homogeneous medium, a
uniformly expanding medium with an overdense shell around the source,
a homogeneous medium with a quadratic velocity profile, and a medium
with a self-consistent density and velocity perturbation around the
source. A static medium test problem is also treated. Since these
problems may serve as useful tests of Monte Carlo schemes, we provide
their solutions in machine-readable format
(Appendix~\ref{ap:solutions}).

Following \cite{1975ApJ...202..465M} and \cite{1976ApJ...210..419M},
the first method is based on solving the ray and angular moment forms
of the comoving frame radiative transfer equation in a spherically
symmetric medium subject to a monotonic velocity profile $V(r)$, where
$V \geq 0$ and $V' \geq 0$ for all $r$ (see
Section~\ref{subsect:grid_method}). We described general boundary
conditions required by these methods in order to treat the radiative
transfer of \Lya scattering in a cosmological context. We also
obtained an efficient prescription for the finite difference
representation of the source function for partial frequency
redistribution based on the diffusion approximation.

We compare our results with Monte Carlo solutions to the radiative
transfer equation using an implementation that determines the \Lya
radiation mean intensity from the accumulated path lengths of the
photons within a given volume and frequency range. This greatly
improves the accuracy over a technique that uses spatial and frequency
bin crossings to estimate the intensity. For 200
logarithmically-spaced grid zones covering three decades in radius,
the noise level in the scattering rate is kept under 10 per cent per
radial bin using $2\times10^5$ photon packets. The code computes the
Doppler redistribution of the frequencies upon scattering both
directly and by interpolating on the RII redistribution function. The
latter speeds the computations by a factor of a few, but requires a
grid tailored to a specific application, so lacks generality.

In Section~\ref{sect:HomogTest} we validated our schemes using various
test problems for resonance line photon scattering in spherical
symmetry, including:\ (i) \Lya scattering in an optically thick static
sphere \citep{DHS_06} and (ii) the \Lya scattering halo of a
monochromatic \Lya point source in a uniformly expanding homogeneous
medium \citep{1999ApJ...524..527L}. We found the ray and moment method
results agreed with the analytic solutions, at least in the optically
thick regime where the analytic solutions are valid, while in case
(ii) our solution matched the Monte Carlo solution found by Loeb \&
Rybicki. A final test problem was described for a continuum source in
a uniformly expanding homogeneous medium. We utilised the analytic
solution of \cite{1999ApJ...524..527L}, valid in the zero-temperature
diffusion approximation, as a Green's function to obtain the
equivalent solution for a flat source spectrum in the diffusion
approximation. The solution gives a radial profile for the \Lya
scattering rate that varies as $r^{-7/3}$, steeper than the
free-streaming dependence $r^{-2}$, and the same dependence found by
other authors in Monte Carlo studies of this problem. Our analytic
solution also allows for a maximum frequency cutoff from the source;
both of our numerical methods recover the effects of the cutoff. The
photon production rate between \Lya and \Lyb required to couple the
spin temperature to the gas kinetic temperature is found to be ${\dot
  N}_{\alpha, {\rm th}}^{\rm cont}\simeq4.39\times10^{55}[(1+z)/11]
r_{\rm Mpc}^{7/3}\,{\rm s}^{-1}$, where $r_{\rm Mpc}$ is the distance
from the source in megaparsecs. This is less demanding than for an
emission-line source by several orders of magnitude at a distance of
1~Mpc.

We solve the test problem of a continuum source in a uniformly
expanding homogeneous IGM numerically, outside of the diffusion limit,
using our ray/moment and Monte Carlo methods for spherical
symmetry. In suitable spherically symmetric problems such as this test
problem, the combined ray/moment equation solution method may be used
to quickly produce noise-free results, in contrast to the Monte Carlo
approach for a practical number of photon packets. For more general
situations, the Monte Carlo method is more readily extended to treat
cartesian grids with general configurations of sources and arbitrary
density, temperature and peculiar velocity fields within the
scattering medium.

For precision results, however, a grid-based scheme such as the method
presented here may be desirable. In this case, the diffusion
approximation may be used as an inner boundary condition on the
surface of an inner core region. For non-monotonic velocity fields,
the problem would need to be divided into monotonic flow regions and
pieced together. It may not, however, be necessary to solve the
coupled moment and ray equations to obtain the scattering rates. We
find for all solutions that the linear approximation $I_\nu\simeq
J_\nu + 3H_\nu\mu$ holds to high accuracy over a very broad frequency
region (at least 100 Doppler widths for $T=10$~K gas) across the line
centre, giving the Eddington approximation value $f_\nu\simeq1/3$, for
linear flow fields before converging to the free-streaming value
$f_\nu=1$ far in the wings. Allowing for a perturbed velocity flow
gives close to the expected value $g_\nu=3/5$ over a similarly broad
region. This considerably simplifies the radiative transfer
computation, as solving the ray equation may be circumvented,
requiring only solutions to the moment equations.

We found frequency redistribution produces solutions with different
features across the line centre compared with the coherent scattering
case near the source, resulting in variations in the scattering rate
of up to $\sim50$ percent about the coherent scattering
results. Further from the source the solutions for a homogeneous
expanding IGM were found to agree closely with the solution for
coherent scattering, and roughly follow the analytically predicted
$r^{-7/3}$ radial dependence out to $\tilde{r}\simeq1$, beyond which
the diffusion approximation breaks down. For $\tilde{r}>1$, the
scattering rate more nearly approaches the free-streaming value
$1/4\pi\tilde{r}^2$ before becoming causally truncated.

Recoils are found to suppress the scattering rate by approximately 20
percent for a medium at a temperature of $T=10$~K, in good agreement
with estimates based on the diffusion approximation solution for the
radiation produced by a uniform and isotropic distribution of sources
in a uniformly expanding homogeneous medium. Our computations extend
the result beyond the diffusion approximation, and show that nearly
the same suppression factor applies for an isolated source. Very near
the source, frequency redistribution modifies the suppression factor
by up to 15 percent.

In Section~\ref{sect:InhomogTest} we examined the continuum source
\Lya scattering problem allowing for inhomogeneities in the
surrounding scattering medium, namely an overdense shell and a
quadratic velocity profile. We found substantial deviations in the
profile of the \Lya scattering rate in each case compared with a
homogeneous medium. The overdense shell produces not only an
enhancement of the \Lya scattering rate within the shell, but boosts
the scattering rate between the shell and the source as well as a
consequence of backscattering. A shadowed region with a deficit in the
scattering rate compared with the homogeneous medium solution extends
beyond the shell some distance before recovering to the homogeneous
medium value. The scattering rate produced by the quadratic velocity
profile is enhanced over the rate for the corresponding linear
velocity profile, except near the outer boundary, where the rate is
lower.

As a less-contrived example of an inhomogeneous medium we considered a
spherically symmetric density and velocity perturbation in a uniformly
expanding homogeneous IGM that satisfies the linear continuity
equation. We found that the resulting scattering rate increases
non-linearly with increasing values of the perturbation amplitude for
perturbations beginning to become nonlinear, and grows exponentially
with the amplitude once nonlinear as the velocity profile flattens
near the source. We infer that the \Lya scattering rate will depend
sensitively on the velocity structure of the IGM.

\begin{appendix}

\section{\Lya Scattering Source Function for the Moment Equations}
\label{ap:RMsource}

A particularly useful representation of the source function for
resonance-line scattering results from assuming the diffusion
approximation or Fokker-Planck approximation
\citep{1994ApJ...427..603R}, in which a Taylor expansion of
$J_{\nu}(r)$ under the integral in equation~(\ref{eq:S_res}) gives
\begin{equation}
\begin{split}
  S_{\nu}(r) = J_{\nu}(r) + \frac{(\Delta \nu_{\rm D})^2}{2 \varphi(\nu)} \frac{\partial}{\partial \nu} \left[\varphi(\nu) \frac{\partial J_{\nu}(r)}{\partial \nu} \right] \\
  + \epsilon \frac{\Delta \nu_{\rm D}}{\varphi(\nu)}
  \frac{\partial}{\partial \nu} \left[\varphi(\nu) J_{\nu}(r) \right]
\end{split}
\label{eq:S_diff}
\end{equation}
where we have included an additional correction term proportional to
the recoil parameter $\epsilon = h\nu_{\alpha}/(\sqrt{2k_{\rm
    B}Tmc^2}) = 0.025(T/{\rm K})^{-1/2}$, arising from the effect of
atomic recoil on the frequency redistribution function
\citep{1959ApJ...129..551F, Basko_1978}. In this section we will
describe how this form of the source function is represented in the
discretised system of moment equations.

We assign a frequency grid given by the discrete set $\{\nu_k\}$, $k =
1, 2, 3, ..., {\rm NF}$, where $\nu_1 = \nu_{\rm max}$ is the bluest
frequency and successive values of $k$ denote redder frequencies:
\begin{equation}
\nu_1 = \nu_{\rm max} > \nu_2 > \nu_3 > ... > \nu_{\rm NF}.
\label{eq:nu_grid}
\end{equation}
The radius grid is given by $\{r_d\}$, $d = 1, 2, 3, ..., {\rm ND}$,
where $r_1 = R$ and $r_{\rm ND} = R_{\rm C}$; increasing values of the
depth index $d$ denote greater depths with respect to the `surface' of
the system at $r = R$:
\begin{equation}
r_1 = R > r_2 > r_3 > ... > r_{\rm ND} = R_{\rm C}.
\label{eq:r_grid}
\end{equation}
In terms of the discrete grids in $\nu$ and $r$, we write the source
function $S_{k,d} \equiv S_{\nu_k}(r_d)$ as a quadrature sum:
\begin{equation}
S_{k,d} = \displaystyle\sum_{k'=1}^{\rm NF} \mathcal{R}_{k',k,d} J_{k',d}
\label{eq:S_quad}
\end{equation} 
where $J_{k',d} \equiv J_{\nu_{k'}}(r_d)$. We require the coefficients
$\mathcal{R}_{k',k,d}$ for $k, k' = 1, 2, ..., {\rm NF}$ and $d = 1,
2, ..., {\rm ND}$, the values of which are determined by a suitable
discretised description of equation~(\ref{eq:S_diff}).

In a system with a uniform temperature it is preferable to work with
the dimensionless frequency variable $x = (\nu - \nu_{\alpha})/\Delta
\nu_{\rm D}$, the offset from line centre in Doppler widths, and
rewrite radiation quantities as e.g. $J(x, r) = (\Delta \nu_{\rm D})
J_{\nu}(r)$. Similar relations hold for higher order moments and the
source function. In these units we define $R(x', x) = (\Delta \nu_{\rm
  D})^2 R(\nu', \nu)$, while the line absorption profile is expressed
as $\phi(x) = (\Delta \nu_{\rm D}) \varphi(\nu)$. The source function
and its diffusion approximation expansion may then be expressed as
\begin{equation}
\begin{split}
  S(x, r) = \frac{1}{\phi(x)} \int R(x', x)
  J(x', r) \, \dd x' = J(x, r) \\
  + \frac{1}{2\phi(x)} \frac{\partial}{\partial x}
  \left[\phi(x)\frac{\partial J(x, r)}{\partial x} + 2 \epsilon
    \phi(x) J(x, r) \right].
\end{split}
\label{eq:S_diff_x}
\end{equation}
The frequency derivative terms are represented using a centred finite
difference scheme, and the coefficients multiplying values of $J_{k,d}
\equiv J(x_k, r_d)$ are compared with equation~(\ref{eq:S_quad}) to
obtain the values of $\mathcal{R}_{k',k,d}$:
\begin{eqnarray}
  \mathcal{R}_{k',k} &=& 
  \left[1-\frac{\phi_{k-1}+\phi_{k+1}}{8\phi_k(\Delta x)^2} \right]
  \delta_{k',k} \nonumber \\
  &+& \left[\frac{\phi_{k-1}}{8 \phi_k (\Delta
      x)^2}\right] \delta_{k',k-2} + \left[\frac{\phi_{k+1}}{8\phi_k(\Delta x)^2} \right]
  \delta_{k',k+2} \nonumber \\
  &+& \epsilon \left[\frac{\phi_{k-1}}{2 \phi_k \Delta x}
  \right] \delta_{k',k-1} - \epsilon \left[\frac{\phi_{k+1}}{2 \phi_k \Delta x}
  \right] \delta_{k',k+1},
\label{eq:Rfunk_diff_easy}
\end{eqnarray}
where we have noted that any dependence on radius is removed for a
uniform temperature medium, so that $\mathcal{R}_{k',k,d} \rightarrow
\mathcal{R}_{k',k}$. Otherwise a radial dependence must be introduced
to the discretised values of $\phi$.

\section{Monte Carlo Method}
\label{ap:MC}

We will assume a uniform temperature medium to allow use of the scaled
frequency offset $x$ defined for a unique value of the Doppler width
$\Delta \nu_{\rm D}$, although the method is easily adapted to a
variable temperature $T(r)$ by formulating it in terms of the
frequency variable $\nu$. We consider a photon packet
emitted/scattered with comoving frequency $x_{\rm em}$ at position
$\mathbf{r}_{\rm em}$ in a direction denoted by unit vector
$\mathbf{\hat k}$. In spherical symmetry it is sufficient to designate the
radius of emission $r_{\rm em}$ and the angle to the local normal
$\theta_{\rm em}$, the latter having cosine $\mu_{\rm em} \equiv
\cos{\theta_{\rm em}} = \mathbf{\hat k} \cdot \mathbf{r}_{\rm em}/ r_{\rm
  em}$. The path of the photon packet prior to the next scattering
event is given by $\mathbf{r} = \mathbf{r}_{\rm em} + \lambda
\mathbf{\hat k}$, where $\lambda$ is the distance along the path of the
packet. We consider a `projected photon packet path' corresponding to
the range $\lambda = -\infty \rightarrow \infty$, having impact
parameter $r_{\rm min}$ defined as the minimum distance from the
origin:\ $r_{\rm min} = r_{\rm em}\sin{\theta_{\rm em}} = r_{\rm
  em}(1-\mu_{\rm em}^2)^{1/2}$. A spherical boundary of radius $r >
r_{\rm min}$ will be intersected by the projected photon packet path
at distances along the path $\lambda_{\pm}$ determined from solving
$\mathbf{r}\cdot\mathbf{r} = r^2$ with $\mathbf{r} = \mathbf{r}_{\rm
  em} + \lambda \mathbf{\hat k}$:
\begin{eqnarray}
  \lambda_{\pm}(r) &=& -\mathbf{\hat k}\cdot\mathbf{r}_{\rm em} \pm \left[r^2-r_{\rm em}^2 + (\mathbf{\hat k}\cdot\mathbf{r}_{\rm em})^2 \right]^{1/2} \nonumber \\
  &=& -\mu_{\rm em}r_{\rm em} \pm (r^2 - r_{\rm min}^2)^{1/2}
\label{eq:lambda_pm}
\end{eqnarray}
where obviously only positive values denote possible intersections of
the actual photon packet path with the sphere. At the radius $r$, the
corresponding angle to the local normal is given by $\mu =
\mathbf{\hat k}\cdot\mathbf{r}/r$ where $\mathbf{r} = \mathbf{r}_{\rm em} +
\lambda_{\pm} \mathbf{\hat k}$:
\begin{eqnarray}
  \mu_{\pm}(r) &=& (\mu_{\rm em}r_{\rm em} + \lambda_{\pm})/r \nonumber \\
  &=& \pm \left[1-\left(r_{\rm min}/r\right)^2\right]^{1/2}.
\label{eq:mu_pm}
\end{eqnarray}
At any point $\mathbf{r}$ along its path, the packet will have a
comoving frequency corresponding to the same lab frame frequency as
$x_{\rm em}$, and thus
\begin{eqnarray}
  x(r) &=& x_{\rm em} - \mathbf{V}(\mathbf{r})\cdot\mathbf{\hat k}/b + \mathbf{V}(\mathbf{r}_{\rm em})\cdot\mathbf{\hat k}/b, \nonumber \\
  &=& x_{\rm em} - \mu(r) V(r)/b + \mu_{\rm em}V(r_{\rm em})/b
\label{eq:x_shift}
\end{eqnarray}
where $\mu$ describes the angle to the local normal of the sphere at
$\mathbf{r}$. For the special case of a Hubble-flow velocity field,
i.e. $V(r) = Hr$, this reduces to
\begin{eqnarray}
  x(r) &=& x_{\rm em} - \left(\mu r - \mu_{\rm em} r_{\rm em}\right)H/b \nonumber \\
  &=& x_{\rm em} - \lambda H/b
\label{eq:x_shift_Hubble}
\end{eqnarray}
where the content of the brackets simplifies following
equation~(\ref{eq:mu_pm}); the frequency shift is linearly dependent
on the distance $\lambda$ travelled along the path. We use these
equations to record the frequency, path length and the angle to the
local normal of the photon packet as it crosses a series of spherical
boundaries of radius $\{r_{d}\}$, the values of which comprise a
radial grid across which we seek the mean intensity $J_{\nu}(r)$ and
the scattering rate $P_{\alpha}(r)$.

The packet is scattered when the accumulated optical depth $\tau$
reaches a sufficiently large value. This optical depth is chosen
according to an exponential probability distribution $e^{-\tau}$,
i.e. $\tau = -\ln{R}$ for a uniform deviate $R$. For a \Lya scattering
opacity $\chi_{\nu} = n_{\rm H} \sigma \varphi(\nu)$, the differential
optical depth accumulated along a length $\dd \lambda$ may be
expressed as $\dd \tau = n_{\rm H}\sigma [\phi(x) /\Delta \nu_{\rm D}]
\, \dd \lambda$. The total optical depth is the integral along the
path. For a static medium of constant temperature and neutral hydrogen
density $n_{\rm H}$, this integral is trivial:\ $\lambda = \tau \Delta
\nu_{\rm D}/[n_{\rm H} \sigma \phi(x)]$; however, for a non-static
medium the frequency $x$ varies as in equation~(\ref{eq:x_shift}) and
no simple $\tau$-dependence may be obtained as a function of
$\lambda$.

A less rigid approach is to assume the density, temperature and
velocity of the medium are essentially constant across shells, as in
the Monte Carlo code of \cite{DHS_06}. Under these circumstances the
comoving frequency of the packet changes in steps as the path crosses
shell boundaries, and the optical depth between two boundaries behaves
as in the uniform static case. The optical depth between boundaries at
distances $\lambda_d$ and $\lambda_{d+1}$ along the packet path is
\begin{equation}
  \Delta \tau = \frac{\sigma n_{\rm H}(r_d) \phi(x_{d})} {\Delta
    \nu_{\rm D}} |\lambda_{d + 1} - \lambda_d |
\label{eq:tau_shell}
\end{equation} 
where the subscript $d$ indicates evaluation at the intersection with
the boundary of radius $r_d$, using equations~(\ref{eq:lambda_pm}) and
(\ref{eq:x_shift}). The packet is permitted to progress along its path
until the sum of the values of $\Delta \tau$ exceeds the randomly
selected total optical depth, at which stage the packet is
scattered. If the optical depth corresponds to a total distance
$\lambda_{\rm max}$ along the path of the packet, the packet is moved
to a new radius $r$ given by applying the cosine rule to the triangle
of sides $r_{\rm em}$, $r$ and $\lambda_{\rm max}$:
\begin{equation}
  r = (r_{\rm em}^2 + \lambda_{\rm max}^2 + 2\mu_{\rm em} r_{\rm em} \lambda_{\rm max})^{1/2}.
\label{eq:radius_update}
\end{equation}
Upon scattering, the photon packet is re-emitted with a new direction
$\mu_{\rm em}$ and a new frequency $x_{\rm em}$.

We generally assume isotropic scattering, for which the emitted
direction is obtained using $\mu_{\rm em} = 2R-1$ for a uniform random
deviate $R$. We determined the emitted frequency from the well-known
frequency redistribution function $R_{\rm II}(x', x)$
\citep{1978stat.book.....M} using a numerical lookup table method. The
method uses the probability distribution $p(x \, | \, x') = R_{\rm
  II}(x', x)/\phi(x)$ for the output/emitted frequency $x$ given the
input/absorbed frequency $x'$, and is described in some detail in
\citet{2012HigginsPhD}. An alternative method that is more adaptable
and has been widely adopted by other authors is based on selecting the
scattering atom thermal velocity components and calculating the output
frequency directly from the resulting Doppler shifts
\citep{2004ApJ...602....1C}. We found our lookup table method to speed
up the Monte Carlo computations by a factor of approximately 3
compared with the direct Doppler shift computation method, however the
required pre-calculation can become cumbersome.

We estimate the mean intensity using the method of \cite{Lucy_1999},
based on the time-averaged number density of photons in a cell of a
given volume. We tailor the method to solve problems in spherical
symmetry by adopting a cell of volume $V$ corresponding to the volume
of the shell between two radial grid-spheres of radius $r$ and
$r+\Delta r$, given by $V \simeq 4\pi r^2 \Delta r$. If a Monte Carlo
photon packet travels a distance $\delta \lambda$ in the shell between
two scattering events, the packet occupies the shell for a time
$\delta t = \delta \lambda/c$ and contributes to the specific number
density within the shell, at the frequency of the packet, the
increment $V^{-1} \delta t/ \Delta t$ where the quantity $\Delta t$ is
the physical time represented by the entire Monte Carlo run. The total
number density $n_{\nu}(r)\, \Delta \nu$ comes from the sum of all the
values of $\delta \lambda$ between any two general `events', which may
be a scattering event, a boundary crossing or redshifting/blueshifting
between adjacent frequency bins, arising from packets travelling
within the shell while having frequency within the range $[\nu,
\nu+\Delta \nu]$. The mean intensity is then computed from $J_{\nu} =
[c/(4\pi)]n_{\nu}$ as
\begin{equation}
J_{\nu}(r) = \frac{1}{4\pi V (\Delta \nu) (\Delta t)} \ \sum_{\rm events} \,\delta \lambda(\nu, r).
\label{eq:J_path}
\end{equation}

The distances $\delta \lambda$ between scattering events and/or
boundary crossings are computed by following the intersections of the
path of a photon packet with the radial grid. For a packet that
travels a short distance $\lambda_{\rm max}$ before scattering, so
that the entire path is contained within the shell at $r$ and the
packet frequency remains within the frequency bin at $\nu$, we take
$\delta \lambda = \lambda_{\rm max}$. However, if the path of the
packet crosses shell boundaries or causes redshifting/blueshifting
between adjacent frequency bins, it is necessary to separate the total
path length $\lambda_{\rm max}$ into multiple values of $\delta
\lambda(\nu, r)$ at the appropriate radii and/or frequencies. For
packets crossing shell boundaries the $\delta \lambda$ values are
obtained from differencing the distances along the path of the packet
at which the shell boundary intersections defined by
equation~(\ref{eq:lambda_pm}) occur. For a packet path defined by
$r_{\rm em}$ and $\mu_{\rm em}$ the choice of $\lambda_+$ or
$\lambda_-$ is given by either of two sequences, distinguished by the
sign of $\mu_{\rm em}$:\ (i) for a packet emitted in an `outward'
direction with $\mu_{\rm em} > 0$, only $\lambda_+$ potentially
represents a boundary crossing as $\lambda_- < 0$ and the packet may
potentially cross every boundary with radius $r > r_{\rm em}$, and
(ii) for a packet emitted `inwards' with $\mu_{\rm em} < 0$, the
solution $\lambda_+ > 0$ for all $r > r_{\rm min}$ and $\lambda_- > 0$
for $r_{\rm min} < r < r_{\rm em}$ and thus the potential path of the
packet will cross shell boundaries of decreasing radii from $r_{\rm
  em}$ down to $r_{\rm min}$ with $\lambda$ given by $\lambda_-$, then
increasing radii from $r_{\rm min}$ upwards with $\lambda$ given by
$\lambda_+$. The values of $\delta \lambda$ are given by the
differences in sequential values of $\lambda$, although we only follow
the sequence in each case until the packet is scattered. Some
additional calculation is needed for non-static media in order to
treat redshifting/blueshifting between adjacent frequency bins while a
packet is `in transit,' i.e. in between scattering events and/or
boundary crossings. We discuss this in applications to particular
problems within the main text.

As an alternative to the summed path-length estimator for the mean
intensity $J_{\nu}(r)$, we describe an estimator for the specific
intensity $I_{\nu}(r, \mu)$. From the definition of specific intensity
we write the differential number of photons in the frequency range
$[\nu,\nu+\dd \nu]$ crossing the boundary at $\mathbf{r}$ within solid
angle $\dd \omega$ about direction $\mathbf{\hat n}$, normal to area
$\dd A$, in time $\dd t$ as $\dd N = I_{\nu}(\mathbf{r},\mathbf{\hat
  n}) \, \dd A \, \dd \omega \, \dd \nu \, \dd t$. The differential
solid angle is simply $\dd \omega = |\dd \mu| \, \dd \phi$ and the
area $\dd A$ normal to $\mathbf{\hat n}$ is related to the
differential area of the boundary $\dd S$ by $\dd A = |\mu| \, \dd
S$. The total number $N(r, \mu, \nu)$ of photons within frequency
range $[\nu, \nu + \dd \nu]$ crossing the boundary of radius $r$
within angular range $[\mu, \mu + \dd \mu]$, is then equal to the
integral of $\dd N$ over the azimuthal angle and the boundary
surface:\ $N(r,\mu,\nu) = I_{\nu}(r,\mu) |\mu| (\oint \dd S) \, (\int
\dd \phi) \, \dd \mu \, \dd \nu \, \dd t$, or
\begin{equation}
  I_{\nu}(r,\mu) = \frac{N(r, \mu, \nu)}{8 \pi^2 r^2 |\mu| (\Delta \mu) (\Delta \nu) (\Delta t)}
\label{eq:I_shell}
\end{equation}
where $\Delta \mu$ and $\Delta \nu$ are the bin sizes in angle and
frequency, and $\Delta t$ is again the physical time represented by
the Monte Carlo run which is related to the total number of photons
emitted by the source. The quantity $N(r,\mu,\nu)$ is determined
simply by incrementing an appropriate counter every time a photon
packet crosses the shell at $r$. This method provides an alternative
means of determining the mean intensity $J_{\nu}$, which follows from
angular integration, however, as it samples only those packets that
cross radial shells, we found it to be a less efficient means of
estimating the mean intensity at line centre, where the path lengths
are short, compared with the summed path-length method. This method
is, however, required to determine higher-order angular moments of the
intensity.

\section{Analytic Solutions for a Homogeneous Expanding Medium}
\label{ap:hem}

\subsection{\Lya Source}

In this section we review the derivation of the solution for a \Lya
point source in a homogeneous expanding medium
\citep{1999ApJ...524..527L}. This solution provides the basis for the
derivation of the corresponding solution for a continuum source.

We rewrite equation~(\ref{eq:sph_comoving}) in a dimensionless form
using the frequency and radius variables previously described in
Section \ref{subsect:LR_sols}:
\begin{equation}
\begin{split}
  \mu \frac{\partial \tilde{I}}{\partial \tilde{r}} + \frac{1-\mu^2}{\tilde{r}} \frac{\partial \tilde{I}}{\partial \mu} + \tilde{\alpha} \left[1-\mu^2+\mu^2 \left( \frac{\dd \ln{V}}{\dd \ln{r}} \right) \right]\frac{\partial \tilde{I}}{\partial \tilde{\nu}} \\
  = -\tilde{\chi} \tilde{I} + \tilde{\eta} + \delta(\tilde{\nu})
  \frac{\delta(\tilde{r})}{(4\pi \tilde{r})^2}
\end{split}
\label{eq:RT_gen_scaled}
\end{equation}
where $\tilde{\alpha} \equiv V/[H(z)r]$, $\tilde{\chi} \equiv r_*
\chi_{\nu}$ and $\tilde{\eta} \equiv (r_*/I_*^l) \eta_{\nu}$. In the
idealised zero-temperature limit assumed by Loeb \& Rybicki, atoms
have zero thermal velocity and RII redistribution becomes coherent in
the frame of the observer, while the line profile is that of natural
broadening alone. Coherent scattering in an expanding medium means
photons emitted at $\nu = \nu_{\alpha}$ can only get redder, and thus
the wing form of the opacity applies while the emissivity reduces to
$\eta_{\nu} = \chi_{\nu}J_{\nu}$. They assumed a velocity profile for
the \HI medium given by the Hubble expansion, $V(r) = H r$; note that
for this linear scaling the expression in square brackets in
equation~(\ref{eq:RT_gen_scaled}) reduces to unity and $\tilde{\alpha}
= 1$. The zeroth and first order angular moment equations are given
by:
\begin{equation}
  \frac{1}{\tilde{r}^2} \frac{\partial (\tilde{r}^2 \tilde{H})}{\partial \tilde{r}} + \frac{\partial \tilde{J}}{\partial \tilde{\nu}}  = \delta(\tilde{\nu}) \frac{\delta(\tilde{r})}{(4\pi \tilde{r})^2},  
\label{eq:mom_zero_hem}
\end{equation}
\begin{equation}
\frac{\partial \tilde{K}}{\partial \tilde{r}} + \frac{3\tilde{K} - \tilde{J}}{\tilde{r}} + \frac{\partial \tilde{H}}{\partial \tilde{\nu}} = -\frac{\tilde{H}}{\tilde{\nu}^2}.
\label{eq:mom_one_hem}
\end{equation}
A solution may be obtained in the Eddington approximation $\tilde{K} =
\tilde{J}/3$. In the diffusion limit the intensity is written as the
Legendre expansion $I_{\nu} = J_{\nu} + 3H_{\nu}\mu$ with $H_{\nu}
\ll J_{\nu}$, corresponding to the large scattering limit
$\tilde{r}\ll\tilde{\nu}$, and thus we neglect the derivative
$\partial \tilde{H}/\partial \tilde{\nu}$ in the first order angular
moment equation. The resulting diffusion relation between $\tilde{H}$
and $\tilde{J}$ is substituted into equation~(\ref{eq:mom_zero_hem})
to obtain a single equation for $\tilde{J}$; if we do this and change
frequency variable to $\sigma = \tilde{\nu}^3/9$ we obtain
\begin{eqnarray}
\frac{\partial \tilde{J}}{\partial \sigma} - \frac{1}{\tilde{r}^2}
\frac{\partial}{\partial \tilde{r}} \left( \tilde{r}^2 \frac{\partial
  \tilde{J}}{\partial \tilde{r}} \right) &=& \frac{\delta(\tilde{r})}{(4\pi \tilde{r})^2} \times  \frac{\delta(\tilde{\nu})}{(3\sigma^2)^{1/3}}
\nonumber\\
&=& \frac{\delta(\tilde{r})}{(4\pi \tilde{r})^2} \times \delta(\sigma)
\label{eq:diff_combined}
\end{eqnarray}
where we have multiplied by the factor $\dd \tilde{\nu}/ \dd \sigma =
(3\sigma^2)^{-1/3}$ in the first line and used $\delta(\tilde{\nu}) =
\delta(\sigma)/|\dd \tilde{\nu}/ \dd \sigma |$ in the second. This
equation is the diffusion equation in three dimensions with $\sigma$
acting as the time variable and a point source impulse at $\sigma =
0$. The solution is given by (see Loeb \& Rybicki's equation~21)
\begin{equation}
  \tilde{J}(\tilde{r}, \tilde{\nu}) = \frac{1}{4\pi} \left(\frac{9}{4\pi
      \tilde{\nu}^3} \right)^{3/2} \exp{\left(-\frac{9 \tilde{r}^2}{4
        \tilde{\nu}^3} \right)}.
\end{equation}
The corresponding flux, after using Loeb \& Rybicki's solution for the
mean intensity in the first order angular moment equation, is
\begin{equation}
  \tilde{H}(\tilde{r}, \tilde{\nu}) = -\frac{\tilde{\nu}^2}{3}
  \frac{\partial \tilde{J}}{\partial \tilde{r}} = \frac{3\tilde{r}}{8\pi\tilde{\nu}} \left(\frac{9}{4\pi \tilde{\nu}^3} \right)^{3/2} \exp{\left(-\frac{9 \tilde{r}^2}{4 \tilde{\nu}^3} \right)}.
\end{equation}

\subsection{Continuum Source}

It is useful to consider the radiative transfer moment equations
applicable in the zero temperature and diffusion limits for a source
term of general frequency dependence $s(\tilde{\nu})$. The combined
moment equation is given by equation~(\ref{eq:diff_combined}) with
$\delta(\tilde{\nu}) \rightarrow s(\tilde{\nu})$. We use a Green's
function approach to solve the combined moment equation by first
considering a monochromatic source emitting at a single frequency
$\tilde{\nu}_{\rm s}$ with $s(\tilde{\nu}) =
\delta(\tilde{\nu}-\tilde{\nu}_{\rm s}) =
(3\sigma^2)^{1/3}\delta(\sigma-\sigma_{\rm s})$ where $\sigma_{\rm s}
= \tilde{\nu}_{\rm s}^3/9$. We label the solution as
$\tilde{G}(\tilde{r}, \sigma, \sigma_{\rm s})$. The Green's function
satisfies
\begin{equation}
\frac{\partial \tilde{G}}{\partial \sigma} - \frac{1}{\tilde{r}^2}
\frac{\partial}{\partial \tilde{r}} \left( \tilde{r}^2 \frac{\partial
  \tilde{G}}{\partial \tilde{r}} \right) = \frac{\delta(\tilde{r})}{(4\pi \tilde{r})^2} \times  \delta(\sigma-\sigma_{\rm s}).
\label{eq:diff_combined_green}
\end{equation}
The solution is then the 3D diffusion solution for a point source
impulse at $\sigma = \sigma_{\rm s}$:
\begin{equation}
\tilde{G}(\tilde{r},\sigma-\sigma_{\rm s}) = \begin{cases} \frac{1}{4\pi}
\left[\frac{1}{4\pi(\sigma - \sigma_{\rm s})} \right]^{3/2}
  \exp{\left[-\frac{\tilde{r}^2}{4(\sigma - \sigma_{\rm s})}\right]} ; & \sigma > \sigma_{\rm s} \\
0 ; & \sigma < \sigma_{\rm s} \end{cases}
\label{eq:green_diff}
\end{equation}
where the factor $1/(4\pi)$ ensures unit source normalisation. Given
$\tilde{G}$, the solution to the problem for a source of general
frequency dependence satisfying equation~(\ref{eq:diff_combined}) with
$\delta(\tilde{\nu}) \rightarrow s(\tilde{\nu})$ is given by
\begin{equation}
\tilde{J}(\tilde{r}, \sigma) = \int_{-\infty}^{\infty}
\tilde{G}(\tilde{r},\sigma-\sigma_{\rm s}) \frac{s(\sigma_{\rm
    s})}{(3\sigma_{\rm s}^2)^{1/3}} \, \dd \sigma_{\rm s}.
\label{eq:J_diff_gen}
\end{equation}
Thus we may construct the solution for a source of arbitrary frequency
dependence from this result and equation~(\ref{eq:green_diff}).
 
We now return to the case of a flat source spectrum as an
approximation to the spectrum of a continuum source in the IGM, for
which the dimensionless source term has frequency dependence
$s(\tilde{\nu}) \rightarrow 1$. It will be useful to specify a blue
cutoff frequency $\tilde{\nu}_{\rm m} < 0$, the maximum frequency (or
minimum value of $\tilde{\nu}$) at which the source emits photons
capable of interacting with the \Lya line, giving a step-function form
for the source frequency distribution:
\begin{equation}
  s(\sigma) = \Theta(\sigma - \sigma_{\rm m}) \equiv \begin{cases} 1 & ; \sigma > \sigma_{\rm m} \\
    0 & ; \sigma \leq \sigma_{\rm m} \end{cases}
%\label{eq:step_source}
\end{equation}
where $\sigma_{\rm m} = \tilde{\nu}_{\rm m}^3/9$. We use the Green's
function and the source frequency distribution to obtain the solution
according to equation~(\ref{eq:J_diff_gen}), valid for $\sigma >
\sigma_{\rm m}$:
\begin{equation}
\begin{split}
  \tilde{J}(\tilde{r},\sigma) = \frac{1}{3^{1/3}} \int_{\sigma_{\rm
      m}}^{\sigma}  \sigma_{\rm s}^{-2/3} \tilde{G}(\tilde{r},\sigma-\sigma_{\rm s}) \, \dd \sigma_{\rm s} \hspace{2.3cm} \\
  = \frac{1}{3^{1/3}(4\pi)^{5/2}} \int_{\sigma_{\rm m}}^{\sigma}
  \frac{1}{\sigma_{\rm s}^{2/3} (\sigma-\sigma_{\rm s})^{3/2}}
  \exp{\left[-\frac{\tilde{r}^2}{4(\sigma-\sigma_{\rm s})} \right]} \,
  \dd \sigma_{\rm s},
\end{split}
\label{eq:J_flat_analytic}
\end{equation}
\begin{equation}
\begin{split}
  \tilde{H}(\tilde{r},\sigma) = -\frac{\tilde{\nu}^2}{3^{4/3}}
  \frac{\partial}{\partial \tilde{r}} \int_{\sigma_{\rm
      m}}^{\sigma}  \sigma_{\rm s}^{-2/3} \tilde{G}(\tilde{r},\sigma-\sigma_{\rm s}) \, \dd \sigma_{\rm s} \hspace{1.5cm} \\
  = \frac{\sigma^{2/3} \tilde{r}}{2(4\pi)^{5/2}} \int_{\sigma_{\rm
      m}}^{\sigma} \frac{1}{\sigma_{\rm s}^{2/3} (\sigma-\sigma_{\rm
      s})^{5/2}} \exp{\left[-\frac{\tilde{r}^2}{4(\sigma-\sigma_{\rm
          s})} \right]} \, \dd \sigma_{\rm s}.
\end{split} 
\label{eq:H_flat_analytic}
\end{equation} 
The forms given by equations~(\ref{eq:J_analytic_calc}) and
(\ref{eq:H_analytic_calc}) are obtained with the change of variable $u
= 9\tilde{r}^2/[4(\tilde{\nu}^3-\tilde{\nu}_{\rm s}^3)]$; these latter
forms are much easier to numerically integrate.

For $\nu_m\rightarrow-\infty$, equations~(\ref{eq:J_flat_analytic})
and (\ref{eq:H_flat_analytic}) have the asymptotic series
representations in $t=-4\sigma/{\tilde r}^2$
\begin{eqnarray}
  \tilde{J}(\tilde{r},t)  &\sim&
  \left(\frac{2^7}{3}\right)^{1/3}\frac{\Gamma(7/6)}{(4\pi)^{5/2}}
  {\tilde r}^{-7/3}\nonumber\\
&&\times\left[1+\sum_{n=1}^\infty\frac{\Gamma(n+7/6)}{\Gamma(7/6)}b(-2/3,n)t^n\right]
\label{eq:J_flat_series_small_t}
\end{eqnarray}
and
\begin{eqnarray}
  \tilde{H}(\tilde{r},t)  &\sim&
  \left(\frac{2^5}{9}\right)^{2/3}\frac{\Gamma(13/6)}{(4\pi)^{5/2}}
{\tilde r}^{-10/3}{\tilde\nu}^2\nonumber\\
&&\times\left[1+\sum_{n=1}^\infty\frac{\Gamma(n+13/6)}{\Gamma(13/6)}b(-2/3,n)t^n\right],
\label{eq:H_flat_series_small_t}
\end{eqnarray}
for $\vert t\vert\ll1$, noting the general series expansion
\begin{eqnarray}
I_p(\tilde{r}, t) &\equiv&
\int_{-\infty}^\sigma\,d\sigma_s\sigma_s^{-2/3}(\sigma-\sigma_s)^{-p}e^{\sigma/[t(\sigma-\sigma_s)]}\nonumber\\
&\sim&\left(\frac{4}{{\tilde r}^2}\right)^{p-\frac{1}{3}}\sum_{n=0}^{\infty}b\left(-\frac{2}{3},n\right)\Gamma\left(p+n-\frac{1}{3}\right)t^n,
\label{eq:Ip_series_small_t}
\end{eqnarray}
where $b(-2/3,n)$ is a binomial coefficent
(${1,\,-2/3,\,5/9,\,\dots}$). For $\vert t\vert\gg1$,
\begin{equation}
I_p(\tilde{r},t)\sim\left(\frac{4}{{\tilde
      r}^2}\right)^{p-\frac{1}{3}}\Gamma(p-1)t^{-2/3}\left(1-\frac{2}{3}\frac{p-1}{t}\pm\dots\right),
\label{eq:Ip_series_large_t}
\end{equation}
where the number of terms retained in the bracketed sum does not
exceed $p$. This gives the leading behaviours
\begin{equation}
  \tilde{J}(\tilde{r},t) \sim\left(\frac{2^7}{3}\right)^{1/3}\frac{1}{2(4\pi)^2}
  {\tilde r}^{-7/3}\frac{1}{t^{2/3}}
\label{eq:J_flat_series_large_t}
\end{equation}
and
\begin{equation}
  \tilde{H}({\tilde{r},t})\sim\frac{1}{(4\pi\tilde{r})^2}\left(1-\frac{1}{t}\right).
\label{eq:H_flat_series_large_t}
\end{equation}

\begin{figure}
\begin{center}
\leavevmode
\epsfig{file = 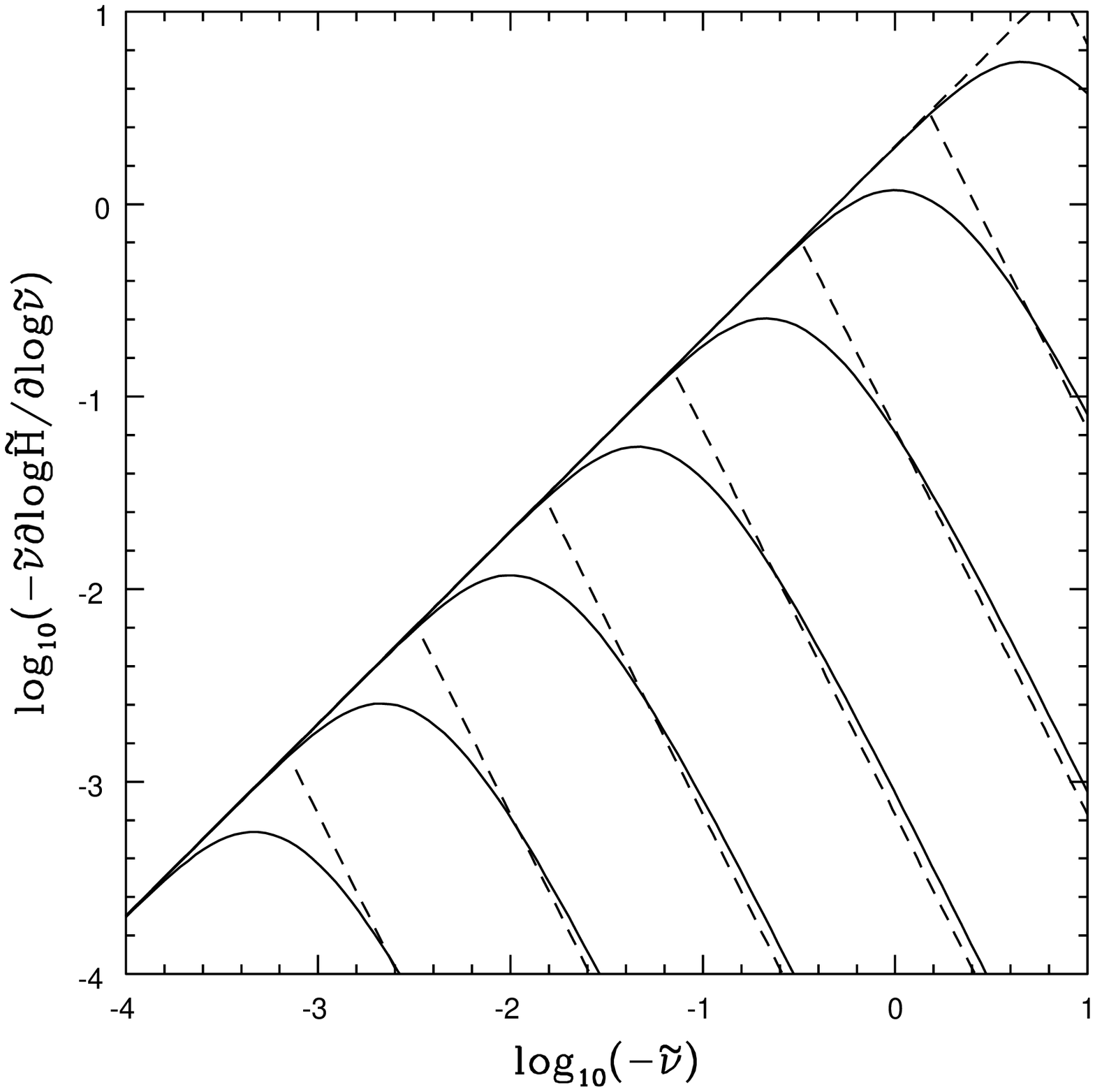, height = 7cm}
\end{center}
\caption{Solid lines:\ The ratio
  $\tilde{\nu}\partial\log\tilde{H}/\partial\log\tilde{\nu}$ for
  various $\tilde{r}$ ranging in decades from $10^{-5}$ to 10 from
  bottom to top. Dashed lines:\ The ratio using the asymptotic
  expansions for $\vert4\tilde{\nu}^3/9\tilde{r}^2\vert\ll1$ and
  $\vert4\tilde{\nu}^3/9\tilde{r}^2\vert\gg1$.}
\label{figure:HnuRat}
\end{figure}

The diffusion approximation breaks down when $\partial {\tilde
  H}/\partial\tilde\nu$ is no longer negligible compared with ${\tilde
  H}/{\tilde\nu}^2$. The approximations
equations~(\ref{eq:H_flat_series_small_t}) and
(\ref{eq:H_flat_series_large_t}) may be used to estimate the radius at
which the solution will fail. For $\vert t\vert\ll1$,
${\tilde\nu}\partial\log {\tilde
  H}/\partial\log{\tilde\nu}\sim2\tilde{\nu}$, while for $\vert
t\vert\gg1$, ${\tilde\nu}\partial\log {\tilde
  H}/\partial\log{\tilde\nu}\sim-(27/4)(\tilde{r}/\tilde{\nu})^2$. Equating
these gives as an approximation for the value of the frequency at
which $\vert{\tilde\nu}\partial\log {\tilde
  H}/\partial\log{\tilde\nu}\vert$ peaks of $\tilde{\nu}_{\rm
  peak}\sim-(3/2){\tilde r}^{2/3}$, with a peak value of
$\vert{\tilde\nu}\partial\log {\tilde
  H}/\partial\log{\tilde\nu}\vert_{\rm peak}\simeq3{\tilde
  r}^{2/3}$. Accordingly, the approximation is valid only for ${\tilde
  r}\ll1/3^{3/2}$. Comparison with Fig.~\ref{figure:HnuRat} shows the
peak of $\vert{\tilde\nu}\partial\log {\tilde
  H}/\partial\log{\tilde\nu}\vert$ exceeds 0.25 for $\tilde{r}>0.1$,
at which the diffusion approximation will begin to break down.

The solution obtained in Section~\ref{subsect:hmecont_num} shows that
while at large $\tilde r$, $\vert{\tilde\nu}\partial\log {\tilde
  H}/\partial\log{\tilde\nu}\vert$ approaches unity at large values of
$\vert{\tilde\nu}\vert$, for $\vert{\tilde\nu}\vert\ll1$ the ratio is
still small, so that $\partial{\tilde H}/\partial{\tilde\nu}$ may
continue to be neglected. Equations~(\ref{eq:mom_zero_hem}) and
(\ref{eq:mom_one_hem}) then admit the similarity solution in $t$ for
$\vert t\vert\lll1$
\begin{equation}
  \tilde{J}(\tilde{r},t) = \frac{1}{(4\pi{\tilde r})^2}j(t),
\label{eq:J_flat_series_large_r}
\end{equation}
\begin{equation}
  \tilde{H}({\tilde{r},t}) = \frac{2}{3(4\pi)^2}
  \frac{\tilde{\nu}^2}{\tilde{r}^3}\left[j(t) + t\frac{dj(t)}{dt}\right]
\label{eq:H_flat_series_large_r}
\end{equation}
and ${\tilde K}={\tilde J}/3$, where $j(t)\sim\left(1 - \frac{1}{2}t +
  \frac{3}{4}t^2 - \frac{45}{24}t^3 \pm \dots\right)$. This is found
to agree well with the numerical solution obtained from the ray/moment
method.

\section{Tables of test suite solutions}
\label{ap:solutions}

Tables of the angle-averaged intensity of the radiation field computed
using the ray and moment method are available on-line for six test
problems, as summarised here. Unless stated otherwise, the maximum
frequency adopted for the continuum sources is $x_m=1000$.

\subsection{Test 1:\ Emission line source in a static medium}

An analytic solution in the diffusion approximation is provided by
\citet{DHS_06}. We provide two solutions for RII scattering, (a)\ in
the Eddington approximation and (b)\ solving the full ray and moment
equations. The solution using the full ray and moment equations is
tabulated in Table~\ref{tab:Test1}. The results in the Eddington
approximation are very similar.

\begin{table*}
\centering
\begin{minipage}{180mm}
  \caption{Test 1:\ $J(r,x)$ (in units of $10^{-9}$) for \Lya source
    in homogeneous, static medium ($T=10$~K, $\kappa_0=100$,
    $r_S=5$):\ RII redistribution}
\begin{tabular}{lrrrrrrrrr}
\hline\hline
$x$ & $r=600$& $r=650$&  $r=700$&  $r=750$&  $r=800$&  $r=850$&
$r=900$&  $r=950$&  $r=1000$\\
\hline
0.000 & 10.08642 & 7.96955 & 6.26884 & 4.87326 & 3.70389 & 2.70188 & 1.81926 & 1.00640 & 0.00030\\
0.100 & 10.08641 & 7.96955 & 6.26883 & 4.87325 & 3.70389 & 2.70188 & 1.81926 & 1.00640 & 0.00030\\
0.200 & 10.08642 & 7.96955 & 6.26884 & 4.87326 & 3.70389 & 2.70188 & 1.81926 & 1.00640 & 0.00031\\
0.300 & 10.08641 & 7.96955 & 6.26883 & 4.87325 & 3.70389 & 2.70188 & 1.81926 & 1.00640 & 0.00033\\
0.400 & 10.08642 & 7.96955 & 6.26884 & 4.87326 & 3.70389 & 2.70188 & 1.81926 & 1.00640 & 0.00035\\
0.500 & 10.08641 & 7.96955 & 6.26883 & 4.87325 & 3.70389 & 2.70188 & 1.81926 & 1.00640 & 0.00038\\
\hline
\end{tabular}
\begin{tabular}{l}
Note:\ The full table is published in the electronic version of the paper. A portion is shown here only for guidance regarding its form and content.
\end{tabular}
\label{tab:Test1}
\end{minipage}
\end{table*}

\subsection{Test 2:\ Emission line source in a uniformly expanding
  homogeneous medium}

An analytic solution in the diffusion approximation is provided by
\citet{1999ApJ...524..527L} as well as Monte Carlo code results. (Also
see Appendix~\ref{ap:hem}.) We provide a solution for coherent
scattering solving the full ray and moment equations. The results are
tabulated in Table~\ref{tab:Test2}. (Vanishingly small values of
$\log_{10}\tilde J$ are indicated by a value of $-100$.)

\begin{table*}
\centering
\begin{minipage}{180mm}
  \caption{Test 2:\ $\log_{10}\tilde J({\tilde r},{\tilde\nu})$ for
    \Lya source in uniformly expanding homogeneous medium:\ coherent
    scattering}
\begin{tabular}{lrrrrrrr}
\hline\hline
$\log_{10}{\tilde r}$ & $\log_{10}{\tilde \nu}=-1.5$& $\log_{10}{\tilde\nu}=-1.0$& $\log_{10}{\tilde\nu}=-0.5$& $\log_{10}{\tilde\nu}=0.0$& $\log_{10}{\tilde\nu}=0.5$& $\log_{10}{\tilde\nu}=1.0$& $\log_{10}{\tilde\nu}=1.5$\\
\hline
-3.000  &   5.4131  &   3.2324  &   1.0482  &  -1.0900  &  -3.1419  &  -5.1395  &  -7.1773\\
-2.987  &   5.4119  &   3.2327  &   1.0483  &  -1.0899  &  -3.1419  &  -5.1395  &  -7.1773\\
-2.975  &   5.4103  &   3.2328  &   1.0484  &  -1.0899  &  -3.1419  &  -5.1395  &  -7.1773\\
-2.963  &   5.4085  &   3.2328  &   1.0484  &  -1.0899  &  -3.1418  &  -5.1395  &  -7.1773\\
-2.950  &   5.4064  &   3.2329  &   1.0485  &  -1.0898  &  -3.1418  &  -5.1395  &  -7.1773\\
\hline
\end{tabular}
\begin{tabular}{l}
Note:\ The full table is published in the electronic version of the paper. A portion is shown here only for guidance regarding its form and content.
\end{tabular}
\label{tab:Test2}
\end{minipage}
\end{table*}

\subsection{Test 3:\ Continuum line source in a uniformly expanding
  homogeneous medium}

An analytic solution in the diffusion approximation is provided in
Appendix~\ref{ap:hem}. We provide three solutions, solving the full
ray and moment equations for (a)\ coherent scattering, (b)\ RII
redistribution, and (c)\ RII redistribution with recoils, all for a
source upper frequency cutoff of $x_m=1000$. An additional solution
for RII redistribution with recoil is also provided with
$x_m=1.4\times10^5$, corresponding to Ly$\beta$. A temperature of
$T=10$~K is assumed for all solutions. The results are tabulated in
Tables~\ref{tab:Test3a}-\ref{tab:Test3d}.

\begin{table*}
\centering
\begin{minipage}{180mm}
  \caption{Test 3a:\ $\log_{10}\tilde J({\tilde r},x)$ for continuum
    source in uniformly expanding homogeneous medium:\ coherent
    scattering}
\begin{tabular}{lrrrrrrr}
\hline\hline
$x$ & $\log_{10}{\tilde r}=-4.2$& $\log_{10}{\tilde r}=-3.9$& $\log_{10}{\tilde r}=-3.6$& $\log_{10}{\tilde r}=-3.3$& $\log_{10}{\tilde r}=-3.0$& $\log_{10}{\tilde r}=-2.7$& $\log_{10}{\tilde r}=-2.4$\\
\hline
-0.624 & 7.53670 & 6.84844 & 6.15478 & 5.45860 & 4.76086 & 4.05933 & 3.35097\\
-0.374 & 7.53669 & 6.84844 & 6.15478 & 5.45860 & 4.76086 & 4.05933 & 3.35097\\
-0.125 & 7.53668 & 6.84844 & 6.15478 & 5.45860 & 4.76086 & 4.05933 & 3.35097\\
 0.125 & 7.53668 & 6.84844 & 6.15478 & 5.45860 & 4.76086 & 4.05933 & 3.35097\\
 0.374 & 7.53667 & 6.84844 & 6.15478 & 5.45860 & 4.76086 & 4.05933 & 3.35097\\
 0.624 & 7.53667 & 6.84844 & 6.15478 & 5.45860 & 4.76086 & 4.05933 & 3.35097\\
\hline
\end{tabular}
\begin{tabular}{l}
Note:\ The full table is published in the electronic version of the paper. A portion is shown here only for guidance regarding its form and content.
\end{tabular}
\label{tab:Test3a}
\end{minipage}
\end{table*}

\begin{table*}
\centering
\begin{minipage}{180mm}
\caption{Test 3b:\ $\log_{10}\tilde J({\tilde r},x)$ for continuum source in
  uniformly expanding homogeneous medium:\ RII redistribution}
\begin{tabular}{lrrrrrrr}
\hline\hline
$x$ & $\log_{10}{\tilde r}=-4.2$& $\log_{10}{\tilde r}=-3.9$& $\log_{10}{\tilde r}=-3.6$& $\log_{10}{\tilde r}=-3.3$& $\log_{10}{\tilde r}=-3.0$& $\log_{10}{\tilde r}=-2.7$& $\log_{10}{\tilde r}=-2.4$\\
\hline
-1.248 & 7.58208 & 6.99331 & 6.36254 & 5.63254 & 4.82461 & 4.07154 & 3.36266\\
-0.749 & 7.58208 & 6.99330 & 6.36254 & 5.63255 & 4.82461 & 4.07154 & 3.36266\\
-0.250 & 7.58208 & 6.99331 & 6.36253 & 5.63254 & 4.82461 & 4.07154 & 3.36266\\
 0.250 & 7.58208 & 6.99330 & 6.36254 & 5.63255 & 4.82461 & 4.07154 & 3.36266\\
 0.749 & 7.58208 & 6.99331 & 6.36253 & 5.63254 & 4.82461 & 4.07154 & 3.36266\\
 1.248 & 7.58208 & 6.99329 & 6.36254 & 5.63254 & 4.82461 & 4.07154 & 3.36266\\
\hline
\end{tabular}
\begin{tabular}{l}
Note:\ The full table is published in the electronic version of the paper. A portion is shown here only for guidance regarding its form and content.
\end{tabular}
\label{tab:Test3b}
\end{minipage}
\end{table*}

\begin{table*}
\centering
\begin{minipage}{180mm}
\caption{Test 3c:\ $\log_{10}\tilde J({\tilde r},x)$ for continuum source in
  uniformly expanding homogeneous medium:\ RII redistribution with recoils}
\begin{tabular}{lrrrrrrr}
\hline\hline
$x$ & $\log_{10}{\tilde r}=-4.2$& $\log_{10}{\tilde r}=-3.9$& $\log_{10}{\tilde r}=-3.6$& $\log_{10}{\tilde r}=-3.3$& $\log_{10}{\tilde r}=-3.0$& $\log_{10}{\tilde r}=-2.7$& $\log_{10}{\tilde r}=-2.4$\\
\hline
-1.248 & 7.56362 & 6.94959 & 6.28331 & 5.51825 & 4.70934 & 3.97269 & 3.26682\\
-0.749 & 7.56013 & 6.94608 & 6.27983 & 5.51476 & 4.70586 & 3.96920 & 3.26333\\
-0.250 & 7.55665 & 6.94262 & 6.27634 & 5.51128 & 4.70237 & 3.96572 & 3.25985\\
 0.250 & 7.55316 & 6.93911 & 6.27286 & 5.50780 & 4.69889 & 3.96223 & 3.25636\\
 0.749 & 7.54968 & 6.93565 & 6.26937 & 5.50431 & 4.69541 & 3.95875 & 3.25288\\
 1.248 & 7.54619 & 6.93215 & 6.26589 & 5.50083 & 4.69192 & 3.95526 & 3.24939\\
\hline
\end{tabular}
\begin{tabular}{l}
Note:\ The full table is published in the electronic version of the paper. A portion is shown here only for guidance regarding its form and content.
\end{tabular}
\label{tab:Test3c}
\end{minipage}
\end{table*}

\begin{table*}
\centering
\begin{minipage}{180mm}
\caption{Test 3d:\ $\log_{10}\tilde J({\tilde r},x)$ for continuum
  source in uniformly expanding homogeneous medium:\ RII
  redistribution with recoil for $x_m=1.4\times10^5$}
\begin{tabular}{lrrrrrrr}
  \hline\hline
$x$ & $\log_{10}{\tilde r}=-2.9$& $\log_{10}{\tilde r}=-2.2$& $\log_{10}{\tilde r}=-1.5$& $\log_{10}{\tilde r}=-0.8$& $\log_{10}{\tilde r}=-0.1$& $\log_{10}{\tilde r}=0.6$& $\log_{10}{\tilde r}=1.3$\\
  \hline
-2.488 & 4.46782 & 2.82518 & 1.21451 &-0.40622 &-1.94001 &-3.42438 &-4.87538\\
-1.493 & 4.46080 & 2.81814 & 1.20748 &-0.41326 &-1.94704 &-3.43141 &-4.88241\\
-0.498 & 4.45394 & 2.81130 & 1.20064 &-0.42010 &-1.95388 &-3.43825 &-4.88925\\
 0.498 & 4.44692 & 2.80426 & 1.19359 &-0.42714 &-1.96093 &-3.44530 &-4.89630\\
 1.493 & 4.44006 & 2.79743 & 1.18676 &-0.43398 &-1.96776 &-3.45213 &-4.90313\\
 2.488 & 4.43304 & 2.79038 & 1.17971 &-0.44102 &-1.97481 &-3.45917 &-4.91018\\
  \hline
\end{tabular}
\begin{tabular}{l}
Note:\ The full table is published in the electronic version of the paper. A portion is shown here only for guidance regarding its form and content.
\end{tabular}
\label{tab:Test3d}
\end{minipage}
\end{table*}

\subsection{Test 4:\ Continuum line source in a uniformly expanding
  medium with an overdense shell}

The density profile is described by equation~(\ref{eq:dens_clump}). We
provide a solution for coherent scattering solving the full ray and
moment equations. The results are tabulated in Table~\ref{tab:Test4}.

\begin{table*}
\centering
\begin{minipage}{180mm}
  \caption{Test 4:\ $\log_{10}\tilde J({\tilde r},x)$ for continuum
    source in uniformly expanding medium with overdense shell:\
    coherent scattering}
\begin{tabular}{lrrrrrrr}
\hline\hline
$x$ & $\log_{10}{\tilde r}=-4.2$& $\log_{10}{\tilde r}=-3.9$& $\log_{10}{\tilde r}=-3.6$& $\log_{10}{\tilde r}=-3.3$& $\log_{10}{\tilde r}=-3.0$& $\log_{10}{\tilde r}=-2.7$& $\log_{10}{\tilde r}=-2.4$\\
\hline
-0.855 & 7.55695 & 6.94865 & 6.53139 & 5.85581 & 5.07235 & 4.25632 & 3.03729\\
-0.513 & 7.55693 & 6.94865 & 6.53139 & 5.85581 & 5.07235 & 4.25632 & 3.03729\\
-0.171 & 7.55692 & 6.94864 & 6.53139 & 5.85581 & 5.07235 & 4.25632 & 3.03729\\
 0.171 & 7.55692 & 6.94864 & 6.53139 & 5.85581 & 5.07235 & 4.25632 & 3.03729\\
 0.513 & 7.55691 & 6.94864 & 6.53139 & 5.85581 & 5.07235 & 4.25632 & 3.03729\\
 0.855 & 7.55691 & 6.94864 & 6.53139 & 5.85581 & 5.07235 & 4.25632 & 3.03729\\
\hline
\end{tabular}
\begin{tabular}{l}
Note:\ The full table is published in the electronic version of the paper. A portion is shown here only for guidance regarding its form and content.
\end{tabular}
\label{tab:Test4}
\end{minipage}
\end{table*}

\subsection{Test 5:\ Continuum line source in an expanding homogeneous
  medium with a quadratic velocity profile}

The velocity profile is given by equation~(\ref{eq:V_quad}). We
provide a solution for coherent scattering solving the full ray and
moment equations. The results are tabulated in Table~\ref{tab:Test5}.

\begin{table*}
\centering
\begin{minipage}{180mm}
  \caption{Test 5:\ $\log_{10}\tilde J({\tilde r},x)$ for continuum
    source in homogeneous medium with quadratic velocity profile:\
    coherent scattering}
\begin{tabular}{lrrrrrrr}
\hline\hline
$x$ & $\log_{10}{\tilde r}=-4.1$& $\log_{10}{\tilde r}=-3.8$& $\log_{10}{\tilde r}=-3.5$& $\log_{10}{\tilde r}=-3.2$& $\log_{10}{\tilde r}=-2.9$& $\log_{10}{\tilde r}=-2.6$& $\log_{10}{\tilde r}=-2.3$\\
\hline
-0.997 & 7.80416 & 7.26099 & 6.64381 & 5.91760 & 5.09385 & 4.21712 & 3.30071\\
-0.598 & 7.80408 & 7.26095 & 6.64379 & 5.91760 & 5.09385 & 4.21712 & 3.30071\\
-0.199 & 7.80405 & 7.26094 & 6.64379 & 5.91760 & 5.09385 & 4.21712 & 3.30071\\
 0.199 & 7.80403 & 7.26093 & 6.64378 & 5.91760 & 5.09385 & 4.21712 & 3.30071\\
 0.598 & 7.80401 & 7.26092 & 6.64378 & 5.91760 & 5.09385 & 4.21712 & 3.30071\\
 0.997 & 7.80399 & 7.26091 & 6.64378 & 5.91759 & 5.09385 & 4.21712 & 3.30071\\
\hline
\end{tabular}
\begin{tabular}{l}
Note:\ The full table is published in the electronic version of the paper. A portion is shown here only for guidance regarding its form and content.
\end{tabular}
\label{tab:Test5}
\end{minipage}
\end{table*}

\subsection{Test 6:\ Continuum line source in a medium with a
  self-consistent linear density and velocity perturbation around the
  source}

The density and velocity perturbations are described by
equations~(\ref{eq:delta_sph}) and (\ref{eq:V_sph}). We provide a
solution for coherent scattering solving the full ray and moment
equations, illustrated for both a linear perturbation with
$\Delta_0=0.5$ and extended into the non-linear regime with
$\Delta_0=2.9$ (the latter case is not a fully self-consistent
cosmological perturbation). The results are tabulated in
Tables~\ref{tab:Test6a} and \ref{tab:Test6b}.

\begin{table*}
\centering
\begin{minipage}{180mm}
  \caption{Test 6a:\ $\log_{10}\tilde J({\tilde r},x)$ for continuum
    source in perturbed expanding medium ($\Delta_0=0.5$):\ coherent
    scattering}
\begin{tabular}{lrrrrrrr}
  \hline\hline
$x$ & $\log_{10}{\tilde r}=-3.0$& $\log_{10}{\tilde r}=-2.5$& $\log_{10}{\tilde r}=-2.0$& $\log_{10}{\tilde r}=-1.5$& $\log_{10}{\tilde r}=-1.0$& $\log_{10}{\tilde r}=-0.5$& $\log_{10}{\tilde r}=0.0$\\
  \hline
-2.488 & 4.86634 & 3.70669 & 2.54773 & 1.39233 & 0.21623 &-0.93978 &-2.12461\\
-1.493 & 4.86612 & 3.70667 & 2.54773 & 1.39233 & 0.21623 &-0.93978 &-2.12461\\
-0.498 & 4.86612 & 3.70667 & 2.54773 & 1.39233 & 0.21623 &-0.93978 &-2.12461\\
 0.498 & 4.86612 & 3.70667 & 2.54773 & 1.39233 & 0.21623 &-0.93978 &-2.12461\\
 1.493 & 4.86612 & 3.70667 & 2.54773 & 1.39233 & 0.21623 &-0.93978 &-2.12461\\
 2.488 & 4.86612 & 3.70667 & 2.54773 & 1.39233 & 0.21623 &-0.93978 &-2.12461\\
\hline
\end{tabular}
\begin{tabular}{l}
Note:\ The full table is published in the electronic version of the paper. A portion is shown here only for guidance regarding its form and content.
\end{tabular}
\label{tab:Test6a}
\end{minipage}
\end{table*}

\begin{table*}
\centering
\begin{minipage}{180mm}
  \caption{Test 6b:\ $\log_{10}\tilde J({\tilde r},x)$ for continuum
    source in perturbed expanding medium ($\Delta_0=2.9$):\ coherent
    scattering}
\begin{tabular}{lrrrrrrr}
\hline\hline
$x$ & $\log_{10}{\tilde r}=-3.0$& $\log_{10}{\tilde r}=-2.5$& $\log_{10}{\tilde r}=-2.0$& $\log_{10}{\tilde r}=-1.5$& $\log_{10}{\tilde r}=-1.0$& $\log_{10}{\tilde r}=-0.5$& $\log_{10}{\tilde r}=0.0$\\
\hline
-2.488 & 5.92320 & 4.76664 & 3.60367 & 2.43260 & 1.21809 &-0.20383 &-1.97424\\
-1.493 & 5.92113 & 4.76643 & 3.60365 & 2.43260 & 1.21809 &-0.20383 &-1.97424\\
-0.498 & 5.92111 & 4.76642 & 3.60365 & 2.43260 & 1.21809 &-0.20383 &-1.97424\\
 0.498 & 5.92111 & 4.76642 & 3.60365 & 2.43260 & 1.21809 &-0.20383 &-1.97424\\
 1.493 & 5.92111 & 4.76642 & 3.60365 & 2.43260 & 1.21809 &-0.20383 &-1.97424\\
 2.488 & 5.92111 & 4.76642 & 3.60365 & 2.43260 & 1.21809 &-0.20383 &-1.97424\\
\hline
\end{tabular}
\begin{tabular}{l}
Note:\ The full table is published in the electronic version of the paper. A portion is shown here only for guidance regarding its form and content.
\end{tabular}
\label{tab:Test6b}
\end{minipage}
\end{table*}

\end{appendix}

%%%%%%%%%%%%%%%%%%%%%%%%%%%%%%%%%
\bigskip
\section*{acknowledgments}

%%%%%%%%%%%%%%%%%%%%%%%%%%%%%%%%%

\bibliographystyle{mn2e-eprint}
\bibliography{hm}

\begin{thebibliography}{}

\bibitem[\protect\citeauthoryear{{Auer}}{{Auer}}{1971}]{Auer_1971}
{Auer} L.~H.,  1971, Journal of Quantitative Spectroscopy and Radiative
  Transfer, 11, 573

\bibitem[\protect\citeauthoryear{{Baek}, {Di Matteo}, {Semelin}, {Combes} \&
  {Revaz}}{{Baek} et~al.}{2009}]{2009A&A...495..389B}
{Baek} S.,  {Di Matteo} P.,  {Semelin} B.,  {Combes} F.,    {Revaz} Y.,  2009,
  \aap, 495, 389

\bibitem[\protect\citeauthoryear{{Basko}}{{Basko}}{1978}]{Basko_1978}
{Basko} M.~M.,  1978, Zhurnal Eksperimental noi i Teoreticheskoi Fiziki, 75,
  1278

\bibitem[\protect\citeauthoryear{{Bouwens}, {Illingworth}, {Labbe}, {Oesch},
  {Trenti}, {Carollo}, {van Dokkum}, {Franx}, {Stiavelli}, {Gonz{\'a}lez},
  {Magee} \& {Bradley}}{{Bouwens} et~al.}{2011}]{2011Natur.469..504B}
{Bouwens} R.~J.,  {Illingworth} G.~D.,  {Labbe} I.,  {Oesch} P.~A.,  {Trenti}
  M.,  {Carollo} C.~M.,  {van Dokkum} P.~G.,  {Franx} M.,  {Stiavelli} M.,
  {Gonz{\'a}lez} V.,  {Magee} D.,    {Bradley} L.,  2011, \nat, 469, 504

\bibitem[\protect\citeauthoryear{{Chen} \& {Miralda-Escud{\'e}}}{{Chen} \&
  {Miralda-Escud{\'e}}}{2004}]{2004ApJ...602....1C}
{Chen} X.,  {Miralda-Escud{\'e}} J.,  2004, \apj, 602, 1

\bibitem[\protect\citeauthoryear{{Chuzhoy} \& {Zheng}}{{Chuzhoy} \&
  {Zheng}}{2007}]{2007ApJ...670..912C}
{Chuzhoy} L.,  {Zheng} Z.,  2007, \apj, 670, 912

\bibitem[\protect\citeauthoryear{{Dijkstra}, {Haiman} \& {Spaans}}{{Dijkstra}
  et~al.}{2006}]{DHS_06}
{Dijkstra} M.,  {Haiman} Z.,    {Spaans} M.,  2006, \apj, 649, 14

\bibitem[\protect\citeauthoryear{{Field}}{{Field}}{1958}]{1958PROCIRE.46..240F}
{Field} G.~B.,  1958, Proc. I.R.E., 46, 240

\bibitem[\protect\citeauthoryear{{Field}}{{Field}}{1959a}]{1959ApJ...129..536F}
{Field} G.~B.,  1959a, \apj, 129, 536

\bibitem[\protect\citeauthoryear{{Field}}{{Field}}{1959b}]{1959ApJ...129..551F}
{Field} G.~B.,  1959b, \apj, 129, 551

\bibitem[\protect\citeauthoryear{{Furlanetto} \& {Pritchard}}{{Furlanetto} \&
  {Pritchard}}{2006}]{2006MNRAS.372.1093F}
{Furlanetto} S.~R.,  {Pritchard} J.~R.,  2006, \mnras, 372, 1093

\bibitem[\protect\citeauthoryear{{Harrington}}{{Harrington}}{1973}]{1973MNRAS.162...43H}
{Harrington} J.~P.,  1973, \mnras, 162, 43

\bibitem[\protect\citeauthoryear{{Higgins}}{{Higgins}}{2012}]{2012HigginsPhD}
{Higgins} J.,  2012, PhD thesis, Univ. Edinburgh

\bibitem[\protect\citeauthoryear{{Higgins} \& {Meiksin}}{{Higgins} \&
  {Meiksin}}{2009}]{2009MNRAS.393..949H}
{Higgins} J.,  {Meiksin} A.,  2009, \mnras, 393, 949

\bibitem[\protect\citeauthoryear{{Komatsu}, {Smith}, {Dunkley}, {Bennett},
  {Gold}, {Hinshaw}, {Jarosik}, {Larson}, {Nolta}, {Page}, {Spergel} \&
  {Halpern}}{{Komatsu} et~al.}{2011}]{2011ApJS..192...18K}
{Komatsu} E.,  {Smith} K.~M.,  {Dunkley} J.,  {Bennett} C.~L.,  {Gold} B.,
  {Hinshaw} G.,  {Jarosik} N.,  {Larson} D.,  {Nolta} M.~R.,  {Page} L.,
  {Spergel} D.~N.,    {Halpern} M.,  2011, \apjs, 192, 18

\bibitem[\protect\citeauthoryear{{Loeb} \& {Rybicki}}{{Loeb} \&
  {Rybicki}}{1999}]{1999ApJ...524..527L}
{Loeb} A.,  {Rybicki} G.~B.,  1999, \apj, 524, 527

\bibitem[\protect\citeauthoryear{{Lucy}}{{Lucy}}{1999}]{Lucy_1999}
{Lucy} L.~B.,  1999, \aap, 344, 282

\bibitem[\protect\citeauthoryear{{Madau}, {Meiksin} \& {Rees}}{{Madau}
  et~al.}{1997}]{MMR97}
{Madau} P.,  {Meiksin} A.,    {Rees} M.~J.,  1997, \apj, 475, 429

\bibitem[\protect\citeauthoryear{{McLure}, {Dunlop}, {de Ravel}, {Cirasuolo},
  {Ellis}, {Schenker}, {Robertson}, {Koekemoer}, {Stark} \& {Bowler}}{{McLure}
  et~al.}{2011}]{2011MNRAS.418.2074M}
{McLure} R.~J.,  {Dunlop} J.~S.,  {de Ravel} L.,  {Cirasuolo} M.,  {Ellis}
  R.~S.,  {Schenker} M.,  {Robertson} B.~E.,  {Koekemoer} A.~M.,  {Stark}
  D.~P.,    {Bowler} R.~A.~A.,  2011, \mnras, 418, 2074

\bibitem[\protect\citeauthoryear{{Mihalas}}{{Mihalas}}{1978}]{1978stat.book.....M}
{Mihalas} D.,  1978, {Stellar atmospheres (2nd edition)}.
San Francisco, W.~H.~Freeman and Co., 1978.

\bibitem[\protect\citeauthoryear{{Mihalas}, {Kunasz} \& {Hummer}}{{Mihalas}
  et~al.}{1975}]{1975ApJ...202..465M}
{Mihalas} D.,  {Kunasz} P.~B.,    {Hummer} D.~G.,  1975, \apj, 202, 465

\bibitem[\protect\citeauthoryear{{Mihalas}, {Kunasz} \& {Hummer}}{{Mihalas}
  et~al.}{1976}]{1976ApJ...210..419M}
{Mihalas} D.,  {Kunasz} P.~B.,    {Hummer} D.~G.,  1976, \apj, 210, 419

\bibitem[\protect\citeauthoryear{{Mihalas}, {Kunasz} \& {Hummer}}{{Mihalas}
  et~al.}{1977}]{1977ApJ...214..337M}
{Mihalas} D.,  {Kunasz} P.~B.,    {Hummer} D.~G.,  1977, \apj, 214, 337

\bibitem[\protect\citeauthoryear{{Ono}, {Ouchi}, {Mobasher}, {Dickinson},
  {Penner}, {Shimasaku}, {Weiner}, {Kartaltepe}, {Nakajima}, {Nayyeri},
  {Stern}, {Kashikawa} \& {Spinrad}}{{Ono} et~al.}{2012}]{2012ApJ...744...83O}
{Ono} Y.,  {Ouchi} M.,  {Mobasher} B.,  {Dickinson} M.,  {Penner} K.,
  {Shimasaku} K.,  {Weiner} B.~J.,  {Kartaltepe} J.~S.,  {Nakajima} K.,
  {Nayyeri} H.,  {Stern} D.,  {Kashikawa} N.,    {Spinrad} H.,  2012, \apj,
  744, 83

\bibitem[\protect\citeauthoryear{{Pritchard} \& {Loeb}}{{Pritchard} \&
  {Loeb}}{2011}]{2011arXiv1109.6012P}
{Pritchard} J.~R.,  {Loeb} A.,  2011, ArXiv e-prints, 1109.6012

\bibitem[\protect\citeauthoryear{{Roy}, {Xu}, {Qiu}, {Shu} \& {Fang}}{{Roy}
  et~al.}{2009a}]{2009ApJ...694.1121R}
{Roy} I.,  {Xu} W.,  {Qiu} J.-M.,  {Shu} C.-W.,    {Fang} L.-Z.,  2009a, \apj,
  694, 1121

\bibitem[\protect\citeauthoryear{{Roy}, {Xu}, {Qiu}, {Shu} \& {Fang}}{{Roy}
  et~al.}{2009b}]{2009ApJ...703.1992R}
{Roy} I.,  {Xu} W.,  {Qiu} J.-M.,  {Shu} C.-W.,    {Fang} L.-Z.,  2009b, \apj,
  703, 1992

\bibitem[\protect\citeauthoryear{{Rybicki} \& {dell'Antonio}}{{Rybicki} \&
  {dell'Antonio}}{1994}]{1994ApJ...427..603R}
{Rybicki} G.~B.,  {dell'Antonio} I.~P.,  1994, \apj, 427, 603

\bibitem[\protect\citeauthoryear{{Schenker}, {Stark}, {Ellis}, {Robertson},
  {Dunlop}, {McLure}, {Kneib} \& {Richard}}{{Schenker}
  et~al.}{2012}]{2012ApJ...744..179S}
{Schenker} M.~A.,  {Stark} D.~P.,  {Ellis} R.~S.,  {Robertson} B.~E.,  {Dunlop}
  J.~S.,  {McLure} R.~J.,  {Kneib} J.-P.,    {Richard} J.,  2012, \apj, 744,
  179

\bibitem[\protect\citeauthoryear{{Semelin}, {Combes} \& {Baek}}{{Semelin}
  et~al.}{2007}]{2007A&A...474..365S}
{Semelin} B.,  {Combes} F.,    {Baek} S.,  2007, \aap, 474, 365

\bibitem[\protect\citeauthoryear{{Tasitsiomi}}{{Tasitsiomi}}{2006}]{2006ApJ...645..792T}
{Tasitsiomi} A.,  2006, \apj, 645, 792

\bibitem[\protect\citeauthoryear{{Vanzella}, {Pentericci}, {Fontana},
  {Grazian}, {Castellano}, {Boutsia}, {Cristiani}, {Dickinson}, {Gallozzi},
  {Giallongo}, {Giavalisco}, {Maiolino}, {Moorwood}, {Paris} \&
  {Santini}}{{Vanzella} et~al.}{2011}]{2011ApJ...730L..35V}
{Vanzella} E.,  {Pentericci} L.,  {Fontana} A.,  {Grazian} A.,  {Castellano}
  M.,  {Boutsia} K.,  {Cristiani} S.,  {Dickinson} M.,  {Gallozzi} S.,
  {Giallongo} E.,  {Giavalisco} M.,  {Maiolino} R.,  {Moorwood} A.,  {Paris}
  D.,    {Santini} P.,  2011, \apjl, 730, L35

\bibitem[\protect\citeauthoryear{{Vonlanthen}, {Semelin}, {Baek} \&
  {Revaz}}{{Vonlanthen} et~al.}{2011}]{2011A&A...532A..97V}
{Vonlanthen} P.,  {Semelin} B.,  {Baek} S.,    {Revaz} Y.,  2011, \aap, 532,
  A97

\bibitem[\protect\citeauthoryear{{Wouthuysen}}{{Wouthuysen}}{1952}]{1952AJ.....57R..31W}
{Wouthuysen} S.~A.,  1952, \aj, 57, 31

\bibitem[\protect\citeauthoryear{{Zheng} \& {Miralda-Escud{\'e}}}{{Zheng} \&
  {Miralda-Escud{\'e}}}{2002}]{2002ApJ...578...33Z}
{Zheng} Z.,  {Miralda-Escud{\'e}} J.,  2002, \apj, 578, 33

\end{thebibliography}

\label{lastpage}

\end{document}